\documentclass[preprint]{elsarticle}
\clearpage{}\usepackage{xspace}

\newcommand{\range}[2] {\rangec{#1}{#2}{,}}
\newcommand{\rangec}[3] {\ensuremath{#1#3...#3#2}}

\newcommand{\srange}[3] {\range{#1_{#2}}{#1_{#3}}}

\newcommand{\ion}[2]{{\mathsf{#1}}_{#2}}
\newcommand{\ionP}[3]{{\mathsf{#1(}#3\mathsf{)}}_{#2}}

\newcommand{\rrP}[2]{{\mathtt{#1(}#2\mathsf{)}}}
\newcommand{\rr}[1]{\mathtt{#1}}
\clearpage{}

\usepackage{amsmath,amsthm,amssymb,amsfonts}
\usepackage{mathtools}
\usepackage{marvosym}
\usepackage{stmaryrd}

\usepackage[ruled,vlined,linesnumbered]{algorithm2e}
\SetKwFor{ForEach}{for each}{do}{end}
\SetKwFor{ForAll}{for all}{do}{end}
\SetKwProg{Fn}{function}{}{}
\SetKwProg{Pr}{procedure}{}{}
\DontPrintSemicolon
\usepackage{booktabs}
\usepackage{makecell}
\usepackage{float}
\usepackage{textcomp}
\usepackage[capitalise, nameinlink, sort]{cleveref}

\theoremstyle{plain}

\newtheorem{theorem}{Theorem}
\newtheorem{lemma}{Lemma}
\theoremstyle{remark}

\usepackage{xspace}
\usepackage{comment}
\newcommand{\eg}{e.g.~}
\newcommand{\ie}{i.e.~}
\makeatletter
\newcommand*{\etc}{\@ifnextchar{.}{etc}{etc.\@\xspace}}
\makeatother

\usepackage{fontawesome}
\usepackage{tikz}
\usetikzlibrary{calc}
\usetikzlibrary{fit}
\usetikzlibrary{calc}
\usetikzlibrary{shapes}
\usetikzlibrary{shapes.geometric}
\usetikzlibrary{shapes.multipart}
\usetikzlibrary{shapes.gates.logic.US}
\usetikzlibrary{arrows}
\usetikzlibrary{arrows.meta}

\tikzset{
  big edge/.style={
    green,
    thick,
  },
  big edgep/.style={
    big edge,
    -{Circle[fill=black,black,width=2,length=2,sep=-1]}
  },
  big pedge/.style={
    big edge,
    {Circle[fill=black,black,width=2,length=2,sep=-1]}-
  },
  big pedgep/.style={
    big edge,
    {Circle[fill=black,black,width=2,length=2,sep=-1]}-{Circle[fill=black,black,width=2,length=2,sep=-1]}
  },
  big edgec/.style={
    big edge,
    -{Bar[fill=green,green,width=4,length=0,sep=0]}
  },
  big pedgec/.style={
    big edge,
    {Circle[fill=black,black,width=2,length=2,sep=-1]}-{Bar[fill=black,black,width=4,length=0,sep=0]}
  },
  big region/.style={
    draw,
    rectangle,
    rounded corners=1.5,
    dashed,
    dash pattern=on 1pt off 1pt,
    thin,
    gray,
  },
  big site/.style={
    big region,
    fill=gray!60,
    text=black,
  },
  big react/.style={
    black,
    thick,
    -stealth,
    line width=3,
    shorten <=3,
    shorten >=3,
  },
  big react rev/.style={
    black,
    thick,
    stealth-stealth,
    line width=3,
    shorten <=3,
    shorten >=3,
  },
  lbl/.style={
    font=\tiny\sf,
    inner sep=1,
  },
  lbl conc/.style={
    font=\tiny,
    inner sep=1,
  }
}
 \tikzset{
  event/.style={
isosceles triangle,
    shape border rotate=90,
    draw,
    inner sep=1.2,
  },
  reducing/.style={
    red
  },
  reducing site/.style={
    big site,
    red,
    thick,
    fill=gray!60,
  },
}
 
\usepackage{subcaption}
\newcommand\id{\mathsf{id}}

\DeclareMathOperator{\react}{\mathrel{\frac{\raisebox{0.75mm}{\begin{scriptsize}\ensuremath{\hspace*{1mm}\ \hspace*{1mm}}\end{scriptsize}}}{}} \joinrel{\!\!\vartriangleright}}
\newcommand{\xreact}[1]{\operatorname{\mathrel{\frac{\raisebox{0.75mm}{\begin{scriptsize}\ensuremath{\hspace*{1mm}\ #1 \hspace*{1mm}}\end{scriptsize}}}{}} \joinrel{\!\!\vartriangleright}}}
\newcommand{\xreacts}[1]{\operatorname{\mathrel{\frac{\raisebox{0.75mm}{\begin{scriptsize}\ensuremath{\hspace*{1mm}\ #1 \hspace*{1mm}}\end{scriptsize}}}{}} \joinrel{\!\!\vartriangleright^*}}}

\DeclareMathOperator{\rrul}{\mathrel{\frac{\raisebox{0.75mm}{\begin{scriptsize}\ensuremath{\hspace*{1mm}\ \hspace*{1mm}}\end{scriptsize}}}{}} \joinrel{\!\!\blacktriangleright}}

\newcommand*{\defeq}{\stackrel{\text{def}}{=}}

\newcommand{\enc}[1]{\llbracket #1 \rrbracket}
\newcommand*{\ltribrace}{\llbracket \mskip-5mu\llbracket}
\newcommand*{\rtribrace}{\rrbracket\mskip-5mu\rrbracket}
\newcommand{\encR}[1]{\ltribrace #1 \rtribrace}

\newcommand{\mprod}{\mid}
\newcommand{\pprod}{\parallel}

\newcommand{\CAN}{\textsc{Can}\xspace}

\usepackage[epsilon]{backnaur}
\makeatletter

\renewcommand\bnf@tsfont[1]{\texttt{#1}}
\makeatother

\usepackage[disable]{todonotes}

\begin{document}
	\title{Modelling and Verifying BDI Agents with Bigraphs}

\author[1]{Blair Archibald}
\ead{blair.archibald@glasgow.ac.uk}
\author[1]{Muffy Calder}
\ead{muffy.calder@glasgow.ac.uk}
\author[1]{Michele Sevegnani}
\ead{michele.sevegnani@glasgow.ac.uk}
\author[1]{Mengwei Xu\corref{cor1}}
\ead{mengwei.xu@glasgow.ac.uk}

\cortext[cor1]{Corresponding Author}
\address[1]{School of Computing Science, University of Glasgow, UK}

	\begin{abstract}

   The Belief-Desire-Intention (BDI) architecture is a
popular framework for rational agents;   most verification approaches
are based on reasoning about implementations of BDI programming languages. 
We investigate an alternative approach based on reasoning about BDI agent semantics, through a model of the execution of an agent program.    
We employ Milner’s bigraphs as the modelling   framework and present an encoding 
for the Conceptual Agent Notation (CAN) language – a superset
of AgentSpeak featuring declarative goals, concurrency, and
failure recovery.
 
We provide an encoding of   the syntax and semantics of \CAN agents, and  give  a rigorous proof that  the encoding is faithful.
Verification  is based on the  use of  mainstream software tools including BigraphER, and a small case study verifying several properties of Unmanned Aerial Vehicles (UAVs) illustrates the framework in action. 
The {\em executable} framework is a foundational step  that will enable   more advanced reasoning such as plan preference,  intention priorities and trade-offs, and  interactions with an environment under uncertainty.

\end{abstract}


\maketitle

\section{Introduction}\label{sec:introduction}

The Belief-Desire-Intention (BDI)~\cite{bratman:intention} architecture is a popular and well-studied rational agent framework  and forms the basis of,  among others, AgentSpeak~\cite{s:agentspeak}, Artificial Autonomous Agents Programming Language~(3APL)~\cite{hindriks1999agent}, A Practical Agent Programming Language~(2APL)~\cite{dastani:practical}, Jason~\cite{h:programming}, and Conceptual Agent Notation~(\CAN)~\cite{sardina:hierarchical}.
	In a BDI agent, the {\em (B)eliefs} represent what the agent knows, the {\em (D)esires} what the agent wants to bring about, and the {\em (I)ntentions} those desires the agent has chosen to act upon. 
	BDIs have been very successful in many areas such as business~\cite{s:making}, healthcare~\cite{braubach:negotiation}, and  engineering~\cite{dj:multi}.

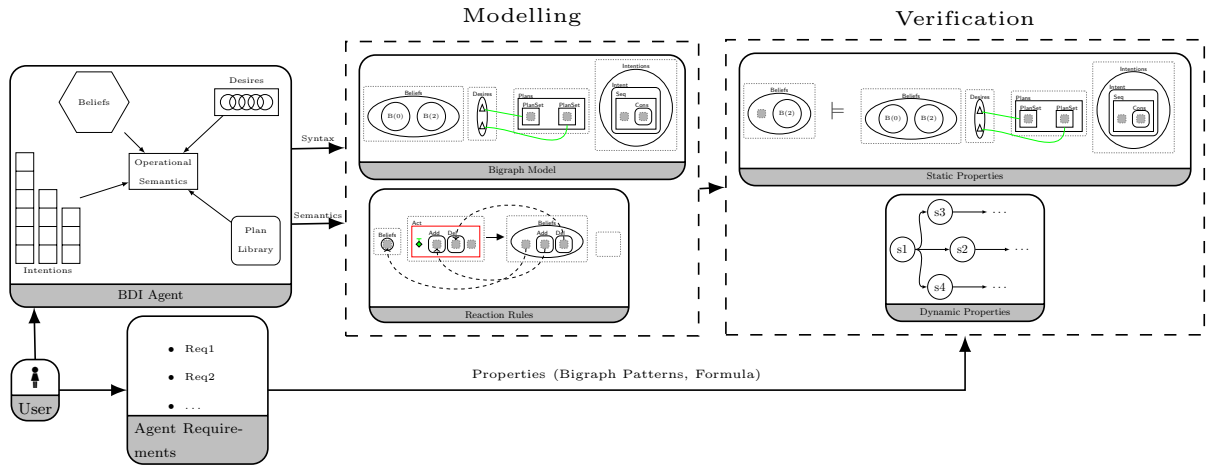
\begin{figure}
\hspace{-60px}\resizebox{1.4\linewidth}{!}{
	\newsavebox\actrr
\sbox{\actrr}{\begin{tikzpicture}[]
  \begin{scope}[local bounding box=lhs]
\node[big site] (s1_l) {};
    \node[ellipse, inner sep=0, draw, fit=(s1_l)] (bs) {};
    \node[anchor=south, inner sep=0.3] at (bs.north) (bs_lbl) {\tiny \sf Beliefs};
    \node[big region, fit=(bs)(bs_lbl)] (r1) {};

\node[diamond, draw, fill=green, inner sep=1.2, right=0.7 of s1_l] (cr) {};
    \draw[big edgec] (cr.north) to[in=-90, out=90] ($(cr.north) + (0,0.08)$) {};

    \node[big site, right=0.3 of cr] (s2_l) {};
    \node[draw, rounded corners, fit=(s2_l)] (add) {};
    \node[anchor=south west, inner sep=0.3] at (add.north west) (add_lbl) {\tiny \sf Add};

    \node[big site, right=0.3 of s2_l] (s3_l) {};
    \node[draw, rounded corners, fit=(s3_l)] (del) {};
    \node[anchor=south west, inner sep=0.3] at (del.north west) (del_lbl) {\tiny \sf Del};

    \node[big site, right=0.2 of s3_l] (s4_l) {};
    \node[draw, red, fit=(cr)(add)(add_lbl)(del)(del_lbl)(s4_l)] (act) {};
    \node[anchor=south west, inner sep=0.3] at (act.north west) (act_lbl) {\tiny \sf Act};
    \node[big region, fit=(act)(act_lbl)] (r2) {};
  \end{scope}

  \begin{scope}[shift={(4,0)}, local bounding box=rhs]
\node[big site] (s1_r) {};

    \node[big site, right=0.3 of s1_r] (s2_r) {};
    \node[draw, rounded corners, fit=(s2_r)] (add) {};
    \node[anchor=south west, inner sep=0.3] at (add.north west) (add_lbl) {\tiny \sf Add};

    \node[big site, right=0.3 of s2_r] (s3_r) {};
    \node[draw, rounded corners, fit=(s3_r)] (del) {};
    \node[anchor=south west, inner sep=0.3] at (del.north west) (del_lbl) {\tiny \sf Del};

    \node[ellipse, inner sep=0, draw, fit=(s1_r)(add)(add_lbl)(del)(del_lbl)] (bs) {};
    \node[anchor=south, inner sep=0.3] at (bs.north) (bs_lbl) {\tiny \sf Beliefs};
    \node[big region, fit=(bs)(bs_lbl)] (r1) {};

\node[big region, minimum width=20, minimum height=20, right=0.8 of s3_r] (r1) {};
  \end{scope}

  \draw[->, dashed] (s1_r) to[out=-90, in=-90] (s1_l);
  \draw[->, dashed, looseness=1] (s2_r) to[out=-90, in=-90] (s2_l);
  \draw[->, dashed, looseness=1.1] (s3_r) to[out=90, in=90] (s3_l);

  \node[] at ($(lhs.east)!0.5!(rhs.west) + (0,0)$) {$\rrul$} ;
\end{tikzpicture}
 }\newsavebox\bigmodel
\sbox{\bigmodel}{\begin{tikzpicture}
\begin{scope}
          \node[circle, draw] (bn) {\tiny B(0)};
          \node[circle, draw, right=0.2 of bn] (bn2) {\tiny B(2)};
          \node[ellipse, inner sep=0, draw, fit=(bn)(bn2)] (bs) {};
          \node[anchor=south, inner sep=0.3] at (bs.north) (bs_lbl) {\tiny \sf Beliefs};
          \node[big region, fit=(bs)(bs_lbl)] (rb) {};
        \end{scope}

\begin{scope}[shift={(2.5,0.2)}]
          \node[event] (e1) {};
          \node[event, below=0.3 of e1] (e2) {};
          \node[ellipse, inner sep=1, draw, fit=(e1)(e2)] (ds) {};
          \node[anchor=south, inner sep=0.3] at (ds.north) (ds_lbl) {\tiny \sf Desires};
          \node[big region, fit=(ds)(ds_lbl)] (rd) {};
        \end{scope}

\begin{scope}[shift={(6.6,0)}]
          \node[big site] (s1_l) {};
          \node[big site, right=0.3 of s1_l] (s2_l) {};
          \node[draw, rounded corners, fit=(s2_l)] (cons) {};
          \node[anchor=south west, inner sep=0.3] at (cons.north west) (cons_lbl) {\tiny \sf Cons};
          \node[draw, fit=(s1_l)(cons)(cons_lbl)] (seq) {};
          \node[anchor=south west, inner sep=0.3] at (seq.north west) (seq_lbl) {\tiny \sf Seq};
          \node[draw, rounded corners, fit=(seq)(seq_lbl)] (intent) {};
          \node[anchor=south west, inner sep=0.3] at (intent.north west) (intent_lbl) {\tiny \sf Intent};

          \node[ellipse, inner sep=1, draw, fit=(intent)(intent_lbl)] (is) {};
          \node[anchor=south, inner sep=0.3] at (is.north) (is_lbl) {\tiny \sf Intentions};
          \node[big region, fit=(is)(is_lbl)] (ri) {};
        \end{scope}

\begin{scope}[shift={(3.9,0)}]
          \node[big site] (s1) {};
          \node[big site, right=0.8 of s1] (s2) {};
          \node[draw, fit=(s1)] (ps1) {};
          \node[anchor=south west, inner sep=0.3] at (ps1.north west) (ps1_lbl) {\tiny \sf PlanSet};

          \node[draw, fit=(s2)] (ps2) {};
          \node[anchor=south west, inner sep=0.3] at (ps2.north west) (ps2_lbl) {\tiny \sf PlanSet};

          \node[draw, fit=(ps1)(ps1_lbl)(ps2)(ps2_lbl)] (plans) {};
          \node[anchor=south west, inner sep=0.3] at (plans.north west) (plans_lbl) {\tiny \sf Plans};

          \node[big region, fit=(plans)(plans_lbl)] (r1) {};
        \end{scope}

        \draw[big edge] (ps1.west) to[] (e1.east);
        \draw[big edge] (ps2.south) to[out=-90, in=0] (e2);

      \end{tikzpicture}
}

\newsavebox\agent
\sbox{\agent}{\begin{tikzpicture}[]
\node[draw, align=center] (spirals) {\tiny Operational \\ \tiny Semantics};

  \begin{scope}[shift={($(spirals) + (-2,-2)$)}, local bounding box=intentions]
    \node[draw, rectangle split, rectangle split parts=3, anchor=south] (n1) {};
    \node[draw, rectangle split, rectangle split parts=4, left=0.5 of n1.south, anchor=south] (n2) {};
    \node[draw, rectangle split, rectangle split parts=6, left=0.5 of n2.south, anchor=south] (n3) {};
  \end{scope}
  \begin{scope}[shift={($(intentions.south) + (0,-0.15)$)}]
    \node [] (os) {\tiny Intentions};
  \end{scope}

  \begin{scope}[shift={($(spirals) + (-1.5,1.5)$)}, local bounding box=beliefs]
    \node[draw, regular polygon, regular polygon sides=6] (n) {\tiny Beliefs};
  \end{scope}

  \begin{scope}[shift={($(spirals) + (1.2,1.5)$)}, local bounding box=desires]
    \foreach \i in {1,...,5}{
      \node[draw, circle] at (\i*0.2,0) (c_\i) {};
    }
    \node[draw, fit=(c_1)(c_2)(c_3)(c_4)(c_5)] (border) {};
    \node[anchor=south] at (border.north) {\tiny Desires};
  \end{scope}

  \begin{scope}[shift={($(spirals) + (2,-1.5)$)}, local bounding box=plans]
    \node[draw, rounded corners, minimum width=30, minimum height=30, align=center] (n) {\tiny Plan \\ \tiny Library};
  \end{scope}

  \draw[-latex] (plans) -- (spirals);
  \draw[-latex] (desires) -- (spirals);
  \draw[-latex] (beliefs) -- (spirals);
  \draw[-latex] (intentions) -- (spirals);
\end{tikzpicture}
}

\newsavebox\static
\sbox{\static}{\begin{tikzpicture}
    \begin{scope}[local bounding box=pattern]
      \node[big site] (bn) {};
      \node[circle, draw, right=0.2 of bn] (bn2) {\tiny B(2)};
      \node[ellipse, inner sep=0, draw, fit=(bn)(bn2)] (bs) {};
      \node[anchor=south, inner sep=0.3] at (bs.north) (bs_lbl) {\tiny \sf Beliefs};
      \node[big region, fit=(bs)(bs_lbl)] (rb) {};
    \end{scope}
    \node[right=0.3 of pattern.east] (models) {\Large $\models$};
    \node[right=1.2 of pattern.east] (bigraph) {
      \usebox{\bigmodel}
    };
  \end{tikzpicture}
}

\newsavebox\dynamic
\sbox{\dynamic}{\begin{tikzpicture}[remember picture]
    \node[draw, circle] (s1) {s1};
    \node[draw, circle, right=of s1] (s2) {s2};
    \node[draw, circle, above right=0.8 of s1] (s3) {s3};
    \node[draw, circle, below right=0.8 of s1] (s4) {s4};
    \node[circle, right=of s2] (s5) {$\dots$};
    \node[circle, right=of s3] (s6) {$\dots$};
    \node[circle, right=of s4] (s7) {$\dots$};

    \draw[-latex] (s1) to[out=0, in=-180] (s2);
    \draw[-latex] (s1) to[out=0, in=-180] (s3);
    \draw[-latex] (s1) to[out=0, in=-180] (s4);
    \draw[-latex] (s2) to[out=0, in=-180] (s5);
    \draw[-latex] (s3) to[out=0, in=-180] (s6);
    \draw[-latex] (s4) to[out=0, in=-180] (s7);
  \end{tikzpicture}
}

\begin{tikzpicture}
  \tikzset{
    grayb/.style = {
      draw,
      rounded corners,
      rectangle split,
      rectangle split parts=2,
      rectangle split part fill={white,gray!50}}
  }

  \begin{scope}[local bounding box=usr, scale=0.5, every node/.append style={transform shape}]
    \node[grayb] {\Large \Ladiesroom \nodepart{two} \footnotesize User};
  \end{scope}

  \begin{scope}[shift={($(usr.north) + (1, 1.5)$)}, scale=0.4, every node/.append style={transform shape}, local bounding box=agent]
    \node[grayb] {
      \usebox{\agent}
      \nodepart{two} \footnotesize BDI Agent};
  \end{scope}

  \begin{scope}[shift={($(usr.east) + (1.2, 0)$)}, scale=0.5, every node/.append style={transform shape}, local bounding box=reqs]
    \node[grayb, text width=2.2cm] {
      {
      \tiny
      \begin{itemize}
        \item Req1
        \item Req2
        \item \dots
      \end{itemize}}
      \nodepart{two} \scriptsize Agent Requirements};
  \end{scope}

\begin{scope}[shift={($(agent.east) + (1.8, -0.6)$)}, scale=0.3, every node/.append style={transform shape}, local bounding box=rrs]
    \node[grayb] (r_ruls) {
      \usebox{\actrr}
      \nodepart{two} \footnotesize Reaction Rules};
  \end{scope}

\begin{scope}[shift={($(agent.east) + (2, 0.6)$)}, scale=0.3, every node/.append style={transform shape}, local bounding box=big]
    \node[grayb] (big_model) {
      \usebox{\bigmodel}
      \nodepart{two} \footnotesize Bigraph Model};
  \end{scope}

\node[draw, dashed, fit=(rrs)(big)] (modelling) {};
  \node[anchor=south] at (modelling.north) (modelling_lbl) {\tiny Modelling};

  \begin{scope}[shift={($(modelling.east) + (2.3, 0.6)$)}, scale=0.3, every node/.append style={transform shape}, local bounding box=verification_static]
    \node[grayb] (big_model) {
      \usebox{\static}
      \nodepart{two} \footnotesize Static Properties};
  \end{scope}

  \begin{scope}[shift={($(modelling.east) + (2.3, -0.6)$)}, scale=0.3, every node/.append style={transform shape}, local bounding box=verification_dynamic]
    \node[grayb] (dynamic) {
      \usebox{\dynamic}
      \nodepart{two} \footnotesize Dynamic Properties};
  \end{scope}

\node[draw, dashed, fit=(verification_static)(verification_dynamic)] (verification) {};
  \node[anchor=south] at (verification.north) (verification_lbl) {\tiny Verification};

  \draw[-latex] (usr) -- (reqs);
  \draw[-latex] (usr.north) -- (agent.226);
  \draw[-latex] (agent.15) -- (modelling.167) node[midway, above, scale=0.4] {\tiny Syntax};
  \draw[-latex] (agent.345) -- (modelling.191) node[midway, above, scale=0.4] {\tiny Semantics};

  \draw[-latex] (modelling.east) -- (verification.west);
  \draw[-latex] (reqs.east) -| (verification.south) node[near start, above, scale=0.6] {\tiny Properties (Bigraph Patterns, Formula)};

\end{tikzpicture}
   }
  \caption{Modelling and Verification Framework for BDI Agents.}
  \label{fig:bdimodelliverificationframework}
\end{figure}

	The deployment of autonomous systems in real-world  applications raises  concerns of trustworthiness and safety, for example in scenarios such as autonomous  control in space~\cite{1656029} and human-robot interaction in healthcare~\cite{lestingi2020formal}.
	There is a growing demand for verification techniques to aid analysis of  behaviours   in increasingly complex and critical domains, 
and there has been a  proliferation of techniques and languages supporting BDI agent verification.
  Most of these approaches, however, focus on verifying the \emph{implementations} of BDI programming languages.
  For example, recent work implements a BDI agent programming language as a set of Java classes -- the Agent Infrastructure Layer (AIL) \cite{dennis2008flexible} -- that can be verified using the Java PathFinder~\cite{visser2003model} program model checker.
  While verifying an implementation tells you how the system \emph{will} operate, it might not correspond to how it \emph{should} operate with respect to the semantics of the given BDI programming language.

  We   present an approach that reasons on the \emph{semantics} of the BDI programming language (rather than the implementation of it) through a mathematical model of the execution of the agent program, i.e.~verified executable semantics.

Our approach models and verifies BDI agents, specified in the \CAN language, by encoding them as an instance of Milner's Bigraphical Reactive Systems (BRS)~\cite{milner2008bigraphs}.
\CAN features a high-level agent programming language that captures the essence of BDI concepts without describing implementation details such as data structures.
As a superset of AgentSpeak, \CAN includes advanced BDI agent behaviours such as reasoning with \emph{declarative goals}, \emph{concurrency}, and \emph{failure recovery}. 
Importantly, although we focus on \CAN, the language features are similar to those of other mainstream BDI languages and the same modelling techniques would apply to other BDI programming languages.

Bigraphs provide a meta-modelling framework that has been developed as a unifying theory for existing calculi, \eg $\pi$-calculus~\cite{bundgaard2006typed}. 
As a graph-based rewriting formalism, over rules called reaction rules, Bigraphs provide an intuitive diagrammatic representation, which is ideal for visualising the execution process of \CAN. Support for sharing, probabilistic, and conditional rewriting extensions~\cite{TCS2015,prob2020,archibald2020conditional} in Bigraphs further enables the increases of expressiveness (when required) for more advanced reasoning, e.g. probabilistic reasoning. For analysis, BigraphER~\cite{sevegnani2016bigrapher} is a freely available tool for working with bigraph models including rewriting, verification based on bigraph patterns, and transition system export to model checking tools~\cite{benford2016lions}.

Our  bigraph encoding of the \CAN language  includes: i) a structural encoding that maps the syntax of \CAN (\eg beliefs, plans, and intentions) into equivalent bigraphs, and ii) an encoding of the operational semantics of \CAN as a set of reaction rules.
We provide a correctness proof that the translation of \CAN semantics into reaction rules is \emph{faithful}.

The framework is  depicted in \cref{fig:bdimodelliverificationframework}.
 On the left we have the BDI agents and  agent requirements -- logical formulas.  In the middle, Modelling, the agents are translated 
 into bigraphs that capture the structural elements  and reaction rules that capture their dynamics.
 Once we have encoded the agent into bigraphs, we use the model to perform verification (on the right) of user-specified \emph{agent requirements}. Verification takes two forms: checking   static properties of a state (the current bigraph representing an agent at some point in its execution) through bigraph patterns, and   checking dynamic properties, expressed   as temporal logic properties, against    the   transition system generated by  BigraphER.   
 Finally,  the  user can employ BigraphER simply  to  ``run'' their agent model with different initial settings. 
 
We illustrate the framework with a small case study based on Unmanned Aerial Vehicles (UAVs).

We make the following contributions:
\begin{itemize}
		\item an encoding of the  \CAN language and  operational semantics in bigraphs, using   regions to represent the perspectives of {\em Belief}, {\em Desire},   {\em Intention},   and {\em Plan},
		\item proof that the encoding is faithful by showing  each \CAN semantic rule   is encoded by a (finite) sequence of reaction rules,
		\item an illustration  of our framework in a UAV case study, 
		\item a reflection on   aspects of   \CAN   based on insights gained  from  the encoding,
		\item   
		an overview of how we will build upon this foundation in future to  reason  about plan selection, intention tradeoffs and  priorities, and  interactions with an uncertain environment. 
	\end{itemize} 
	
	The paper is organised as follows: in~\cref{sec:preliminaries}, we recall preliminaries of BDI agents in the \CAN language and bigraphs; in~\cref{sec:structuralEncoding}, we provide the structural encoding that maps the syntax of \CAN   into equivalent bigraphs; in~\cref{sec:core}, we present a comprehensive review of the core semantics of \CAN (excluding concurrency and declarative goals) and, in particular, how the operation of \CAN semantics can be viewed as AND/OR trees.
	In~\cref{sec:corebigraphs}  we encode the semantics given in~\cref{sec:core} and 
	in~\cref{sec:extended}, we present the semantics for concurrency and declarative goals and provide their bigraphical encodings. 
	In~\cref{sec:case} and~\cref{sec:caseproperties}, we illustrate our framework with examples and  
	in~\cref{sec:reflections}, we  reflect  on aspects   of \CAN.
	In \cref{sec:related}  we discuss related work; in \cref{sec:Future Work} we lay out the our plans of for future extensions to this work;    we conclude in \cref{sec:conclusions}.

	\section{Preliminaries}\label{sec:preliminaries}

We give an overview of   BDI agents, described in the Conceptual Agent Notation (\CAN) language, as well as Bigraphs and Bigraphical Reactive Systems (BRS).

\subsection{BDI Agents}
 
A BDI agent has an explicit representation of beliefs, desires, and intentions. 
The beliefs correspond to what the agent believes about the environment,
while the desires are a set of \emph{external} events that the agent can respond to. 
To responds to those events, the agent selects an appropriate plan (given its beliefs) from the pre-defined plan library and commits to the selected plan   by turning it into a new intention. 

\CAN  is a superset of AgentSpeak~\cite{s:agentspeak}  featuring the same core operational semantics, along with several additional appealing features: declarative goals, concurrency, and failure handling. In the following, we introduce the syntax of \CAN,   the semantics are given in \cref{sec:core}.

A \CAN agent  consists of a belief base~$ \mathcal{B} $ and a plan library~$ \Pi $. 
The \emph{belief base} $\mathcal{B} $ is a set of formulas encoding the current beliefs. 
Without loss of generality, we specify our belief base following the logical language in AgentSpeak~\cite{s:agentspeak} that takes the form $ \varphi ::= b \mid \neg b \mid (\varphi_1 \wedge \varphi_2) \mid \mathit{true} \mid \mathit{false} $. 
More complex logics are possible but are outwith the scope of this paper, \ie we show how to encode general BDI agents in bigraphs, not how to encode specific logics.
All that we assume for any chosen logical language is that it has   belief operators  to check whether a belief formula $ \varphi $ follows from the belief base (i.e. $ \mathcal{B} \models \varphi ) $,  to add a belief atom $ b $ to a belief base $ \mathcal{B} $ (i.e. $ \mathcal{B} \cup \{b\}$), and   to delete a belief atom from a belief base (i.e. $\mathcal{B} \setminus \{b\} $). 

A \emph{plan library} $ \Pi  $ contains the operational procedures of an   agent and is a finite collection of plans of the form $ Pl = e : \varphi \leftarrow P$ with $ Pl $ the plan identifier, $ e $ the triggering event, $ \varphi $ the context condition, and $ P $ the plan-body.  
The triggering event $ e  $ specifies why the plan is triggered, the context condition  $ \varphi $ determines \emph{when} the plan-body $ P $ is able to handle the   event.  
We denote the triggering event of a plan $ Pl $ $ \textit{trigger}(Pl) $ and we   call $ E = \{ \textit{trigger}(Pl)\mid Pl \in \Pi \}$ the \emph{event set} that the agent knows how to respond to (\ie it has plans for response -- though it might be the case none are applicable). 
For convenience, we call the set of events  from the external environment the external event set, denoted   $E^{e}$. 
Finally, the remaining events (which occur as a part of the plan-body) are   either sub-events or internal events. 

By convention (e.g. in~\cite{h:programming}), the plan-body $ P $ in a plan $ Pl = e : \varphi \leftarrow P$ may be referred to as the {\em program} or {\em agent program}  
and  has the following syntax: \begin{center}
	\footnotesize
	$  P ::= \  \mathit{act} \ | \ ?\varphi \ | \ +\mathit{b} \ | \ -\mathit{b} \ | \ e  \ | \ \mathit{P}_{1};\mathit{P}_{2} \ | \ \mathit{P}_{1}\parallel\mathit{P}_{2} \ | \ \mathit{goal}(\varphi_{s}, \mathit{P}, \varphi_{f}) $ 
\end{center}
with $ act $ an action,  $ ?\varphi  $ a test for $ \varphi $ entailment in the belief base, $ +b $ and $ -b $ represent belief addition and deletion, and $ e  $ is a sub-event (i.e. internal event).   
Actions $ act $ take the form $act = \varphi \leftarrow \langle \phi^{+}, \phi^{-} \rangle$, where $ \varphi  $ is the pre-condition, and   $\phi^{+} $ and $ \phi^{-} $ are the addition and  deletion sets (resp.) of belief atoms, i.e.  a belief base $ \mathcal{B} $ is revised with    addition and deletion sets     $ \phi^{+}$ and  $ \phi^{-}$ to be $ (\mathcal{B} \setminus \phi^{-}) \cup \phi^{+}$. 
In addition, there are   composite programs   $ \mathit{P}_{1};\mathit{P}_{2} $ for sequence and $ \mathit{P}_{1}\parallel\mathit{P}_{2} $ for interleaved concurrency. 
Finally, a declarative goal program  $ \mathit{goal}(\varphi_{s}, \mathit{P}, \varphi_{f}) $ expresses that the declarative goal $ \varphi_{s} $ should be achieved through   program $ P $, failing if $ \varphi_{f} $ becomes true, and retrying as long as neither $ \varphi_{s} $ nor $ \varphi_{f} $  is true (see in~\cite{sardina:agent} for details).  
Additionally, there are    auxiliary program forms that are used internally when assigning semantics  to programs, namely $ nil $,  the empty program, and $ \mathit{P}_{1} \rhd \mathit{P}_{2} $ that executes $ \mathit{P}_{2}$ if the case that  $\mathit{P}_{1} $ fails. 

When a plan $ Pl = e : \varphi \leftarrow P$ is selected to respond to an event, its plan-body $ P $ is  adopted as an intention in the intention base $ \Gamma$ (a.k.a. the partially executed plan-body).
Finally,    we assume   a plan library does not have recursive plans (thus avoiding potential infinite state space). 

\subsubsection{Running Example -- Conference Travel Agent}

For illustration, we give a classic example -- arranging a conference trip -- as shown in~\cref{fig:running_example_CAN}.

				\begin{figure}
		\begin{align*}
&\textbf{Conference Travel Agent}\\
&1 \ \text{Belief base: } b_1, b_2, b_6, b_7\\
&2  \   \text{External events: }  e_1\\
&3 \ \text{Plan library: }\\
&4 \  Pl_1 = e_1: \varphi_{1} \leftarrow  act_1; act_2 \text{ s.t. }  \varphi_{1} = b_1 \wedge  b_2 \\
&5 \  Pl_2 = e_1: \varphi_{2} \leftarrow  act_3;  e_2; act_4 \text{ s.t. }  \varphi_{2} = b_6 \wedge  b_7 \\
&6 \  Pl_3 =  e_2: \varphi_{3} \leftarrow  act_5;  act_6  \text{ s.t. } \varphi_{3} = b_8\\
&7 \ \text{Actions}\\
&8  \ \ act_1 =  b_3 \leftarrow \langle \{b_4\}, \emptyset\rangle\\
&9  \ \ act_2 =  b_4 \leftarrow \langle \{b_5\}, \emptyset\rangle\\
&10 \  act_3 =  true \leftarrow \langle \{b_8\}, \emptyset\rangle\\
&11  \ act_4 =  b_9 \leftarrow \langle \{b_5\}, \emptyset\rangle\\
&12  \ act_5 =  b_8 \leftarrow \langle \{b_{10}\}, \emptyset\rangle\\
&13  \ act_6=  b_{10}\leftarrow \langle \{b_9\}, \{b_8, b_{10}\}\rangle\\
 & \text{where } e_1  \text{ stands for } \texttt{conference\_travelling},  \ e_2 \text{  for } \texttt{get\_onboard}, \\
 &  act_1 \text{  for } \texttt{start\_car},  \ act_2 \text{  for } \texttt{driving}, \ act_3 \text{  for } \texttt{book\_flight},   \\
  & act_4 \text{  for } \texttt{go\_to\_venue},  \ act_5 \text{  for } \texttt{go\_to\_airport},  \ act_6 \text{  for } \texttt{flying}, \\
    &  b_1 \text{  for } \texttt{own\_car},  \ b_2 \text{  for } \texttt{driving\_distance},  \ b_3 \text{  for } \texttt{car\_functional}, \\
     &  b_4 \text{  for } \texttt{engine\_on},  \ b_5 \text{  for } \texttt{at\_venue},  \ b_6 \text{  for } \texttt{budget\_allowed}, \\
     &  b_7 \text{  for } \texttt{flight\_available}, \  b_8 \text{  for } \texttt{flight\_booked},  \ b_9 \text{  for } \texttt{flight\_landed}, \\
         & \text{and }  b_{10} \text{  for } \texttt{at\_airport}.\\
	\end{align*}
		\caption{A  BDI agent for  conference travelling}
		\label{fig:running_example_CAN}
	\end{figure}

A BDI agent desires to arrange a conference trip, denoted by an external event $ e_{1}$. 
We assume  there are only two ways to travel to the    conference.
The first way is to travel by car, given by the plan $Pl_{1} = e_{1} : \varphi_{1} \leftarrow act_{1}; act_{2}$ s.t. $ \varphi_{1} = b_1 \wedge  b_2$. 
The plan~$Pl_{1}$ expresses that if the agent believes it owns a car (i.e. $ b_1$) and the venue is in the driving distance (i.e. $b_2$), it can start the car and drive all the way to the venue. 
To specify the actions, we have $ act_{1} = b_3 \leftarrow \langle \{b_4\}, \emptyset\rangle$ and $ act_{2} = b_4 \leftarrow \langle \{b_5\}, \emptyset\rangle$. For example, the action~$act_{1}$ expresses  that if the car is functional (i.e. $ b_3$) and after executing $act_{1}$, the belief of the engine being on (i.e. $b_4$) will be added while deleting nothing from the belief base.

The second way is to travel by air, given by the plan $Pl_{2} =e_{1} : \varphi_{2}  \leftarrow act_{3}; e_{2};  act_{4}$ s.t. $ \varphi_{2} = b_6 \wedge  b_7$. 
 This plan  expresses that if the budget allows (i.e. $ b_6$) and there is a flight (i.e. $ b_7$), the agent can book the ticket first, then post internally a sub-event  to  actually travelling by plane, and go to the venue after landing. 
 For actions, we have $ act_{3} = true \leftarrow \langle \{b_8\}, \emptyset\rangle$ and $ act_{4} = b_9 \leftarrow \langle \{b_5\}, \emptyset\rangle$. 
 To address the sub-event $e_{2}$, we have    plan $Pl_{3} = e_{2} : \varphi_{3}  \leftarrow act_{5};  act_{6}$ s.t. $ \varphi_{3} = b_8$.
 $Pl_{3}$ expresses that if the agent believes the flight has been booked,  it can go to the airport and fly by plane.
Also, we have $ act_{5} = b_8 \leftarrow \langle \{b_{10}\}, \emptyset\rangle$ and $ act_{6} = b_{10} \leftarrow \langle \{b_9\}, \{b_8, b_{10}\}\rangle$. 
In particular,   action $act_{6}$ indicates that if at airport (i.e. $b_{10}$ for~\texttt{at\_airport}), after the flight it will add the belief atom $b_9$ for~\texttt{flight\_landed}, and delete both   belief atoms $b_8$ for~\texttt{flight\_booked} and $b_{10}$ for~\texttt{at\_airport}.

  We define the initial belief base  to be  $\mathcal{B} = \{b_1, b_2, b_6, b_7\}$. This expresses   the agent believes that it owns a car~($b_1$), the venue is in the driving distance~($b_2$), the budget is sufficient for flight~($b_6$), and there is a flight available~($b_7$). 

\subsection{Bigraphs}\label{sec:bigraphs}
Bigraphs are a universal modelling language, introduced by Milner~\cite{milner2009space}, for both modelling ubiquitous systems and as a unifying theory for many existing calculi for concurrency and mobility.
A bigraph consists of a pair of relations over the same set of \emph{entities}: a directed forest representing topological space in terms of containment, and a hyper-graph expressing the interactions and (non-spatial) relationships among entities. 
Each entity is assigned a \emph{type}, which determines its \emph{arity} (\ie number of links), and whether it is \emph{atomic} (\ie it cannot contain other entities).
For the purpose of presenting our approach, we provide only an informal overview of bigraphs. A concise semantics can be found elsewhere e.g.~\cite{milner2009space}. 

\begin{table}
	\centering
  \footnotesize
	\caption{Bigraph components and operations.}
	\begin{tabular}{@{}lcc@{}}
		\textbf{Component/Operation} & \textbf{Algebraic Form} & \textbf{Diagrammatic Form} \\
		\toprule
		Entity of arity 1 & $\ion{K}{a}$ &
																					\raisebox{-10pt}{
																									 \begin{tikzpicture}[]
																										 \node[ellipse,draw] (k) {};
																										 \node[anchor=south east, inner sep=0.5] at (k.north west) (k_lbl) {\tiny \sf K};
																										 \node[above=0.3 of k] (a) {$a$};
																										 \draw[big edge] (k.north) to[in=-90] (a.south);
																									 \end{tikzpicture}
																					} \\
		Name closure & $/a\; \ion{K}{a}$ &
																					\raisebox{-2pt}{
																									 \begin{tikzpicture}[]
																										 \node[ellipse,draw] (k) {};
																										 \node[anchor=south east, inner sep=0.5] at (k.north west) (k_lbl) {\tiny \sf K};
																										 \node[above=0.3 of k] (a) {};
																										 \draw[big edgec] (k.north) to[in=-90] (a.south);
																									 \end{tikzpicture}
																					} \\[0.2cm]
		Site & $\id$ &
																									 \begin{tikzpicture}[]
																										 \node[big site] (s1) {};
																										 \node[big region, fit=(s1)] (r1) {};
																									 \end{tikzpicture} \\[0.2cm]
		Region & $1$ &
																									 \begin{tikzpicture}[]
																										 \node[big region] (r1) {};
																									 \end{tikzpicture} \\
		\midrule
		Nesting & $\ion{Act}{}.\ion{B}{}.\id$ &
																					\raisebox{-10pt}{
																									 \begin{tikzpicture}[]
																										 \node[big site] (s1) {};
																										 \node[ellipse,draw, fit=(s1), minimum width=20] (b) {};
																										 \node[anchor=south east, inner sep=0.5] at (b.north west) (b_lbl) {\tiny \sf B};
																										 \node[ellipse,draw, fit=(b), xshift=-1.2] (a) {};
																										 \node[anchor=south east, inner sep=0.5] at (a.north west) (a_lbl) {\tiny \sf Act};
																									 \end{tikzpicture}
																									 } \\
		Parallel product  & $\ion{C}{x}.\id \pprod \ion{D}{x}.\id$ &
																					\raisebox{-10pt}{
																									 \begin{tikzpicture}[]
																										 \node[big site] (s1) {};
																										 \node[ellipse,draw, fit=(s1), minimum width=20] (c) {};
																										 \node[anchor=south east, inner sep=0.5] at (c.north west) (c_lbl) {\tiny \sf C};
																										 \node[big region, fit=(c)(c_lbl)] (r1) {};

																										 \node[big site, right=1.1 of s1] (s2) {};
																										 \node[ellipse,draw, fit=(s2), minimum width=20] (d) {};
																										 \node[anchor=south east, inner sep=0.5] at (d.north west) (d_lbl) {\tiny \sf D};
																										 \node[big region, fit=(d)(d_lbl)] (r2) {};

																										 \coordinate (mid) at ($(r1.east)!0.5!(r2.west)$);
																										 \node[] at ($(mid) + (0,0.6)$) (a) {$x$};

																										 \draw[big edge] (c.east) to[out=0, in=-90] (a);
																										 \draw[big edge] (d.west) to[out=-180, in=-90] (a);
																									 \end{tikzpicture}
																									 } \\
		Merge product  & $\ion{C}{x}.\id \mprod \ion{D}{x}.\id$ &
																					\raisebox{-10pt}{
																									 \begin{tikzpicture}[]
																										 \node[big site] (s1) {};
																										 \node[ellipse,draw, fit=(s1), minimum width=20] (c) {};
																										 \node[anchor=south east, inner sep=0.5] at (c.north west) (c_lbl) {\tiny \sf C};

																										 \node[big site, right=1.1 of s1] (s2) {};
																										 \node[ellipse,draw, fit=(s2), minimum width=20] (d) {};
																										 \node[anchor=south east, inner sep=0.5] at (d.north west) (d_lbl) {\tiny \sf D};

																										 \node[big region, fit=(c)(c_lbl)(d)(d_lbl)] (r1) {};

																										 \coordinate (mid) at ($(c.east)!0.5!(d.west)$);
																										 \node[] at ($(mid) + (0,0.6)$) (a) {$x$};

																										 \draw[big edge] (c.east) to[out=0, in=-90] (a);
																										 \draw[big edge] (d.west) to[out=-180, in=-90] (a);
																									 \end{tikzpicture}
																									 } \\
	\end{tabular}
	\label{tab:bigraph_elems}
\end{table}

Bigraphs can be described in algebraic terms or with an \emph{equivalent}
diagrammatical representation as shown in \cref{tab:bigraph_elems}.
In general, bigraphs permit any kind of shape (sometimes coloured) for typed
entities. We allow entities to be \emph{parameterised}, \ie $\ionP{K}{}{n}$, allowing them to represent \emph{families} of
entities. Entities can be connected through green links. \emph{Names}\footnote{Specifically outer-names.} allow links
(or potential links) to bigraphs in an external environment or context, and are
written above the bigraph. Unconnected links are \emph{closed} and drawn as a
closed-off link. Grey
rectangles are called \emph{sites} that indicate parts of the model that have
been abstracted away. In other words, an entity containing a site can contain
zero or more entities of any kind. Finally, a dashed rectangle denotes a
\emph{region} of adjacent parts of the system.

Topological placement of entities is described using: \emph{nesting} that
defines the containment relation on entities; \emph{merge product} that places
two entities side-by-side at the \emph{same} hierarchical level; and
\emph{parallel product} that places entities in separate regions (allowing them
to be at different levels of the hierarchy). In both merge and parallel product, bigraphs are
linked on common names. An overview of the bigraph components and operations are given in~\cref{tab:bigraph_elems}.

\begin{figure}
	\centering
	\begin{subfigure}[b]{0.45\linewidth}
		\centering
    \resizebox{0.7\linewidth}{!}{
			\begin{tikzpicture}[]
  \begin{scope}[local bounding box=lhs]
    \node[big site] (s1_l) {};
    \node[ellipse,draw, fit=(s1_l), minimum width=20] (b) {};
    \node[anchor=south, inner sep=0.5] at (b.north) (b_lbl) {\tiny \sf Cpy};
    \node[ellipse,draw, fit=(b)(b_lbl), xshift=-1.2] (a) {};
    \node[anchor=south east, inner sep=0.5] at (a.north west) (a_lbl) {\tiny \sf Act};

    \node[draw, fill=green!60, above right=0.5 of b] (enable) {};
    \node[anchor=south, inner sep=0.5] at (enable.north) (en_lbl) {\tiny \sf Enabled};
    \node[big region, fit=(a)(a_lbl)(enable)(en_lbl)] (r1) {};

    \draw[big edge] (b.east) to[out=0, in=-90] (enable);
  \end{scope}

  \begin{scope}[shift={(3,0)}, local bounding box=rhs]
    \node[big site] (s1_r) {};
    \node[big site, right=0.2 of s1_r] (s2_r) {};
    \node[ellipse,draw, fit=(s1_r)(s2_r)] (a) {};
    \node[anchor=south east, inner sep=0.5] at (a.north west) (a_lbl) {\tiny \sf Act};
    \node[big region, fit=(a)(a_lbl)] (r1) {};
  \end{scope}

  \draw[->, dashed] (s1_r) to[out=-90, in=-90] (s1_l);
  \draw[->, dashed] (s2_r) to[out=-90, in=-90] (s1_l);
  \node[] at ($(lhs.east)!0.5!(rhs.west)$) {$\rrul$} ;
\end{tikzpicture}
 		}
		\caption{Reaction rule $\mathtt{copy}$.}\label{fig:example_rr}
	\end{subfigure}
	\begin{subfigure}[b]{0.45\linewidth}
		\centering
    \resizebox{0.9\linewidth}{!}{
			\begin{tikzpicture}[]
  \begin{scope}[local bounding box=lhs]
    \node[draw] (d) {};
    \node[draw, ellipse, fit=(d)] (c1) {};
    \node[anchor=south, inner sep=0.5] at (c1.north) (c1_lbl) {\tiny \sf Cpy};

    \node[draw, right=1.1 of d, diamond, inner sep=0.9] (b) {};
    \node[draw, ellipse, fit=(b)] (c2) {};
    \node[anchor=south, inner sep=0.5] at (c2.north) (c2_lbl) {\tiny \sf Cpy};
    \node[ellipse,draw, fit=(c2)(c2_lbl)] (a) {};
    \node[anchor=south east, inner sep=0.5] at (a.north west) (a_lbl) {\tiny \sf Act};

    \node[draw, fill=green!60, above right=0.8 of b] (enable) {};
    \node[anchor=south, inner sep=0.5] at (enable.north) (en_lbl) {\tiny \sf Enabled};
    \node[big region, fit=(c1)(c1_lbl)(a)(a_lbl)(en_lbl)(enable)] (r1) {};

    \draw[big edge] (c2.east) to[out=0, in=-90] (enable);
    \draw[big edgec] (c1.east) to[out=0, in=-90] ($(c1.east) + (0.15,0.3)$);
  \end{scope}

  \begin{scope}[shift={(4,0)}, local bounding box=rhs]
    \node[draw] (d) {};
    \node[draw, ellipse, fit=(d)] (c1) {};
    \node[anchor=south, inner sep=0.5] at (c1.north) (c1_lbl) {\tiny \sf Cpy};

    \node[draw, right=0.9 of d, diamond, inner sep=0.9] (b) {};
    \node[draw, right=0.2 of b, diamond, inner sep=0.9] (b2) {};
    \node[ellipse,draw, fit=(b)(b2)] (a) {};
    \node[anchor=south east, inner sep=0.5] at (a.north west) (a_lbl) {\tiny \sf Act};

    \node[big region, fit=(c1)(c1_lbl)(a)(a_lbl)] (r1) {};

    \draw[big edgec] (c1.east) to[out=0, in=-90] ($(c1.east) + (0.15,0.3)$);
  \end{scope}
  \node[] at ($(lhs.east)!0.5!(rhs.west)$) {$\react$} ;
\end{tikzpicture}
 		}
		\caption{Result of applying reaction rule $\mathtt{copy}$. }\label{fig:rr_outcome} \end{subfigure}

	\caption{Example bigraph and reaction rule.}
	\label{fig:bigraph_example}
\end{figure}
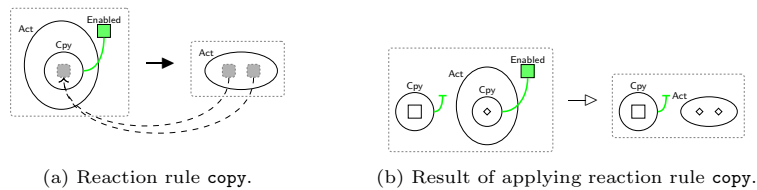

Example bigraphs are shown in \cref{fig:bigraph_example} where the left and
right hand side of arrows represent different bigraphs. Written algebraically the left-side of \cref{fig:rr_outcome} is:
$$(/l\; \ion{Cpy}{l}.\ion{Square}{}.1) \mprod /c\; (\ion{Act}{}.\ion{Cpy}{c}.\ion{Diamond}{}.1 \mprod \ion{Enabled}{c}.1)$$

\subsection{Bigraphical Reactive Systems}\label{sec:brs}

A bigraph represents a system at a single point in time. To allow models to evolve over time we can specify a Bigraphical Reactive System (BRS) that acts as a rewriting system. A BRS consists of a set of \emph{reaction rules} of the form $L \rrul R$, where $L$ and $R$ are bigraphs. Intuitively, a bigraph $B$ evolves to $B'$ by matching and rewriting an occurrence of $L$ in $B$ with $R$. Such a reaction is indicated with $B \react B'$. We use $\react^+$ to denote \emph{one} or more applications of a rule, and $\react^*$ to denote \emph{zero} or more rule applications. 
We also write $\xreact{\mathtt{rule}}$ to identify the reaction rule being applied to generate the transition. If no  name is specified we assume any rule applied.
Reaction rules can be parameterised when they are defined over entities with parameterised types, \ie a rule $\rrP{r}{k}$ for all values of $k$.
The \emph{transition system} of a BRS is a (possibly infinite) graph whose vertices are bigraphs representing the reachable states and whose edges represent reactions over bigraphs.

An example reaction rule, $\mathtt{copy}$, is shown in \cref{fig:example_rr}. This models a reaction which should \emph{copy} elements of a site found within an entity $\ion{Cpy}{}$ \emph{only} if that $\ion{Cpy}{}$ is inside an $\ion{Act}{}$ and connected through a link to an $\ion{Enable}{}$ entity.
The use of a site abstracts from the specific entities to be copied.
To allow copying (and deletion), reaction rules can be augmented with \emph{instantiation maps} that determine a mapping between sites on the left and right-hand side of a reaction rule.
Instantiation maps are denoted graphically as dashed arrows mapping sites in the right-hand side $R$ to sites in left-hand side $L$.
The instantiation map is omitted from a rule definition when it is an identity.
For example, the outcome of applying the reaction rule $\mathtt{copy}$ to a larger starting bigraph is shown in \cref{fig:rr_outcome}. Notice that as the copy on-the-left is not within an $\ion{Act}{}$, there is only one match.

We also use conditional bigraphs~\cite{archibald2020conditional} that allow \emph{application conditions} to specify contextual requirements within the rewrite system. For example, we can exclude certain bigraphs appearing within sites of the left-hand-side of a rule. We write conditions in the form:
\tikz[baseline]{
		  \node[anchor=base] (if1) {if};
		  \node[right=0.02 of if1] (lb1) {$\langle -, $};
		  \node[circle, fill=black, right=0.2 of lb1] (ct) {};
		  \node[big region, fit=(ct)] (ct_r) {};
			\node[right=0.15 of ct] (lb2) {$, \downarrow \rangle$};
}
where the $-$ indicates a negative conditions, the black circle represents an arbitrary bigraph we want to disallow, and $\downarrow$ indicates we should disallow from the site\footnote{Conditional bigraphs also allows positive, and contextual conditions, however we do not use these here.}. Importantly the bigraph in the condition cannot appear \emph{anywhere} in the site, including nested below other entities.
When more than one condition is specified for a reaction (separated by commas) they must \emph{all} hold for the rule to apply.

Furthermore, rule \emph{priorities} can be introduced by defining a partial ordering on the reaction rules of a BRS, as implemented in~\cite{80211_bigraphs}. A reaction rule of lower priority can be applied only if no rule of higher priority is applicable. We write $\mathtt{r_{1}} < \mathtt{r_{2}}$ when $\mathtt{r_{2}}$ has higher priority than $\mathtt{r_1}$. This notation extends to sets in the natural manner, e.g. $\{\mathtt{r_{1}}, \mathtt{r_{3}}\} < \{\mathtt{r_{2}}, \mathtt{r_{4}}\}$, where rules in the same set have the same priority.

A common approach for verifying a BRS is through (bounded) model checking on its transition system. To allow labelling of states, which are themselves bigraphs, we define predicates as \emph{bigraph patterns}.
Informally, a pattern can be seen as a left-hand-side of a reaction rule, \ie the input to the matching problem.
A single state may have multiple labels if multiple patterns occur in it. Patterns can also be combined with standard Boolean operators to form logical formulae.

\section{Encoding BDI Agents in Bigraphs}
\label{sec:structuralEncoding}

We define the structural encoding that maps the  syntax (e.g. plans and actions) of a  \CAN BDI agent     into equivalent bigraphs.

Recall  a BDI agent is specified by a belief base $\mathcal{B}$ consisting of a set of belief atoms, \eg
$ \mathcal{B} = \{\srange{b}{1}{n}\} $, a set of events (i.e. desires) the agent responds to, and a plan library~$\Pi$ containing plans in form of $Pl = e : \varphi \leftarrow P$.
As the agent executes, plan-bodies selected for addressing desires become the intentions of the agent.

  We take a \emph{multi-perspective} approach (as introduced in~\cite{benford2016lions})   in which     perspectives  are represented by  separate and parallel regions. 
Mirroring the core components of a BDI agent, we employ four perspectives: \emph{Belief} that handles knowledge storage and updates; \emph{Desire} that manages the external events; \emph{Intention} that captures the current execution states of plan-bodies; and \emph{Plan} that holds instructions for the agent on how to bring about its desires (\ie how to respond to specific events).   This approach allows us to separate design concerns, to be explicit how and when concerns interact, and to visualise them naturally, as shown in \cref{fig:bdimodelliverificationframework}.
It also facilitates model extension, for example we could in future add perspectives for the external (uncertain) environment, or we could \emph{replace} the Beliefs perspective with one that allows more complex logic formulas. 

The entities in the bigraph model for the syntax of a BDI agent are given in \cref{tbl:ctrl1},  grouped by the four perspectives.
For each entity we give the algebraic form as well as structural information in the form of valid parents and linked entities.
The only atomic entities, \ie that cannot nest other entities, are belief atoms $\ionP{B}{}{n}$, logical constant e.g. $\ion{false}{}$, and events $\ion{E}e$.
Detailed information on the role for each of these entities is given as we introduce the encoding.

\begin{table}
	\caption{Bigraph entities for BDI syntax encoding}
    \scriptsize
	\label{tbl:ctrl1}
	\centering
    \begin{tabular}{@{}lccccr@{}}
\textbf{Description} & \textbf{Entity} & \textbf{Parent(s)} & \textbf{LinksTo} & \makecell{\textbf{Diagrammatic} \\ \textbf{Form}} \\
\toprule
      Belief Base & $\ion{Beliefs}{}$ & & &
                                      \begin{tikzpicture}
                                        \node[big site, opacity=1] (s1) {};
                                        \node[ellipse, inner sep=0, draw, fit=(s1)] (bs) {};
                                        \node[anchor=south, inner sep=0.3] at (bs.north) (bs_lbl) {\tiny \sf Beliefs};
                                      \end{tikzpicture} \\
      Belief Atoms & $\ionP{B}{}{n}$ & $\{\ion{Beliefs}{}, \ion{Pre}{}, \ion{Add}{}, \ion{Del}{}\}$ & &
                                                                                                 \begin{tikzpicture}[scale=0.75, every node/.append style={transform shape}]
                                                                                                   \node[circle, draw] (bn) {\tiny B(n)};
                                                                                                 \end{tikzpicture}
                                                                                                 \\
      Logical False & $\ion{false}{}$ & $\{ \ion{Beliefs}{}, \ion{Pre}{} \}$ & &
                                                                            \begin{tikzpicture}
                                                                              \node[] (false) {\tiny \sf false};
                                                                            \end{tikzpicture} \\
      \midrule
      Desire Set & $\ion{Desires}{}$ & & &
                                      \begin{tikzpicture}
                                        \node[big site] (s1) {};
                                        \node[draw, fit=(s1)] (desires) {};
                                        \node[anchor=south, inner sep=0.3] at (desires.north) (desires_lbl) {\tiny \sf Desires};
                                      \end{tikzpicture} \\
      Event & $\ion{E}{e}$ & $\{\ion{Desires}{}, \ion{PB}{}, \ion{Conc}{}\}$ & $\ion{PlanSet}{e}$ &
                                                                                          \begin{tikzpicture}
                                                                                            \node[event] (e1) {};
                                                                                            \draw[big edgec] (e1.east) to[in=-90,out=0] ($(e1.east) + (0.1, 0.2)$);
                                                                                          \end{tikzpicture} \\
      \midrule
      Intention base & $\ion{Intentions}{}$ & & &
                                         \begin{tikzpicture}
                                           \node[big site] (s1) {};
                                           \node[draw, fit=(s1)] (intents) {};
                                           \node[anchor=south, inner sep=0.3] at (intents.north) (intentions_lbl) {\tiny \sf Intentions};
                                         \end{tikzpicture} \\
      Intention  & $\ion{Intent}{}$ & $\ion{Intentions}{}$ & &
                                                  \begin{tikzpicture}

                                                    \node[big site] (s1) {};
                                                    \node[draw, rounded corners, fit=(s1)] (intent) {};
                                                    \node[anchor=south west, inner sep=0.3] at (intent.north west) (intent_lbl) {\tiny \sf Intent};
                                                  \end{tikzpicture} \\
      \midrule
      Plan library & $\ion{Plans}{}$ & &  &
                                          \begin{tikzpicture}
                                            \node[big site] (s1) {};
                                            \node[draw, fit=(s1)] (plans) {};
                                            \node[anchor=south west, inner sep=0.3] at (plans.north west) (plans_lbl) {\tiny \sf Plans};
                                          \end{tikzpicture} \\
      Relevant Plans & $\ion{PlanSet}{e}$ & $ \{\ion{Plans}{}, \ion{Intent}{}, \ion{Seq}{}, \ion{Cons}{}, \ion{L}{}, \ion{R}{} \} $& $\ion{E}{e}$
                                                                                     & \begin{tikzpicture}
                                                                                       \node[big site] (s1_r) {};
                                                                                       \node[draw, fit=(s1_r)] (rp) {};
                                                                                       \node[anchor=south, inner sep=0.3] at (rp.north) (rp_lbl) {\tiny \sf PlanSet};
																					   \draw[big edgec] (rp.east) to[in=-90,out=0] ($(rp.east) + (0.1, 0.2)$);
                                                                                     \end{tikzpicture} \\
      Plan & $\ion{Plan}{}$ & $\ion{PlanSet}{e}$ & &
                                              \begin{tikzpicture}
                                                \node[big site] (s) {};
                                                \node[draw, fit=(s)] (pl) {};
                                                \node[anchor=south west, inner sep=0.3] at (pl.north west) (pl_lbl) {\tiny \sf Plan};
                                              \end{tikzpicture} \\
      Plan Body & $\ion{PB}{}$ & $\ion{Plan}{}$ & &
                                                  \begin{tikzpicture}
                                                    \node[big site] (s) {};
                                                    \node[draw, rounded corners, fit=(s)] (pbdy) {};
                                                    \node[anchor=south west, inner sep=0.3] at (pbdy.north west) (pbdy_lbl) {\tiny \sf PB};
                                                  \end{tikzpicture} \\
      Action & $\ion{Act}{}$ & $\{\ion{PB}{}, \ion{Seq}{}, \ion{Cons}{}, \ion{L}{}, \ion{R}{} \}$ & &
                                               \begin{tikzpicture}
                                                 \node[big site] (s) {};
                                                 \node[draw, fit=(s)] (act) {};
                                                 \node[anchor=south west, inner sep=0.3] at (act.north west) (act_lbl) {\tiny \sf Act};
                                               \end{tikzpicture} \\
      Precondition & $\ion{Pre}{}$ & $\{\ion{Act}{}$, $\ion{Plan}{}\}$ & &
                                                                       \begin{tikzpicture}
                                                                         \node[big site] (s) {};
                                                                         \node[draw, rounded corners, fit=(s)] (pre) {};
                                                                         \node[anchor=south west, inner sep=0.3] at (pre.north west) (pre_lbl) {\tiny \sf Pre};
                                                                       \end{tikzpicture} \\
      Belief Addition & $\ion{Add}{}$ & $\ion{Act}{}$ &  &
                                                         \begin{tikzpicture}
                                                           \node[big site] (s) {};
                                                           \node[draw, rounded corners, fit=(s)] (add) {};
                                                           \node[anchor=south west, inner sep=0.3] at (add.north west) (add_lbl) {\tiny \sf Add};
                                                         \end{tikzpicture} \\
      Belief Deletion & $\ion{Del}{}$ & $\ion{Act}{}$ & &
                                                        \begin{tikzpicture}
                                                          \node[big site] (s) {};
                                                          \node[draw, rounded corners, fit=(s)] (del) {};
                                                          \node[anchor=south west, inner sep=0.3] at (del.north west) (del_lbl) {\tiny \sf Del};
                                                        \end{tikzpicture} \\
      Sequence $ ; $ & $\ion{Seq}{}$ & $\{\ion{PB}{}, \ion{Try}{}\}$ & &
                                                         \begin{tikzpicture}
                                                           \node[big site] (s) {};
                                                           \node[draw, fit=(s)] (seq) {};
                                                           \node[anchor=south west, inner sep=0.3] at (seq.north west) (seq_lbl) {\tiny \sf Seq};
                                                         \end{tikzpicture} \\
      Plan Choice $ \rhd $ & $\ion{Try}{}$ & $\{\ion{Intent}{}, \ion{Seq}{}, \ion{Goal}{}, \ion{L}{}, \ion{R}{} \}$ & &
                                                            \begin{tikzpicture}
                                                              \node[big site] (s) {};
                                                              \node[draw, fit=(s)] (tri) {};
                                                              \node[anchor=south west, inner sep=0.3] at (tri.north west) (tri_lbl) {\tiny \sf Try};
                                                            \end{tikzpicture} \\
      Next Pointer & $\ion{Cons}{}$ & $\{\ion{Seq}{}, \ion{Try}{} \}$ & &
                                                                          \begin{tikzpicture}

                                                                            \node[big site] (s) {};
                                                                            \node[draw, rounded corners, fit=(s)] (cons) {};
                                                                            \node[anchor=south west, inner sep=0.3] at (cons.north west) (cons_lbl) {\tiny \sf Cons};

                                                                          \end{tikzpicture} \\

      Concurrency $ \parallel $ & $\ion{Conc}{}$ & $\{\ion{Seq}{}, \ion{Try}{}, \ion{PB}{} \}$ & &
                                                             \begin{tikzpicture}
                                                               \node[big site] (s) {};
                                                               \node[draw, fit=(s)] (conc) {};
                                                               \node[anchor=south west, inner sep=0.3] at (conc.north west) (conc_lbl) {\tiny \sf Conc};
                                                             \end{tikzpicture} \\
      Concurrency Markers & $\{\ion{L}{}, \ion{R}{}\}$ & $\ion{Conc}{}$ & &
                                                                            \begin{tikzpicture}
                                                                              \node[big site] (s) {};
                                                                              \node[draw, rounded corners, fit=(s)] (l) {};
                                                                              \node[anchor=south west, inner sep=0.3] at (l.north west) (l_lbl) {\tiny \sf L};
                                                                            \end{tikzpicture} \\
      Declarative Goal & $\ion{Goal}{}$ & $\{\ion{Seq}{}, \ion{Try}{}, \ion{L}{}, \ion{R}{}, \ion{PB}{}\}$ & &
                                                            \begin{tikzpicture}
                                                              \node[big site] (s) {};
                                                              \node[draw, fit=(s)] (goal) {};
                                                              \node[anchor=south west, inner sep=0.3] at (goal.north west) (goal_lbl) {\tiny \sf Goal};
                                                            \end{tikzpicture} \\
      Success Condition & $\ion{SC}{}$ & $\ion{Goal}{}$ & &
                                                            \begin{tikzpicture}
                                                              \node[big site, right=0.8 of s1_l] (s2_l) {};
                                                              \node[draw, rounded corners, fit=(s2_l)] (sc) {};
                                                              \node[anchor=south west, inner sep=0.3] at (sc.north west) (sc_lbl) {\tiny \sf SC};
                                                            \end{tikzpicture} \\
      Failure Condition & $\ion{FC}{}$ & $\ion{Goal}{}$ & &
                                                            \begin{tikzpicture}
                                                              \node[big site, right=0.2 of s3_l] (s4_l) {};
                                                              \node[draw, rounded corners, fit=(s4_l)] (fc) {};
                                                              \node[anchor=south west, inner sep=0.3] at (fc.north west) (fc_lbl) {\tiny \sf FC};
                                                            \end{tikzpicture} \\

	\end{tabular}
\end{table}
We define an encoding $\enc{\cdot} : BDI \to \mathbf{Bg}(\mathcal{K})$ that maps the syntax of a BDI agent  -- including beliefs, desires, intentions, and plans -- to an equivalent bigraph, where $\mathcal{K}$ denotes the set of all entity types in
\cref{tbl:ctrl1}. 
No information is lost through~$\enc{\cdot}$ and it is possible to define the inverse encoding~$\enc{\cdot}^{-1}$ establishing  an equivalence.
Although the inverse is easy to define, some cases are context dependent, \eg rules \ref{enc-precondition} and \ref{enc-effects} related to belief atoms in~\cref{fig:bdi-big} have the same bigraph representation but always appear in distinct contexts (pre-conditions and action outcomes respectively)\footnote{The inverse of true is a special case as we may map either to the truth term or an empty context set. However both options give rise to behaviourally equivalent agents.}. For brevity we omit the details of the inverse encoding.

The encoding is defined inductively as shown in \cref{fig:bdi-big}. To aid
explanation, we give the encoding in two parts.
In~\cref{fig:enc-agent}, the encoding of agent belief, desire and intention structures are given. In the second part, the encoding of plans, in particular the plan-bodies, of an agent are provided in~\cref{fig:enc-programs}. The parts are not distinct \eg the
plans within the plan library are encoded using the encoding of plan-bodies. We use 
$\prod \ion{M}{} \defeq \ion{M}{} \mprod \dots \mprod \ion{M}{}$ to denote iterated
merge product.
In the next few sections, we explain the (numbered) rules in~\cref{fig:bdi-big}. 

\begin{figure}
	\centering

	\begin{subfigure}{1.0\linewidth}
	\centering
	\footnotesize
	\begin{align}
		\label{enc-full}
		\enc{\langle \mathcal{B}, E^{e}, \Gamma,  \Pi\rangle} &=
		\ion{Beliefs}{}.\enc{\mathcal{B}} \pprod \ion{Desires}{}.{\enc{E^{e}}} \\ &\pprod \ion{Intentions}{}.\enc{\Gamma} \pprod \ion{Plans}{}.\enc{\Pi} \\
		\label{enc-atoms}
		\enc{b_{n}} &= \ionP{B}{}{n} \\
		\label{enc-false}
		\enc{false} &= \ion{false}{} \\
		\label{enc-true}
		\enc{true} &= 1 \\
		\label{enc-beliefSet}
		\enc{\mathcal{B} = \{b_{1} \dots b_{n}\}} &= \enc{b_{1}} \mprod \dots \mprod \enc{b_{n}} \\
		\label{enc-eventSet}
			\enc{E^e = \{e_{1} \dots e_{n}\}} &= \enc{e_{1}} \mprod \dots \mprod \enc{e_{n}} \\
		\label{enc-event}
		\enc{e} &= \ion{E}{e} \\
		\label{enc-intentionSet}
		\enc{\Gamma = \{\srange{P}{1}{n}\}} &= \ion{Intent}{}.\enc{P_{1}} \mid \dots \mid \ion{Intent}{}.\enc{P_{n}} \\
		\label{enc-planlib}
		\enc{\Pi = \{Pl_{1} \dots Pl_{n}\}} &= \prod\limits_{
			e \in E}  \underset{\text{where} \ trigger(Pl_{j})=trigger(Pl_k)=e}{\ion{PlanSet}{e}.(
			\enc{Pl_{j}}  \mprod \dots
			\mprod \enc{Pl_{k}})}
	\end{align}
	\caption{Beliefs, desire, intention, and plan library encoding.}\label{fig:enc-agent}
	\end{subfigure}

	\begin{subfigure}{1.0\linewidth}
	\centering
	\footnotesize
	\begin{align}
		\label{enc-nil}
		\enc{\mathit{nil}} &= \mathsf{1} \\
		\label{enc-actions}
		\enc{act = \varphi \leftarrow \langle \phi^{+}, \phi^{-} \rangle} &= \ion{Act}{}.(\ion{Pre}{}.\enc{\varphi} \mprod \ion{Add}{}.\enc{\phi^{+}} \mprod \ion{Del}{}.\enc{\phi^{-}}) \\
		\label{enc-precondition}
		\enc{\varphi = b_{1} \land \dots \land b_{n}} &= \enc{b_{1}} \mprod \dots \mprod \enc{b_{n}} \\
		\label{enc-effects}
		\enc{\phi^{\pm} = \{b_{1} \dots b_{n}\}} &= \enc{b_{1}} \mprod \dots \mprod \enc{b_{n}} \\
		\label{enc-seq}
		\enc{P_{1} ; P_{2}} &= \ion{Seq}{}.(\enc{P_{1}} \mprod \ion{Cons}{}.\enc{P_{2}}) \\
		\label{enc-concurrency}
		\enc{P_{1} \parallel P_{2}}&= \ion{Conc}{}.(\ion{L}{}.\enc{P_{1}} \mid \ion{R}{}.\enc{P_{2}})\\
		\label{enc-declarativegoal}
		\enc{goal(\varphi_{s}, P, \varphi_{f})}&= \ion{Goal}.(\ion{SC}{}.\enc{\varphi_{s}}{} \mid \enc{P} \mid \ion{FC}{}.\enc{\varphi_{f}}{})  \\
		\label{enc-backup}
		\enc{P_{1} \rhd P_{2}}&= \ion{Try}{}.(\enc{P_{1}} \mid \ion{Cons}{}.\enc{P_{2}}) \\
		\label{enc-plans}
		\enc{e: (| \varphi_{1} : P_{1}, \dots, \varphi_{2} : P_{2}|)} &= \ion{PlanSet}{e}.(\enc{\varphi_{1} : P_{1}} \mprod \dots \mprod \enc{\varphi_{2} : P_{2}}) \\
		\label{enc-relplan}
		\enc{\varphi : P} &= \ion{Plan}.(\ion{Pre}{}.\enc{\varphi} \mprod \ion{PB}.\enc{P})  \\
		\label{enc-planstructure}
		\enc{Pl = e : \varphi \leftarrow P} &= \enc{\varphi : P}
		\end{align}
	\caption{Plan and plan-body encoding.}\label{fig:enc-programs}
	\end{subfigure}
	\caption{Encoding $\enc{\cdot}$ from BDI agents to bigraphs.}\label{fig:bdi-big}
\end{figure}

\subsection{Encoding of Beliefs, Desires, and Intentions}

Rule~\ref{enc-full} ensures   the top-level components of an agent -- beliefs, desires, intentions, and plan library -- are mapped to  separate perspective (region) in the bigraph.

We assume all belief formulas $\varphi$ are expressed in propositional logic.
Recall that the convention in AgentSpeak for the belief base is   $ \varphi ::= b \mid \neg b \mid (\varphi_1 \wedge \varphi_2) \mid \mathit{true} \mid \mathit{false} $. 
For convenience, the \emph{parameterised entities} $\ionP{B}{}{i}$ (rule~\ref{enc-atoms}) are used for both positive and negative atoms: $ b$ or $ \neg b$ (\eg $\enc{b} = \enc{b_0} = \ionP{B}{}{0}, \enc{\neg b} = \enc{b_1} = \ionP{B}{}{1}$). 
Using this, all formulas can be constructed in   pure conjunctive form, i.e. $\varphi = b_{1} \land \dots \land b_{n}$.
We allow logical constants for $true$ and $false$ representing formulas that are always/never entailed, \eg an action with pre-condition $false$ never executes.
In the bigraph model, we only assign an entity $\ion{false}{}$ to represent logical $false$ (rule~\ref{enc-false}), while the logical constant $true$ is mapped to the empty bigraph (rule~\ref{enc-true}) as we assume an empty formula is always true, \eg there may be no pre-condition for some action.

Encoding the belief base $\mathcal{B}$ (and any set concept in general) from a BDI agent to bigraphs leverages the bag-like nature of nesting (rule~\ref{enc-beliefSet}).
For empty sets, we have $\enc{\emptyset} = 1$, \ie the bigraph with one empty region.

To encode desires, an entity of $\ion{E}{e}$ is created for each possible event (that an agent \emph{desires} to respond to) as seen in rules~\ref{enc-eventSet} and~\ref{enc-event}.
Importantly, $\ion{E}{e}$ exports a \emph{name} $e$ that allows us to identify specific events using links.
Recall that a set of relevant plans is the set of plans which have the same triggering event. 
We use this when encoding the plan library $\Pi$ (rule~\ref{enc-planlib}) by having it contain sets of relevant plans $\ion{PlanSet}{e}$ with $e$ connecting the event $e$ with the set of plans that respond to it.
This differs from typical BDI agents where the plan library is a flat set of plans.
This use of   indexing by   event name   through relevant plans
decreases the likelihood of some potential human errors, \eg misspelling of event names
 and also simplifies   agent reasoning by avoiding repetitive searching for relevant plans.

Finally, for intentions, we utilise the same set-like structure as beliefs, this time encoding individual (partially executed) plan-bodies as required (rule~\ref{enc-intentionSet}), which will be discussed   in the next section.

\subsection{Encoding Plans and Plan-Bodies}

\begin{figure}
{	\footnotesize
\begin{bnf*}
	\bnfprod{Plan} { \bnfts{e} : \bnfpn{Pre} \bnfsp \bnfts{$\leftarrow$} \bnfsp \bnfpn{UserP}} \\
	\bnfprod{UserP} {
		\bnfpn{BasicP} \bnfor \bnfpn{UserP} \bnfts{;} \bnfpn{UserP} \bnfor \bnfpn{UserP} \bnfts{$\parallel$} \bnfpn{UserP} \bnfor \\
		\bnfmore
		  goal( \bnfts{$\varphi_{s}$}, \bnfts{e} , \bnfts{$\varphi_{f}$})
	} \\
		\bnfprod{P} {
		\bnfts{nil} \bnfor \bnfpn{BasicP} \bnfor \bnfpn{P} \bnfts{;} \bnfpn{P}  \bnfor \bnfpn{P} \bnfts{$\parallel$} \bnfpn{P} \bnfor \\
		\bnfmore
			goal( \bnfts{$\varphi_{s}$}, \bnfpn{P} , \bnfts{$\varphi_{f}$}) \bnfor \bnfpn{P} \bnfts{$\rhd$} \bnfpn{P} \bnfor \\
			\bnfmore
			\bnfts{e} : (| \bnfts{$\varphi_1$} : \bnfpn{UserP},\dots,\bnfts{$\varphi_n$} : \bnfpn{UserP} |)
	} \\
		\bnfprod{BasicP} {
			\bnfts{e}
			\bnfor \bnfpn{Act}
			\bnfor \bnfts{$+b$}
			\bnfor \bnfts{$-b$}
			\bnfor \bnfts{$?\varphi$}
			} \\
	\bnfprod{Act} { \bnfpn{Pre} \bnfsp \bnfts{$\leftarrow$} \bnfsp \bnfts{$\langle \phi^+,\phi^- \rangle$} }	\\
	\bnfprod{Pre} { \bnfts{$\varphi$} \bnfor \bnfts{$false$} \bnfor \bnfts{$truth$}} 
\end{bnf*}}
\caption{Grammar for Plans and Plan-bodies.}
\label{fig:grammar}
\end{figure}
  
Plans and plan-bodies are specified with the language given in~\cref{fig:grammar}, which includes
includes two forms of plan-bodies:    ${\langle \text{UserP}\rangle}$
that the user writes,  and the more comprehensive  ${\langle \text{P}\rangle}$   that can occur during any execution.
  
A plan $\text{\texttt{e}} : \text{Pre} \leftarrow \langle \text{UserP}\rangle$ consists of a triggering event $ e $, the context (pre-condition) $\langle \text{Pre} \rangle$, and a user-defined plan-body specified by $\langle \text{UserP} \rangle$.
The user-defined plan-body $\langle \text{UserP} \rangle$  may be  the basic building block $\langle \text{BasicP} \rangle$ including handling an internal event \texttt{e}, or executing an action $\langle \text{Act}\rangle$.
Actions also have the pre-condition $\langle \text{Pre} \rangle$, which indicates when an action is valid for execution given in the current belief state.
After executing an action, $\phi^{+}$ and $\phi^{-}$ are sets of beliefs to be added and removed from the belief state, respectively.
The user-defined plan-body $\langle \text{UserP} \rangle$  can also be combined in the   three ways: $\langle \text{UserP} \rangle;\langle \text{UserP} \rangle$ executing those two $ \langle \text{UserP} \rangle$ in sequence, $\langle \text{UserP} \rangle \parallel \langle \text{UserP} \rangle$ pursing those two $\langle \text{UserP} \rangle$ concurrently, and $goal(\varphi_{s}, e, \varphi_{f})$ achieving the state $ \varphi_{s} $ through addressing an internal event $e$, failing when $ \varphi_{f} $ holds, and retrying as long as neither $ \varphi_{s} $ nor $ \varphi_{f} $  is believed to be true.
Internally (i.e.~during execution) programs may have an additional three forms: $nil$ is the empty program that is always successful, $\langle \text{P} \rangle \rhd \langle \text{P} \rangle$ represents \emph{trying} the first $\langle \text{P} \rangle$  while keeping the second $\langle \text{P} \rangle$ as a backup in case the first $\langle \text{P} \rangle$ fails, and $\mathtt{e} :(|\varphi_1 : \langle \text{UserP} \rangle \dots \varphi_n : \langle \text{UserP} \rangle|)$ is a set of backup plans which are all triggered by the event $\mathtt{e}$.

The bigraph encoding of plans and plan-bodies
(\cref{fig:enc-programs}) mirrors the grammar given in~\cref{fig:grammar} by specifying a mapping for each syntactic form. 
Each individual plan  is represented as the pairing of some pre-condition (as encoded belief atoms), nested in the entity~$\ion{Pre}{}$, and an encoded plan-body, nested in entity~$\ion{PB}{}$ (rule~\ref{enc-relplan} and \ref{enc-planstructure} in~\cref{fig:enc-programs}).

Bigraphical entities of $ \langle \text{UserP} \rangle $ are built by introducing additional controls for each form, \eg $\ion{Seq}{}$.
As the merge product operator of bigraphs is commutative, \eg $A \mprod B \equiv B \mprod A$, we need to add additional entities to force an ordering on the children.
For example, the sequencing $ P_1; P_2$ (rule \ref{enc-seq} in~\cref{fig:enc-programs}) utilises an entity $\ion{Cons}{}$ that identifies $ P_2$ as the next to execute after the successful execution of its predecessor~$ P_1$. 
Likewise, the form $P_{1} \rhd P_{2}$  (rule \ref{enc-backup}), that \emph{tries} $P_{1}$ with $P_{2}$ as a backup, uses $\ion{Cons}{}$ to distinguish between $P_{1}$ and $P_{2}$.
For concurrency (rule~\ref{enc-concurrency}) we further require two additional controls $\ion{L}{}$ and $\ion{R}{}$ to identify the left and right of the concurrency structure~$\parallel$.
Finally, for the form of declarative goals $goal(\varphi_s, P, \varphi_f)$ (rule~\ref{enc-declarativegoal}), we map it to an entity $\ion{Goal}{}$ that nests a success condition $\ion{SC}{}$, failure condition $\ion{FC}{}$ and the current form of the remaining program.

Finally, actions are encoded (rule \ref{enc-actions}) in a similar way.
In particular, raw entailment and belief state update forms, \ie $?\varphi, +b$, and $-b$, may be
seen as special cases of actions that do not update the external environment. We establish the following equivalences to unify them under the same action encoding. 
\begin{align*}
?\varphi &\equiv act : \varphi \leftarrow \langle \emptyset, \emptyset \rangle \\
+b &\equiv act : \emptyset \leftarrow \langle \{b\}, \emptyset \rangle \\
-b &\equiv act : \emptyset \leftarrow \langle \emptyset, \{b\} \rangle
\end{align*}

\subsection{Example of Encoding}
\label{ex:static_encoding}
	To show how our encoding works 
\cref{tab:encodingexample} provides the mapping for a BDI agent  for the travelling example in~\cref{fig:running_example_CAN}. 
 
	\begin{table}
	\centering
	\caption{Example encoding of a conference travel agent in~\cref{fig:running_example_CAN}.}
  \footnotesize
	\begin{tabular}{@{}ll@{}}
		\textbf{Agent} & \textbf{$\enc{\text{Agent}}$} \\
		\toprule
			$\mathcal{B} = \{b_1, b_2, b_6, b_7\}$ &  $\ion{Beliefs}.(\ion{B(1)}{} \mprod \ion{B(2)}{} \mprod \ion{B(6)}{} \mprod \ion{B(7)}{} )$ \\
		$\Pi = \{Pl_1, Pl_2, Pl_3\}$ &  $\ion{Plans}.(\ion{PlanSet}{e_1}.(\enc{Pl_1} \mprod \enc{Pl_2}) \mprod  \ion{PlanSet}{e_2}.\enc{Pl_3})$ \\
		$Pl_1 = e_{1} : \varphi_{1}  \leftarrow act_{1}; act_{2}$ &  $\ion{Plan}.(\ion{Pre}{}.(\enc{\varphi_{1}} \mprod \ion{PB}{}.\ion{Seq}{}.(\enc{act_{1}} \mprod \ion{Cons}{}.\enc{act_{2}}))$ \\
			$Pl_2 = e_{1} : \varphi_{2}  \leftarrow act_{3};  e_2; act_{4}$ &  $\ion{Plan}.(\ion{Pre}{}.(\enc{\varphi_{2}} \mprod \ion{PB}{}.\ion{Seq}{}.(\enc{act_{3}} \mprod \ion{Cons}{}.\ion{Seq}{}.( 	\enc{e_{2}} \mprod \ion{Cons}{}. \enc{act_{4}})))$ \\
		$Pl_3 = e_{2} : \varphi_{3}  \leftarrow act_{5}; act_{6}$ &  $\ion{Plan}.(\ion{Pre}{}.(\enc{\varphi_{3}} \mprod \ion{PB}{}.\ion{Seq}{}.(\enc{act_{5}} \mprod \ion{Cons}{}.\enc{act_{6}}))$ \\
		$\varphi_{1} = b_{1}\wedge b_{2}$ & $\ion{B(1)}{} \mprod \ion{B(2)}{}$ \\
		$act_1 = b_{3} \leftarrow \langle \{b_{4}\}, \emptyset \rangle $ & $\ion{Act}{}.(\ion{Pre}{}.\ion{B(3)}{} \mprod \ion{Add}{}.\ion{B(4)}{} \mprod \ion{Del}{}.\mathsf{1})$ \\
		$act_2 = b_{4} \leftarrow \langle \{b_{5}\}, \emptyset \rangle $ & $\ion{Act}{}.(\ion{Pre}{}.\ion{B(4)}{} \mprod \ion{Add}{}.\ion{B(5)}{} \mprod \ion{Del}{}.\mathsf{1})$ \\
		$\varphi_{2} = b_{6}\wedge b_{7}$ & $\ion{B(6)}{} \mprod \ion{B(7)}{}$ \\
			$act_3 = true \leftarrow \langle \{b_{8}\}, \emptyset \rangle $ & $\ion{Act}{}.(\ion{Pre}{}.\ion{1}{} \mprod \ion{Add}{}.\ion{B(8)}{} \mprod \ion{Del}{}.\mathsf{1})$ \\
					$e_{2} $ & $\ion{E}{e_2} $ \\
		$act_4 = b_9 \leftarrow \langle \{b_{5}\}, \emptyset \} \rangle$ & $\ion{Act}{}.(\ion{Pre}{}.\ion{B(9)}{}  \mprod \ion{Add}{}.\ion{B(5)}{}  \mprod \ion{Del}{}.\mathsf{1})$ \\
			$\varphi_{3} = b_{8} $ & $\ion{B(8)}{} $ \\
			$act_5 = b_{8}\leftarrow \langle \{b_{10}\}, \emptyset \rangle $ & $\ion{Act}{}.(\ion{Pre}{}.\ion{B(8)}{} \mprod \ion{Add}{}.\ion{B(10)}{} \mprod \ion{Del}{}.\mathsf{1})$  \\
			$act_6 = b_{10} \leftarrow \langle \{b_{9}\}, \{b_{8}, b_{10}\} \rangle $ & $\ion{Act}{}.(\ion{Pre}{}.\ion{B(10)}{} \mprod \ion{Add}{}.\ion{B(9)}{} \mprod \ion{Del}{}.(\ion{B(8)}{} \mprod \ion{B(10)}{})  )$  
	\end{tabular}
	\label{tab:encodingexample}
\end{table}

This completes  the structural encoding (i.e. the syntactic specification of a BDI agent),   we now turn our attention to  a
\emph{behavioural} encoding of BDI agents (i.e. the operation semantics of a BDI agent) as a bigraphical reactive system. We do so in an incremental manner in the following three steps:  in  ~\cref{sec:core}
 we define the semantics of a subset of \CAN that we call the core  \CAN. 
 Core \CAN semantics excludes concurrency and declarative goals, and  so   resembles      AgentSpeak~\cite{s:agentspeak}.
 In~\cref{sec:corebigraphs} we  encode   core \CAN   as a BRS and in~\cref{sec:extended} we extend such a BRS for the core \CAN to include concurrency and declarative goals. 
  
\section{Semantics of Core CAN Language}
\label{sec:core}

\subsection{Overview of Core CAN language }\label{sec:overviewsemantics}
The core operation of an agent in response to an (external) event  is as follows. 
    All relevant plans for that event are retrieved from the (pre-defined) plan library. 
 An \emph{applicable} plan is selected (if one exists) and
 its plan-body is added to the intention base. 
 The plan-body consists of discrete steps, \eg actions or sub-events. When executing   
  a sub-event, its
  applicable plan requires to be found, and its plan-body is also   added
 to the intention base   -- this forms an execution
\emph{tree} within the intention. 
A BDI agent continues to execute until there are no pending 
events, and all intentions are completed (either successfully or with failure).

\subsection{Core \CAN Semantics}\label{sec:coresemantics}

We specify the behaviour of an agent as an operational semantics~\cite{plotkin1981structural} 
defined over configurations $ \mathcal{C} $ and transitions $ \mathcal{C} \rightarrow \mathcal{C}'$.
Transitions $ \mathcal{C} \rightarrow \mathcal{C}'$ denote a single execution step between configuration $ \mathcal{C} $ and $ \mathcal{C}' $.
We write $ \mathcal{C} \rightarrow $ (resp. $ \mathcal{C} \nrightarrow $) to state that there is (resp.\ is not) a $\mathcal{C}'$ such that $ \mathcal{C} \rightarrow \mathcal{C}'$.

A derivation rule specifies the necessary conditions for an an agent to transition to a new configuration.
A derivation rule consists of a (possibly empty) set of premises $ p_{i} $ $( i =1, \ldots, n )$ on $\mathcal{C}$, and a conclusion, denoted by
\begin{center}
	$  \dfrac{p_{1} \  \  \  \  p_{2} \  \  \  \ \cdots \  \  \  \ p_{n}}{\mathcal{C} \rightarrow \mathcal{C'}} \  \  \  \ l $
\end{center}
where $ l $ is a rule name.
We write $\mathcal{C} \xrightarrow{l} \mathcal{C}'$ to denote $\mathcal{C}$ evolves to $\mathcal{C}'$ through the application of derivation rule $ l$.

The $\CAN$ semantics were originally defined~\cite{sardina:hierarchical} over the triple $\langle\mathcal{B}, \mathcal{A}, P\rangle$ where $\mathcal{B}$ is the current belief base, $\mathcal{A}$ the \emph{sequence} of actions that have been executed, and $P$ the current partially executed plan-body. 
As the recorded sequence of executed actions is never used to determine the operation of an agent, \ie there are no pre-condition on $\mathcal{A}$, we do not include it here (i.e. $ \langle\mathcal{B}, P\rangle$).
It is trivial to log the action sequence within the bigraph model if required, however we do not do so here because an action log introduces states that would otherwise be isomorphic (resulting in larger transition systems).

The semantics of \CAN language is specified by two types of transitions.
The first transition type, denoted as $\rightarrow$, specifies \emph{intention-level} evolution in terms of configuration $ \langle\mathcal{B}, P\rangle$ where $ \mathcal{B} $
is the current belief set, and $ P $ the plan-body currently being executed (\ie
the next step of the current intention). 
The second type, denoted as $\Rightarrow$, specifies \emph{agent-level} evolution over $\langle E^{e}, \mathcal{B}, \Gamma\rangle$, detailing how to
execute a complete agent where $E^{e}$ stands for the a set of pending external events required to address.

\begin{figure}
	\scriptsize
	
	\begin{center}
		$  \dfrac{act: \psi \leftarrow \langle\phi^{-},\phi^{+}\rangle \  \   \  \ \mathcal{B} \vDash \psi}{\langle\mathcal{B},  act\rangle \rightarrow \langle (\mathcal{B}\setminus\phi^{-}\cup\phi^{+}), nil\rangle} \   act $
		\qquad
		$  \dfrac{\mathcal{B}\models \phi}{\langle \mathcal{B},  ?\phi \rangle     \rightarrow \langle \mathcal{B},  nil\rangle}  \  ? $
	\end{center}
	
	\begin{center}
		$ \dfrac{}{\langle \mathcal{B},  +b \rangle     \rightarrow \langle \mathcal{B}\cup \{b\},  nil\rangle} \  +b $
		\qquad
		$ \dfrac{}{\langle \mathcal{B},  -b \rangle     \rightarrow \langle \mathcal{B}\setminus \{b\},  nil\rangle} \  -b $
	\end{center}

	\begin{center}
		$ \dfrac{\Delta = \{\varphi: P \mid (e' = \varphi \leftarrow P) \in \Pi  \wedge e' = e      \}}{ \langle \mathcal{B},  e\rangle                \rightarrow \langle \mathcal{B},  e:(\mid \Delta\mid)\rangle       }  \  event  $
		
	\end{center}

	\begin{center}
		
		$ \dfrac{\varphi:P \in \Delta \ \ \ \mathcal{B} \models \varphi }{  \langle \mathcal{B},  e:(\mid \Delta\mid) \rangle                \rightarrow \langle \mathcal{B},  P \rhd e:(\mid \Delta \setminus \{\varphi:P\} \mid)\rangle               } \ select $
		
	\end{center}

	\begin{center}
		$ \dfrac{\langle \mathcal{B},  P_{1}\rangle \rightarrow \langle \mathcal{B}',  P'_{1}\rangle}{\langle \mathcal{B},  P_{1} \rhd P_{2}\rangle \rightarrow \langle \mathcal{B}',  P'_{1} \rhd P_{2})\rangle} \rhd_{;}$
		\qquad
		$\dfrac{}{\langle \mathcal{B},  (nil \rhd P_{2})\rangle \rightarrow \langle \mathcal{B}',  nil\rangle} \rhd_{\top} $
	\end{center}

	\begin{center}
		$ \dfrac{P_{1} \neq nil \ \ \ \langle \mathcal{B},  P_{1}   \rangle \nrightarrow \ \ \ \langle \mathcal{B},  P_{2}\rangle \rightarrow \langle \mathcal{B}',  P'_{2}\rangle}{\langle \mathcal{B},  P_{1} \rhd P_{2}\rangle \rightarrow \langle \mathcal{B}',  P'_{2}\rangle} $ \ \ \  $ \rhd_{\bot} $
	\end{center}
	
	\begin{center}
		$ \dfrac{\langle \mathcal{B},  P_{1}\rangle \rightarrow \langle \mathcal{B}',  P'_{1}\rangle}{\langle \mathcal{B},  (P_{1};P_{2})\rangle \rightarrow \langle \mathcal{B}',  (P'_{1};P_{2})\rangle}  ;$
		\qquad 
		$\dfrac{\langle \mathcal{B},  P\rangle \rightarrow \langle \mathcal{B}',  P'\rangle}{\langle \mathcal{B},  (nil;P)\rangle \rightarrow \langle \mathcal{B}',  P' \rangle}  ;_{\top} $
	\end{center}

	\caption{Core \CAN semantics.}
	\label{fig:core_semantics}
\end{figure}

\cref{fig:core_semantics} gives the set of derivation rules for evolving any single intention. 
For example, derivation rule $ act $ handles the execution of an action, when the pre-condition is met,
resulting in a belief state update. Rules $?$, $ +b $ and $ -b $ are special
actions that perform pre-condition check ($?$), adding one belief atom ($+b$)
and deleting atoms ($-b$). 
As in \cref{sec:structuralEncoding}, we assume an
equivalence between $act$ and $?, +b, -b$ and do not directly model these rules.
Rule $ event $ replaces an event with the set of relevant plans, while
rule $ select $ chooses an applicable plan from a set of relevant plans while
retaining un-selected plans as backups. 
With these backup plans, the rules for failure
recovery $\rhd_{;}$, $\rhd_{\top}$, and $\rhd_{\bot}$ enable new plans to be selected if
the current plan fails (due to e.g. the unexpected environment changes).
Finally, rules $;$ and $;_{\top}$ describe executing plan-bodies in sequence.

\begin{figure}
	\scriptsize
	\begin{center}
		$ \dfrac{e \in E^{e}}{\langle E^{e},  \mathcal{B},  \Gamma\rangle  \Rightarrow \langle E^{e} \setminus\{e\}, \mathcal{B},  \Gamma \cup \{e\}    \rangle } A_{event} $
	\end{center}

	\begin{center}
		$ \dfrac{P \in \Gamma \ \ \ \langle \mathcal{B},  P\rangle \rightarrow \langle  \mathcal{B}',  P'\rangle }{\langle E^{e},  \mathcal{B},  \Gamma\rangle  \Rightarrow \langle E^{e}, \mathcal{B}', (\Gamma \setminus \{P\})\cup \{P'\}    \rangle }   A_{step} $
	\end{center}

	\begin{center}
		$ \dfrac{P \in \Gamma \ \ \ \langle \mathcal{B},  P\rangle \nrightarrow}{\langle E^{e}, \mathcal{B},  \Gamma \rangle  \Rightarrow \langle E^{e}, \mathcal{B},  \Gamma \setminus\{P\}    \rangle }  A_{update} $
	\end{center}
	
	\caption{Derivation rules for agent configuration.}
	\label{fig:agentCANSemantics}
\end{figure}

The agent-level semantics are given in \cref{fig:agentCANSemantics}. An agent
configuration is defined by the triple $ \langle E^e, \mathcal{B}, \Gamma\rangle $
consisting of a set of external events $ E^{e} $ to which the agent is required to 
respond, the belief set $\mathcal{B}$, and the intention base $ \Gamma $ -- a set of partially executed plan-bodies $ P $
that the agent has already committed to. 
The derivation rule $ A_{event} $ handles external events,
which originate from the environment\footnote{As we do not model the environment explicitly, we assume any events are waiting in the desire set at the start of an agent execution.}, by adopting them as intentions. Rule
$ A_{step} $ selects an intention from the intention base, and evolves a
single step w.r.t. intention-level transition, while $ A_{update} $ discards
intentions which cannot make any intention-level transition (either because it
has already succeeded, or it failed)\footnote{In the original \CAN semantics there is no way to determine if an event was handled successfully or not,   both cases are treated the same way (by removing the intention when it is done or cannot progress).}.

\subsection{Example of Core \CAN Semantics}
\label{sec:semanticsExample}

To show how an agent evolves in the \CAN semantics we use the  conference travelling example in~\cref{fig:running_example_CAN}.  
Assuming the external event~$ e_{1}$ has already been converted from a desire to an intention, \cref{fig:semanticevolutionflow_new} illustrates the intention-level evolution of this intention according to the rules presented in~\cref{fig:core_semantics}. 
In~\cref{fig:semanticevolutionflow_new} agents evolve from  left to   right,  each line consists of a single step of an intention.
Below each step we  show the sub-rules that applied.  A commentary is as follows.

When  the event  $e_1$ is posted to the agent, the {\em event} rule  in~\cref{fig:core_semantics} transforms     $ e_{1}$ into the   program containing all the relevant plans available~(1).  
 If the agent believes that it owns a car and the venue is within driving distance (i.e. $\varphi_1$ holds holds), then the  {\em select} rule   transforms the set of relevant plans into the  selected plan~(2), which indicates the sequence~$act_{1};act_{2}$ is ready for execution, while  the other  plans are indicated as backup on the right-hand side of the symbol~$\rhd $. 
Next, the agent tries to execute the program $act_{1};act_{2}$. 
Given the belief base in~\cref{fig:running_example_CAN},  the pre-condition $act_{1}$ does not hold (e.g.~the car engine fails to start), thus  $act_1 \nrightarrow$. Meanwhile, the backup plan is applicable shown by the derivation  $select$ from $ e_{1}: (|\varphi_{2} : act_{3}; e_{2}; act_{4}|) $ to $  act_{3}; e_{2}; act_{4} \rhd e_{1}: (|\emptyset|)$.
According to the rule~$\rhd_{\bot} $, the agent can initiate the failure recovery by trying such a backup plan, resulting in the program shown in~(3). 
Since $ act_3$ has $true$ as its pre-condition, it can always be executed shown by $ act_{3} \xrightarrow{act} nil $. 
After   execution of $ act_3$, the rule~$;$ then updates the entire sequence from~$ act_{3};e_2;act_4$ to $ nil ;e_2;act_4$. After the left-hand side of~$\rhd$ is updated, the rule~$\rhd_;$ can then further transform the program to that in~(4).
In order to discard the symbol $ nil$ in a sequence, it requires the part after $ nil$ in a sequence to be progressed, namely $ e_2; act_4$. 
To progress the $ e_2; act_4$, it requires to progress the first part of such a sequence, i.e. $ e_2$. 
To progress the event~$e_2 $, it requires to retrieve a set of its relevant plans. 
Therefore, we have what is shown in the~(5). The rule~$ select$ firstly transforms the event $e_2$ to a set of relevant plans, secondly the rule~$;$ updates the sequence~$ e_2; act_4$, and thirdly the symbol~$nil$ can be removed by the rule~$;_\top$ from the entire sequence~$ nil; e_2; act_4$. 
Finally, the rule~$\rhd_;$ can follow up transforming the entire program  on the left-hand side of~$\rhd$ accordingly.

\begin{figure}
\resizebox{\linewidth}{!}{
		\begin{tikzpicture}
		\node[] at (3,0) (r1) {(1) $ e_1$};
		\node[anchor=west] at (r1.east) (r2) {$\xrightarrow{event}$};
		\node[anchor=west] at (r2.east) (r3) {$ e_{1}: (| \varphi_{1} : act_{1};act_{2}, \varphi_{2} : act_{3}; e_{2}; act_{4}|)$ };  
		\node[anchor=west] at (0,-1)  (r4) { (2)  $e_{1}: (| \varphi_{1} : act_{1};act_{2}, \varphi_{2} : act_{3}; e_{2}; act_{4}|)$};
		\node[anchor=west] at (r4.east) (r5) {$\xrightarrow{select}$};
		\node[anchor=west] at (r5.east) (r6) {$act_{1};  act_{2} \rhd   e_{1}: (|\varphi_{2} : act_{3}; e_{2}; act_{4}|) $};
		\node[anchor=west] at (-2,-2) (r7) {(3)   $act_{1}; act_{2} \rhd   e_{1}: (|\varphi_{2} : act_{3}; e_{2}; act_{4}|) $};
		\node[anchor=west] at ($(r7.east) + (+3.5,+0.4)$) (r8){$ act_1 \nrightarrow$};
		\coordinate (bracketleft) at ($(r7.east) + (-0.1,0)$);
		\node[anchor=west] at (bracketleft) {$ \Bigg [ $};
		\node[anchor=west] at ($(r7.east) + (+0.2,-0.3)$) (r9){$ e_{1}: (|\varphi_{2} : act_{3}; e_{2}; act_{4}|) \xrightarrow{select}  act_{3}; e_{2}; act_{4} \rhd e_{1}: (|\emptyset|)$};
		\coordinate (bracketright) at ($(r9.east) + (-0.3,+0.3)$);
		\node[anchor=west] at (bracketright) {$ \Bigg ]$};
		\node[anchor=west] at ($(r9.east) + (+0.2,+0.3)$) (r9_1){$ act_{3}; e_{2}; act_{3} \rhd e_{1}: (|\emptyset|)$};
		\coordinate (failureleft) at ($(bracketleft.west) + (0.1,-1.5)$);
		\coordinate (failureright) at ($(bracketright.east) + (0.5,-1.5)$);
		
		\draw[] (failureleft) -- (failureright) node[midway, above] {$\xrightarrow{\rhd_{\bot}}$};
		\draw[dashed] (failureleft) -- ($(bracketleft.west) + (0.1,0)$);
		\draw[dashed] (failureright) -- ($(bracketright.east) + (0.5,0)$);
		\draw[] ($(r7.east) + (-0.1,0)$) -- ($(r7.east) + (0.1,0)$) node[midway, above] {};
		\draw[] ($(r9_1.east) + (-4.1,0)$) -- ($(r9_1.east) + (-3.7,0)$) node[midway, above] {};

		\node[anchor=west] at (-1.5,-4) (r10) {(4)  $act_{3}; e_{2}; act_{4} \rhd   e_{1}: (|\emptyset|) $};
		\node[anchor=west] at ($(r10.east) + (+3.5,+0.1)$) (r11) {$ act_{3} \xrightarrow{act} nil $};
\draw[dashed] ($(r11) + (-3,0)$) -- ($(r11) + (-3,-1)$);
		\draw[dashed] ($(r11) + (3,0)$) -- ($(r11) + (3,-1)$);
		
		\node[anchor=west] at ($(r11) + (-2,-1)$) (r12)  {$ act_{3};e_2;act_4 \xrightarrow{;} nil ;e_2;act_4$};
		\draw[] ($(r11) + (-3,-1)$) -- ($(r11) + (-2,-1)$)  node[midway, above] {};
		\draw[] ($(r11) + (2.5,-1)$) -- ($(r11) + (3,-1)$)  node[midway, above] {};
		\draw[dashed] ($(r11) + (-4,0)$) -- ($(r11) + (-4,-2)$);
		\draw[dashed] ($(r11) + (4,0)$) -- ($(r11) + (4,-2)$);
		\draw[] ($(r11) + (-4,-2)$) -- ($(r11) + (4,-2)$) node[midway, above] {$\xrightarrow{\rhd_{;}}$};
		\node[anchor=west] at ($(r11) + (4.8, 0)$) (r13)  {$ nil;e_2;act_4 \rhd   e_{1}: (|\emptyset|) $};
		\draw[] ($(r11) + (-4.5,0)$) -- ($(r11) + (-1,0)$)  node[midway, above] {};
		\draw[] ($(r11) + (1,0)$) -- ($(r11) + (4.8,0)$)  node[midway, above] {};

		\node[anchor=west] at (-2,-6.3) (r14) {(5)  $ nil;e_2;act_4\rhd   e_{1}: (|\emptyset|)  $}; 
		\node[anchor=west] at ($(r14) + (4,0)$) (r15) { $ e_2 \xrightarrow{select} e_{2}: (|\varphi_{3}:act_5;act_6|)  $};    
		\draw[dashed] ($(r15) + (-2.8,0)$) -- ($(r15) + (-2.8,-1)$);
		\draw[dashed] ($(r15) + (3.3,0)$) -- ($(r15) + (3.3,-1)$);
		\node[anchor=west] at ($(r15) + (-2.7,-1)$) (r16) { $ e_2;act_4 \xrightarrow{;} e_{2}: (|\varphi_{3}:act_5;act_6|); act_4  $};   
		\draw[dashed] ($(r15) + (-3.5,0)$) -- ($(r15) + (-3.5,-2)$);
		\draw[dashed] ($(r15) + (4.1,0)$) -- ($(r15) + (4.1,-2)$);
		\node[anchor=west] at ($(r15) + (-3,-2)$) (r17) { $ nil;e_2;act_4 \xrightarrow{;_\top} e_{2}: (|\varphi_{3}:act_5;act_6|); act_4  $};   
		\draw[dashed] ($(r15) + (-4,0)$) -- ($(r15) + (-4,-3)$);
		\draw[dashed] ($(r15) + (4.5,0)$) -- ($(r15) + (4.5,-3)$);
		\draw[] ($(r15) + (-4,-3)$) -- ($(r15) + (4.5,-3)$)  node[midway, above] {$\xrightarrow{\rhd_{;}}$};
		\draw[] ($(r15) + (-2.8,-1)$) -- ($(r15) + (-2.6,-1)$)  node[midway, above] {};
		\draw[] ($(r15) + (3,-1)$) -- ($(r15) + (3.2,-1)$)  node[midway, above] {};
		\draw[] ($(r15) + (-3.5,-2)$) -- ($(r15) + (-3,-2)$)  node[midway, above] {};
		\draw[] ($(r15) + (3.5,-2)$) -- ($(r15) + (3.9,-2)$)  node[midway, above] {};
		\draw[] ($(r15) + (-4.5,0)$) -- ($(r15) + (-2.5,0)$)  node[midway, above] {};
		\draw[] ($(r15) + (2.5,0)$) -- ($(r15) + (4.8,0)$)  node[midway, above] {};
		\node[anchor=west] at ($(r15) + (4.7,0)$) (r18) { $ e_{2}: (|\varphi_{3}:act_5;act_6|);act_4\rhd   e_{1}: (|\emptyset|)  $};   
	\end{tikzpicture}
 }
\caption{Illustration of intention-level evolution of the event $e_{1}$.}
\label{fig:semanticevolutionflow_new}
\end{figure}
For brevity,  we omit the rest of the evolution. 
In practice an agent may execute multiple intentions concurrently.

\subsection{AND/OR Trees}
We can view the semantic evolution of the agent program in terms of reductions over AND/OR trees, and use this
representation to reason about the interactions between events, plans, and
intentions~\cite{logan2017progressing, MXUICTAI19}. 
AND nodes are successful if
\textbf{all} of their children succeed while OR nodes are successful if
\textbf{at least one} child succeeds. 
We make heavy use of such AND/OR trees
in our behavioural encoding of \CAN semantics in bigraphs that can be seen
(in part) as reductions over these trees. 
However, we stress that although the behaviour of a \CAN agent can be visualised via AND/OR trees, in practice, the trees are
not fully realised in memory and are created on-demand as the intention
evolves.

The root of an AND/OR tree is a top-level external event represented as an OR
node, that is, an event succeeds if at least one plan succeeds. The tree is
built implicitly through the syntax of \CAN.
For example, the sequencing symbol $;$ ensures that  execution
must successfully execute all steps in the plan-body to allow the parent AND node to
succeed. 
Meanwhile, the failure recovery symbol $\rhd$ represents choice, with backup plans creating the branching
structure. 
In \cref{sec:concurrency}, an additional form~$\parallel$  will be introduced to
complement~$;$ by identifying branches that can be explored concurrently.

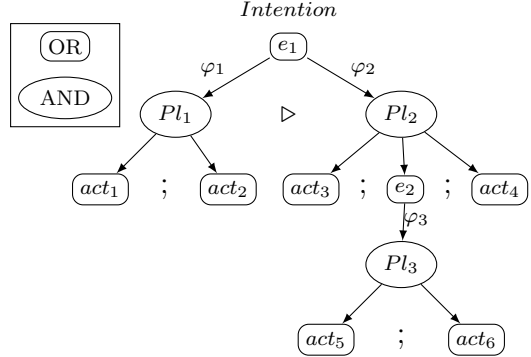
\begin{figure}[!tbh]
	\centering
	\footnotesize
	\begin{tikzpicture}[node distance=4.2em]
  \tikzstyle{ornode}=[draw, rounded corners]
  \tikzstyle{andnode}=[draw, ellipse]
  \tikzstyle{oredge}=[->, -latex]
  \tikzstyle{andedge}=[->, dotted, -latex]
  \tikzstyle{andangle}=[draw, dotted, angle radius=1.3em]

\node[ornode] (G3) at (-1.6, 0) {$ e_{1}$};
  \node[andnode] (P1) at (-3.1, -0.9) {$Pl_{1}$};
  \node[ornode] (a1) at (-4.1, -1.9) {$act_{1}$};
  \node[ornode] (a2) at (-2.4, -1.9) {$act_{2}$};

\node[andnode] (P2) at (-0.1, -0.9) {$Pl_{2}$};
  \node[ornode] (b1) at (-1.3, -1.9) {$act_{3}$};
  \node[ornode,  right of=b1] (b2) {$e_{2}$};
  \node[ornode,  right of=b2] (b3) {$act_{4}$};

\node[andnode] (P4) at (-0.1, -2.9) {$Pl_{3}$};
  \node[ornode] (b4) at (-1.1, -3.9) {$act_{5}$};
  \node[ornode] (b5) at (0.9, -3.9) {$act_{6}$};

\draw[oredge] (G3) edge (P1);
  \draw[oredge] (P1) edge (a1);
  \draw[oredge] (P1) edge (a2);

  \draw[oredge] (G3) edge (P2);
  \draw[oredge] (P2) edge (b1);
  \draw[oredge] (P2) edge (b2);
  \draw[oredge] (P2) edge (b3);

  \draw[oredge] (b2) edge (P4);
  \draw[oredge] (P4) edge (b4);
  \draw[oredge] (P4) edge (b5);

  \node[](1) at (-2.6, -2) {};
  \node[](2) at (-2, -2) {};

  \node[](3) at (-2.6, -2.5) {};
  \node[](4) at (-2, -2.5) {};

  \begin{scope}[xshift=-130, yshift=0.1]
    \node[ornode] (or) {OR};
    \node[below=0.2 of or, andnode] (and) {AND};
    \node[draw, fit=(or)(and)] {};
  \end{scope}

  \node[] at (-1.6, 0.5) {$ Intention$};

  \node[] at (-2.6, -0.3) {$\varphi_{1}$};
  \node[] at (-0.6, -0.3) {$\varphi_{2}$};
  \node[] at (0.1, -2.3) {$\varphi_{3}$};

\node[] at ($(P1.east)!0.5!(P2.west)$) {\large $\triangleright$};
  \node[] at ($(a1.east)!0.5!(a2.west)$) {\large $;$};
  \node[] at ($(b1.east)!0.5!(b2.west)$) {\large $;$};
  \node[] at ($(b2.east)!0.5!(b3.west)$) {\large $;$};
  \node[] at ($(b4.east)!0.5!(b5.west)$) {\large $;$};

\end{tikzpicture}
 	\caption{Snapshot of AND/OR tree representing the intention for the event $e_{1}$ where the agent chose to try $Pl_{1}$ before $Pl_{2}$ during execution.}
	\label{fig:andgraph}
\end{figure}

As an example we revisit the conference travelling example of \cref{fig:running_example_CAN} showing one possible AND/OR tree for the plans $ Pl_{1} = e_{1} : \varphi_{1} \leftarrow act_{1}; act_{2}$,
$ Pl_{2} = e_{1} :\varphi_{2} \leftarrow act_{3}; e_2; act_{4} $, and
$ Pl_{3} = e_{2} :\varphi_{3} \leftarrow act_{5};  act_{6}$. In this case, $Pl_1$ was chosen first and $Pl_2$ kept as a backup plan as shown in~\cref{fig:andgraph}. The top-level
event $ e_{1} $ is achieved if either of the two plans $ Pl_{1} $ or $ Pl_{2} $
are successful. In this case the agent has chosen to do $Pl_{1}$ \emph{before}
$Pl_{2}$, although the ordering is not fixed ahead of execution time. The plan
$ Pl_{1} $ itself involves performing the actions $ act_{1} $ \emph{followed by}
$ act_{2} $, whereas one part of plan-body of plan $ Pl_{2} $ involves achieving the sub-event $ e_{2} $
which can, in turn, be addressed by the plan $ Pl_3$. 

From the point of view of the semantics, the tree is explored in a depth-first
manner with reductions being pushed down the tree. For example, the derivation rule~$;$ reduces a given branch of the tree while the rule $;_{\top}$ moves to the next child at the same AND level. When a node cannot be reduced, \eg if an action pre-condition is unmet,
this failure propagates to the closest branch point (OR-node) where they are
handled by the failure recovery rules (e.g. $\rhd_{\bot}$).

\section{Encoding Core \CAN Semantics in Bigraphs}\label{sec:corebigraphs}

We now encode the core \CAN semantics  (presented in Figs.~\ref{fig:core_semantics} and~\ref{fig:agentCANSemantics}) as a bigraphical reactive system (BRS) and show that the encoding is faithful.  
By faithful we mean that   for each   transition $\xRightarrow{l}$ (resp.\ intention  
$\xrightarrow{l}$)
$\langle E^{e}, \mathcal{B}, \Gamma \rangle \xRightarrow{l} \langle E'^{e}, \mathcal{B}', \Gamma' \rangle$ (resp. $\langle \mathcal{B}, P \rangle \xRightarrow{l} \langle \mathcal{B}', P' \rangle$) there
exists a \textbf{finite} sequence of reaction rules, such that
$\enc{\langle E^{e}, \mathcal{B}, \Gamma \rangle } \react^{+} \enc{\langle E'^{e}, \mathcal{B}', \Gamma'\rangle}$ (resp. $ \enc{\langle \mathcal{B}, P \rangle} \react^{+} \enc{\langle \mathcal{B}', P'\rangle} $) and
 no derivation rules in BDI side is enabled that were not available in the initial state.
 The encoding may introduce  new intermediate states, but  they do not enable additional BDI derivation rules, \ie there is no additional branching.

To encode control flow required for execution, we require additional entities that are not part of the
structural encoding, \ie they do not necessarily have a corresponding agent representation in \CAN.
These additional entities are given in \cref{tab:additionalEntitites} and their
purpose is introduced as they are used.

\begin{table}
	\centering
	\caption{Additional entities for semantics encoding}\label{tab:additionalEntitites}
  \resizebox{\linewidth}{!}{
    \begin{tabular}{@{}lccccc@{}}
      \textbf{Description} & \textbf{Entity} & \textbf{Parent(s)} & \textbf{LinksTo} & \makecell{\textbf{Diagrammatic} \\ \textbf{Form}} \\
      \toprule

      Set of beliefs to check & $\ion{Check}{}$ & $\ion{Beliefs}{}$ & &
                                                                        \begin{tikzpicture}
                                                                            \node[big site] (s) {};
                                                                            \node[ellipse, draw, inner sep=0, fit=(s)] (c) {};
                                                                            \node[anchor=south, inner sep=0.3] at (c.north) (c_lbl) {\tiny \sf Check};
                                                                        \end{tikzpicture} \\
      Unknown check result & $\ion{CheckRes}{}$ & $\{\ion{Act}{},\ion{Plan}{}\}$ & &
                                                                        \begin{tikzpicture}
                                                                            \node[diamond, draw, inner sep=1.2] (cr) {};
                                                                        \end{tikzpicture} \\
      Successful entailment & $\ion{CheckRes}{}.\ion{T}{}$ & $\{\ion{Act}{},\ion{Plan}{}\}$ & &
                                                                        \begin{tikzpicture}
                                                                            \node[diamond, fill=green, draw, inner sep=1.2] (cr) {};
                                                                        \end{tikzpicture} \\
      Failed entailment & $\ion{CheckRes}{}.\ion{F}{}$ & $\{\ion{Act}{},\ion{Plan}{}\}$ & &
                                                                        \begin{tikzpicture}
                                                                            \node[diamond, fill=red, draw, inner sep=1.2] (cr) {};
                                                                        \end{tikzpicture} \\
      Not-yet checked token & $\ion{CheckToken}{}$ & $\ion{Plan}{}$ & &
                                                                        \begin{tikzpicture}
                                                                            \node[circle, fill=black] (ct) {};
                                                                        \end{tikzpicture} \\
      Reduction of entity/site & $\ion{Reduce}{}$ & $\{\ion{Intention}{}, \ion{Try}{}, \ion{Seq}{}, \ion{Conc}{}, \ion{L}{}, \ion{R}{} \}$ & &
                                                                        \begin{tikzpicture}
                                                                            \node[big site] (s) {};
                                                                            \node[draw, red, fit=(s)] (red) {};
                                                                        \end{tikzpicture} \\
      Reduction failure & $\ion{ReduceF}{}$ & $\{\ion{Intention}{}, \ion{Seq}{}, \ion{Try}{}, \ion{L}{}, \ion{R}{}\}$& &
                                                                        \begin{tikzpicture}
                                                                            \node[] (rf) {\large $\nrightarrow$};
                                                                        \end{tikzpicture} \\
	\end{tabular}
  }
\end{table}

For brevity, we give an overview of key aspects of the encoding. The full
\emph{executable} model, for use with BigraphER~\cite{sevegnani2016bigrapher}, is available~\cite{models}.

\subsection{Belief Checks and Updates}
\label{sec:beliefOps}

The \CAN semantics assumes set operations and logic entailment as built-in operators. However, as we want an \emph{executable} semantics, these must be
explicitly encoded in the BRS.

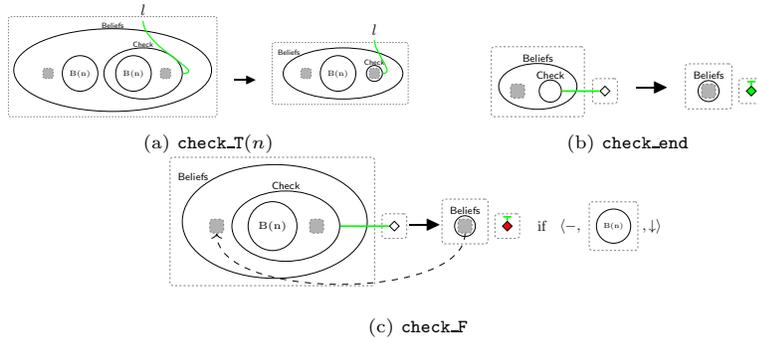
\begin{figure}
  \centering
  \begin{subfigure}[b]{0.45\linewidth}
	\centering
	\resizebox{\linewidth}{!}{
		\begin{tikzpicture}[]
  \begin{scope}[local bounding box=lhs]
    \node[big site] (s1) {};
    \node[circle, draw, right=0.2 of s1] (bn) {\tiny B(n)};
    \node[circle, draw, right=0.4 of bn] (bnc) {\tiny B(n)};
    \node[big site, right=0.2 of bnc] (s2) {};
    \node[ellipse, draw, inner sep=0, fit=(bnc)(s2)] (c) {};
    \node[anchor=south, inner sep=0.3] at (c.north) (c_lbl) {\tiny \sf Check};
    \node[ellipse, inner sep=0, draw, fit=(s1)(bn)(c)(c_lbl)(bn)] (bs) {};
    \node[anchor=south, inner sep=0.3] at (bs.north) (bs_lbl) {\tiny \sf Beliefs};

    \node[big region, fit=(bs)(bs_lbl)] (r1) {};

    \node[above of=c_lbl, yshift=-6] (l) {$l$};
    \draw[big edge] (c.east) to[in=-90, out=0] (l.south);
  \end{scope}

  \begin{scope}[shift={(6,0)}, local bounding box=rhs]
    \node[big site] (s1) {};
    \node[circle, draw, right=0.2 of s1] (bn) {\tiny B(n)};
\node[big site, right=0.3 of bn] (s2) {};
    \node[ellipse, draw, inner sep=0.2, fit=(s2)] (c) {};
    \node[anchor=south, inner sep=0.3] at (c.north) (c_lbl) {\tiny \sf Check};
    \node[ellipse, inner sep=0, draw, fit=(s1)(bn)(c)(c_lbl)(bn)] (bs) {};
    \node[anchor=south east, inner sep=0.3] at (bs.north west) (bs_lbl) {\tiny \sf Beliefs};

    \node[big region, fit=(bs)(bs_lbl)] (r1) {};

    \node[above of=c_lbl, yshift=-7] (l) {$l$};
    \draw[big edge, looseness=1] (c.east) to[in=-90, out=0] (l.south);
  \end{scope}

  \node[] at ($(lhs.east)!0.5!(rhs.west) + (0,-0.5)$) {$\rrul$} ;
\end{tikzpicture}
 	}
	\caption{$\rrP{check\_T}{n}$}
	\label{fig:check_T}
  \end{subfigure}
  \begin{subfigure}[b]{0.45\linewidth}
	\centering
	\resizebox{0.70\linewidth}{!}{
		\begin{tikzpicture}[]
  \begin{scope}[local bounding box=lhs]
    \node[big site] (s1) {};
\node[big site, right=0.3 of s1, opacity=0] (s2) {};
\node[ellipse, draw, inner sep=0.2, fit=(s2)] (c) {};
    \node[anchor=south, inner sep=0.3] at (c.north) (c_lbl) {\tiny \sf Check};
    \node[ellipse, inner sep=0, draw, fit=(s1)(c)(c_lbl)] (bs) {};
    \node[anchor=south, inner sep=0.3] at (bs.north) (bs_lbl) {\tiny \sf Beliefs};

    \node[big region, fit=(bs)(bs_lbl)] (r1) {};

    \node[diamond, draw, inner sep=1.2, right=0.7 of s2] (cr) {};
    \node[big region, fit=(cr)] (r2) {};

    \draw[big edge] (c.east) to[in=180, out=0] (cr.west) {};
  \end{scope}

  \begin{scope}[shift={(3.2,0)}, local bounding box=rhs]
    \node[big site] (s1) {};
    \node[ellipse, inner sep=0, draw, fit=(s1)] (bs) {};
    \node[anchor=south, inner sep=0.3]  at (bs.north) (bs_lbl) {\tiny \sf Beliefs};

    \node[big region, fit=(bs)(bs_lbl)] (r1) {};

    \node[diamond, fill=green, draw, inner sep=1.2, right=0.5 of s1] (cr) {};
    \node[big region, fit=(cr)] (r2) {};

    \draw[big edgec] (cr.north) to[in=-90, out=90] ($(cr.north) + (0,0.08)$) {};
  \end{scope}

  \node[] at ($(lhs.east)!0.5!(rhs.west) + (0,-0.1)$) {$\rrul$};
\end{tikzpicture}
 	}
	\caption{$\mathtt{check\_end}$}
	\label{fig:check_end}
  \end{subfigure}

  \begin{subfigure}[t]{0.8\linewidth}
	\centering
	\resizebox{0.7\linewidth}{!}{
		\begin{tikzpicture}[]
  \begin{scope}[local bounding box=lhs]
    \node[big site] (s1_l) {};
    \node[circle, draw, right=0.4 of s1_l] (bnc) {\tiny B(n)};
    \node[big site, right=0.2 of bnc] (s2_l) {};
    \node[ellipse, draw, inner sep=0, fit=(bnc)(s2_l)] (c) {};
    \node[anchor=south, inner sep=0.3] at (c.north) (c_lbl) {\tiny \sf Check};
    \node[ellipse, inner sep=0, draw, fit=(s1_l)(c)(c_lbl)] (bs) {};
    \node[anchor=south east, inner sep=0.3] at (bs.north west) (bs_lbl) {\tiny \sf Beliefs};

    \node[big region, fit=(bs)(bs_lbl)] (r1) {};

    \node[diamond, draw, inner sep=1.2, right=1.1 of s2_l] (cr) {};
    \node[big region, fit=(cr)] (r2) {};

    \draw[big edge] (c.east) to[in=180, out=0] (cr.west) {};
  \end{scope}

  \begin{scope}[shift={(4.2,0)}, local bounding box=rhs]
    \node[big site] (s1_r) {};
    \node[ellipse, inner sep=0, draw, fit=(s1_r)] (bs) {};
    \node[anchor=south, inner sep=0.3]  at (bs.north) (bs_lbl) {\tiny \sf Beliefs};

    \node[big region, fit=(bs)(bs_lbl)] (r1) {};

    \node[diamond, fill=red, draw, inner sep=1.2, right=0.5 of s1_r] (cr) {};
    \node[big region, fit=(cr)] (r2) {};

    \draw[big edgec] (cr.north) to[in=-90, out=90] ($(cr.north) + (0,0.08)$) {};
  \end{scope}

  \begin{scope}[shift={(5.5,0)}, local bounding box=cond, scale=0.7, transform shape]
    \node[] (if1) {if};
    \node[right=0.08 of if1] (lb1) {$\langle -, $};
    \node[circle, draw, right=0.25 of lb1] (ct) {\tiny B(n)};
    \node[big region, fit=(ct)] (ct_r) {};
    \node[right=0.15 of ct] (lb2) {$, \downarrow \rangle$};
  \end{scope}

  \draw[->, dashed, looseness=0.7] (s1_r) to[out=-90, in=-90] (s1_l);

  \node[] (rr) at ($(lhs.east)!0.5!(rhs.west) + (0,-0.1)$) {$\rrul$};
\end{tikzpicture}
 	}
	\caption{$\mathtt{check\_F}$} \label{fig:check_F}
  \end{subfigure}
  \caption{Reactions for logical entailment.}
  \label{fig:rr_check_set}
\end{figure}

We encode belief updates and checks in the usual recursive manner as shown in
\cref{fig:rr_check_set}. 
For example, the reaction rule $\mathtt{check\_end}$ provides a base-case 
for $\rrP{check\_T}{n}$,
while $\mathtt{check\_F}$ (a conditional rule)  handles the case
when there is no match in the belief base. 
Similar reaction rules (not shown) are provided to perform addition
and deletion of beliefs\footnote{We assume additions/deletions are disjoint (as
 they are in practice) so that there are no race conditions between the
 reactions.}.

The belief check reaction rules use auxiliary entities, \eg $\ion{Check}{l}$ and
$\ion{CheckRes}{l}$ (shown as a diamond). 
These auxiliary entities, which are added from other reaction rules, encode control flow.
As these entities are not part of the \CAN syntax encoding, they
do not enable any additional agent steps. 
After performing the sequence of reaction rules equivalent to
a \CAN derivation rule, no auxiliary entities will be present -- they are only
allowed in intermediate states.

Notice the number of children of $\ion{Check}{}$ decreases on each reaction rule
application suffices to prove that the logical entailment (resp. checks/updates) will
complete in a \emph{finite number} of steps. As such, placing
belief checks/updates into the \emph{highest} rule priority class of the
BRS allows us to assume belief checks/updates are \emph{atomic} with respect to
the other reactions. That is, an agent never sees a part-modified belief
set (as required to model atomic actions).

We use the label
\begin{align*}
\{\rr{set\_{ops}}\} = &\rrP{check\_T}{n}, \rr{check\_end}, \rr{check\_F}, \\ &\rrP{del\_in}{n}, \rrP{del\_notin}{n}, \rr{delete\_end}, \\ &\rr{add\_end}, \rrP{add\_notin}{n}, \rrP{add\_in}{n} 
\end{align*}
to refer to the priority class of set operations.

\subsection{Modelling Reductions}
\label{sec:reductions}

The \CAN semantics assumes a notion of irreducibility. That is, the derivation
rule $\langle \mathcal{B}, P \rangle \nrightarrow $ represents the failure of an agent to perform any
further operation on the program $ P$. For example, $\langle \mathcal{B}, act \rangle \nrightarrow $
holds if the pre-condition of the action $act$ is not met.

While \CAN remains agnostic to such details, we require the notion of
irreducibility to be \emph{explicitly} encoded to obtain an executable
semantics. To encode explicit reduction, we introduce auxiliary
controls~$\ion{Reduce}{}$ (by colouring the entity/site being reduced as red)
and~$\ion{ReduceF}{}$ (representing~$\nrightarrow$).

$\ion{Reduce}{}$ requests the entities nested below are reduced, for example by
executing an action. In the case reduction is not possible, \eg if an action
precondition is not met, $\ion{ReduceF}{}$ represents the failure to reduce,
enabling checks of the premise $\langle \mathcal{B}, P \rangle \nrightarrow$ in derivation rules.

If we view intentions as AND/OR trees, the explicit reductions perform the
\emph{tree search} with $\ion{Reduce}{}$ determining which sub-tree to reduce
next, and $\ion{ReduceF}{}$ indicating a sub-tree could not reduce and
backtracking should be performed.

We define a function
$\encR{} : \langle \mathcal{B}, P \rangle \to \mathbf{Bg}(\mathcal{K \cup \ion{Reduce}{}})$ that,
for belief base $\mathcal{B}$ and intention-level program $P$, requests that the
sub-tree rooted at $P$ be reduced. That is:

$$
\encR{\langle \mathcal{B}, P \rangle} \defeq \enc{\mathcal{B}} \pprod \ion{Reduce}{}.\enc{P}
$$

Where $\mathcal{B}$ is a \emph{mutable}, globally scoped, environment for the
reduction of $P$. This is key benefit of bigraphs for modelling: environments
can be placed in parallel.

The function~$\encR{}$ for reduction plays a key role in our semantic encoding. For example, it forms the
bridge between agent-level steps and intention-level steps, \ie an agent
$\langle\mathcal{B}, E^{e}, \{P \cup \Gamma \}, \Pi \rangle$ can (try to) step intention $P$ using
$\encR{\langle \mathcal{B}, P \rangle}$.

\subsection{Core Semantic Encoding}

Given the atomic set operations and explicit reduction,
We now show how the core \CAN semantics are encoded as a BRS.

\subsubsection{Actions}

The main operation of an agent is to execute actions that update both the external
environment, \eg moving a block, and in-turn revise the internal belief base. Recall that, in the encoding of syntax of \CAN language, we have established entailment and belief state updates
(rules $?\varphi, +b, -b$) as special cases of actions that simply do not update the external environment. As such, we can safely omit the explicit reactions for entailment and belief state updates. 

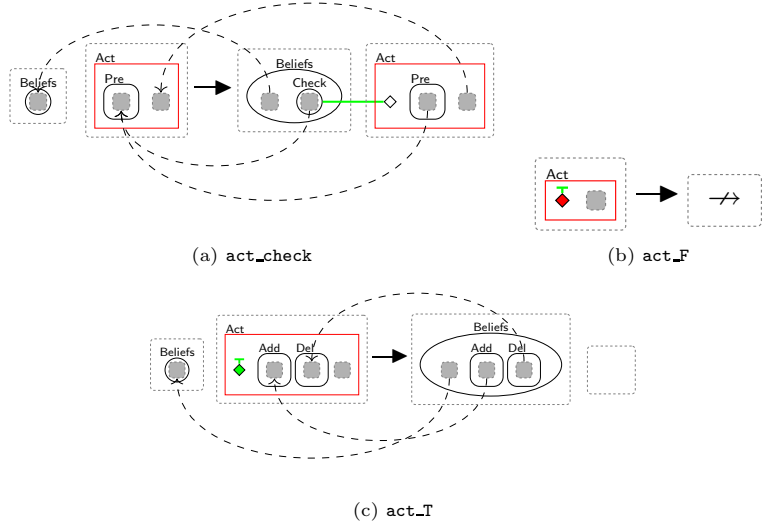
\begin{figure}
  \centering
  \begin{subfigure}[b]{0.55\linewidth}
	\centering
	\resizebox{\linewidth}{!}{
		\begin{tikzpicture}[]
  \begin{scope}[local bounding box=lhs]
\node[big site] (s1_l) {};
    \node[ellipse, inner sep=0, draw, fit=(s1_l)] (bs) {};
    \node[anchor=south, inner sep=0.3] at (bs.north) (bs_lbl) {\tiny \sf Beliefs};
    \node[big region, fit=(bs)(bs_lbl)] (r1) {};

\node[big site, right=0.9 of s1_l] (s2_l) {};
    \node[draw, rounded corners, fit=(s2_l)] (pre) {};
    \node[anchor=south west, inner sep=0.3] at (pre.north west) (pre_lbl) {\tiny \sf Pre};
    \node[big site, right=0.3 of s2_l] (s3_l) {};
    \node[draw, red, fit=(pre)(pre_lbl)(s3_l)] (act) {};
    \node[anchor=south west, inner sep=0.3] at (act.north west) (act_lbl) {\tiny \sf Act};
    \node[big region, fit=(act)(act_lbl)] (r1) {};
  \end{scope}

  \begin{scope}[shift={(3.2,0)}, local bounding box=rhs]
\node[big site] (s1_r) {};
    \node[big site, right=0.3 of s1_r] (s2_r) {};
    \node[ellipse, draw, inner sep=0, fit=(s2_r)] (c) {};
    \node[anchor=south, inner sep=0.3] at (c.north) (c_lbl) {\tiny \sf Check};
    \node[ellipse, inner sep=0, draw, fit=(s1_r)(c)(c_lbl)] (bs) {};
    \node[anchor=south, inner sep=0.3] at (bs.north) (bs_lbl) {\tiny \sf Beliefs};
    \node[big region, fit=(bs)(bs_lbl)] (r1) {};

\node[diamond, draw, inner sep=1.2, right=0.9 of s2_r] (cr) {};
    \draw[big edge] (c.east) to[in=180, out=0] (cr.west) {};
    \node[big site, right=0.3 of cr] (s3_r) {};
    \node[draw, rounded corners, fit=(s3_r)] (pre) {};
    \node[anchor=south west, inner sep=0.3] at (pre.north west) (pre_lbl) {\tiny \sf Pre};
    \node[big site, right=0.3 of s3_r] (s4_r) {};
    \node[draw, red, fit=(cr)(pre)(pre_lbl)(s4_r)] (act) {};
    \node[anchor=south west, inner sep=0.3] at (act.north west) (act_lbl) {\tiny \sf Act};
    \node[big region, fit=(act)(act_lbl)] (r1) {};
  \end{scope}

  \draw[->, dashed] (s1_r) to[out=90, in=90] (s1_l);
  \draw[->, dashed, looseness=1] (s2_r) to[out=-90, in=-90] (s2_l);
  \draw[->, dashed] (s3_r) to[out=-90, in=-90] (s2_l);
  \draw[->, dashed] (s4_r) to[out=90, in=90] (s3_l);

  \node[] at ($(lhs.east)!0.5!(rhs.west) + (0,0)$) {$\rrul$} ;
\end{tikzpicture}
 	}
	\caption{$\mathtt{act\_check}$}
	\label{fig:rr_act_check}
  \end{subfigure}
  \begin{subfigure}[b]{0.3\linewidth}
	\centering
	\resizebox{0.9\linewidth}{!}{
		\begin{tikzpicture}[]
  \begin{scope}[local bounding box=lhs]
    \node[diamond, draw, fill=red, inner sep=1.2] (cr) {};
    \draw[big edgec] (cr.north) to[in=-90, out=90] ($(cr.north) + (0,0.08)$) {};

    \node[big site, right=0.2 of cr] (s1) {};
    \node[draw, red, fit=(cr)(s1)] (act) {};
    \node[anchor=south west, inner sep=0.3] at (act.north west) (act_lbl) {\tiny \sf Act};
    \node[big region, fit=(act)(act_lbl)] (r2) {};
  \end{scope}

  \begin{scope}[shift={(2,0)}, local bounding box=rhs]
\node[] (redf) {\large $\nrightarrow$};
    \node[big region, fit=(redf)] (r1) {};
  \end{scope}

  \node[] at ($(lhs.east)!0.5!(rhs.west) + (0,0)$) {$\rrul$} ;
\end{tikzpicture}
 	}
	\caption{$\mathtt{act\_F}$}
	\label{fig:rr_act_f}
  \end{subfigure}

  \begin{subfigure}[b]{0.55\linewidth}
	\centering
	\resizebox{\linewidth}{!}{
		\begin{tikzpicture}[]
  \begin{scope}[local bounding box=lhs]
\node[big site] (s1_l) {};
    \node[ellipse, inner sep=0, draw, fit=(s1_l)] (bs) {};
    \node[anchor=south, inner sep=0.3] at (bs.north) (bs_lbl) {\tiny \sf Beliefs};
    \node[big region, fit=(bs)(bs_lbl)] (r1) {};

\node[diamond, draw, fill=green, inner sep=1.2, right=0.7 of s1_l] (cr) {};
    \draw[big edgec] (cr.north) to[in=-90, out=90] ($(cr.north) + (0,0.08)$) {};

    \node[big site, right=0.3 of cr] (s2_l) {};
    \node[draw, rounded corners, fit=(s2_l)] (add) {};
    \node[anchor=south west, inner sep=0.3] at (add.north west) (add_lbl) {\tiny \sf Add};

    \node[big site, right=0.3 of s2_l] (s3_l) {};
    \node[draw, rounded corners, fit=(s3_l)] (del) {};
    \node[anchor=south west, inner sep=0.3] at (del.north west) (del_lbl) {\tiny \sf Del};

    \node[big site, right=0.2 of s3_l] (s4_l) {};
    \node[draw, red, fit=(cr)(add)(add_lbl)(del)(del_lbl)(s4_l)] (act) {};
    \node[anchor=south west, inner sep=0.3] at (act.north west) (act_lbl) {\tiny \sf Act};
    \node[big region, fit=(act)(act_lbl)] (r2) {};
  \end{scope}

  \begin{scope}[shift={(4,0)}, local bounding box=rhs]
\node[big site] (s1_r) {};

    \node[big site, right=0.3 of s1_r] (s2_r) {};
    \node[draw, rounded corners, fit=(s2_r)] (add) {};
    \node[anchor=south west, inner sep=0.3] at (add.north west) (add_lbl) {\tiny \sf Add};

    \node[big site, right=0.3 of s2_r] (s3_r) {};
    \node[draw, rounded corners, fit=(s3_r)] (del) {};
    \node[anchor=south west, inner sep=0.3] at (del.north west) (del_lbl) {\tiny \sf Del};

    \node[ellipse, inner sep=0, draw, fit=(s1_r)(add)(add_lbl)(del)(del_lbl)] (bs) {};
    \node[anchor=south, inner sep=0.3] at (bs.north) (bs_lbl) {\tiny \sf Beliefs};
    \node[big region, fit=(bs)(bs_lbl)] (r1) {};

\node[big region, minimum width=20, minimum height=20, right=0.8 of s3_r] (r1) {};
  \end{scope}

  \draw[->, dashed] (s1_r) to[out=-90, in=-90] (s1_l);
  \draw[->, dashed, looseness=1] (s2_r) to[out=-90, in=-90] (s2_l);
  \draw[->, dashed, looseness=1.1] (s3_r) to[out=90, in=90] (s3_l);

  \node[] at ($(lhs.east)!0.5!(rhs.west) + (0,0)$) {$\rrul$} ;
\end{tikzpicture}
 	}
	\caption{$\mathtt{act\_T}$}
	\label{fig:rr_act_t}
  \end{subfigure}

  \caption{Reactions for actions.}
  \label{fig:rr_actions}
\end{figure}

If the pre-condition of an action is true, \ie $\mathcal{B} \models \phi$, performing (or reducing) an action consists of the reactions $\mathtt{act\_check}$ and $\mathtt{act\_T}$ as shown in \cref{fig:rr_actions}.
Firstly, the reaction $\mathtt{act\_check}$ requests the action pre-condition to be checked by nesting a $\ion{Check}{}$ entity within the belief base.
As we have established set operations to be the highest priority class,
we know a belief check operation is finite and applies atomically. Therefore, it does not alter the shape of $\enc{\mathcal{B}}$ (\ie no other \CAN rules are enabled).
After successful entailment of the action pre-condition, the reaction rule $\mathtt{act\_T}$ performs the action by updating the belief base. 
Once again, given the priority of set operations, the set updates will be effectively atomic and no other \CAN derivation rule can interrupt such a belief update.

\begin{lemma}(Faithfulness of $act$)
  When $act$ is applicable it has corresponding finite reaction sequence $\encR{\langle\mathcal{B}, act: \varphi \leftarrow \langle\phi^{+},\phi^{-}\rangle\rangle} \react^{+} \enc{\langle \mathcal{B}', nil\rangle}$.
\end{lemma}
\begin{proof}
$\react^+$ contains only reaction rules
$$\xreact{\mathtt{act\_check}}\xreacts{\{\mathtt{set\_{ops}}\}}\xreact{\mathtt{act\_{T}}}\xreacts{\{\mathtt{set\_{ops}}\}}$$
where all rules are finite, $\mathtt{set\_ops}$ applies atomically, and, by only adding auxiliary variables, all rules do not introduce intermediate states that add new branching points (as required).
\end{proof}

We have discussed what happens when the pre-condition of an action holds. 
However, it is not explicit in \CAN semantics what should be done in the case the pre-condition check fails. 
From the perspective of an AND/OR tree, as actions
are always under AND nodes, the failure needs to be propagated upwards in order
to enable failure recovery to take place. 
As introduced in \cref{sec:reductions},
the entity $\ion{ReduceF}{}$ is provided to explicitly denote the reduction failure. Therefore, we can have the reaction $\mathtt{act\_F}$ to report the failure depicted in~\cref{fig:rr_act_f}. 
Reducing to $\ion{ReduceF}{}$ enables checking
of the premise $\langle B, act \rangle \nrightarrow$ (as is done implicitly in \CAN semantics). 
Once a failure is reported, other reaction can be triggered to, for example, recover from the failure. 
\subsubsection{Plan Selection}

Recall that the agent responds to an event by selecting an applicable plan from a set of pre-defined plans. The following two
derivation rules specify the plan selection. 
The first rule~$event$ converts an event to the set of plans that
respond to that event (i.e. relevant plans), while the second rule~$select$ chooses an applicable plan (if exists) from the set of relevant plans.

The reaction rule corresponding to the derivation rule $event$ is depicted in~\cref{fig:rr_reduce_event}. As
the syntax encoding uses links to connect an event $\ion{E}{e}$ to its set of relevant plans $\ion{PlanSet}{e}$, we can encode the derivation rule $event$ with a single reaction rule by replacing the event entity $\ion{E}{e}$ with  $\ion{PlanSet}{e}$ as shown in~\cref{fig:rr_reduce_event}.

\begin{figure}
\centering
\resizebox{0.5\linewidth}{!}{
	\begin{tikzpicture}[]
  \begin{scope}[local bounding box=lhs]
    \node[regular polygon, regular polygon sides=3, red, draw, inner sep=1.2] (e) {};
    \node[big region, fit=(e)] (r1) {};

    \node[big site, right=0.8 of e] (s1_l) {};
    \node[draw, fit=(s1_l)] (rp) {};
    \node[anchor=south west, inner sep=0.3] at (rp.north west) (rp_lbl) {\tiny \sf PlanSet};

    \node[draw, fit=(rp)(rp_lbl)] (plans) {};
    \node[anchor=south west, inner sep=0.3] at (plans.north west) (plans_lbl) {\tiny \sf Plans};
    \node[big region, fit=(plans)(plans_lbl)] (r2) {};

\coordinate (pmid) at ($(e.east)!0.5!(rp.west)$);
    \node[] at ($(pmid) + (0, 0.9)$) (p) {\tiny $p$};

    \draw[big edge] (e.east) to[out=25, in=180] (pmid) to[out=0, in=180] (rp.west);
    \draw[big edge] (pmid) to[out=80, in=-90] (p.south);
  \end{scope}

  \begin{scope}[shift={(3.4,0)}, local bounding box=rhs]
    \node[big site] (s1_r) {};
    \node[draw, fit=(s1_r)] (rp) {};
    \node[anchor=south west, inner sep=0.3] at (rp.north west) (rp_lbl) {\tiny \sf PlanSet};
    \node[big region, fit=(rp)(rp_lbl)] (r1) {};

    \node[big site, right=0.9 of s1_r] (s2_r) {};
    \node[draw, fit=(s2_r)] (rp2) {};
    \node[anchor=south west, inner sep=0.3] at (rp2.north west) (rp2_lbl) {\tiny \sf PlanSet};

    \node[draw, fit=(rp2)(rp2_lbl)] (plans) {};
    \node[anchor=south west, inner sep=0.3] at (plans.north west) (plans_lbl) {\tiny \sf Plans};
    \node[big region, fit=(plans)(plans_lbl)] (r2) {};

\coordinate (pmid) at ($(rp.east)!0.5!(rp2.west)$);
    \node[] at ($(pmid) + (0, 0.9)$) (p) {\tiny $p$};

    \draw[big edge] (rp.east) to[out=0, in=180] (pmid) to[out=0, in=180] (rp2.west);
    \draw[big edge] (pmid) to[out=80, in=-90] (p.south);
  \end{scope}

  \draw[->, dashed, looseness=1] (s1_r) to[out=-90, in=-90] (s1_l);
  \draw[->, dashed, looseness=1] (s2_r) to[out=-90, in=-90] (s1_l);

  \node[] at ($(lhs.east)!0.5!(rhs.west) + (0,-0.25)$) {$\rrul$} ;
\end{tikzpicture}
 }
\caption{$\mathtt{reduce\_event}$}
\label{fig:rr_reduce_event}
\end{figure}

\begin{lemma}(Faithfulness of $event$)
  $event$ has a corresponding finite reaction sequence $\encR{\langle\mathcal{B}, e\rangle} \react^{+} \enc{\langle \mathcal{B}, e: (\mid\Delta\mid)\rangle}$.
\end{lemma}
\begin{proof}
  $\react^+$ contains only $\xreact{\mathtt{reduce\_{event}}}$. Trivial.
\end{proof}

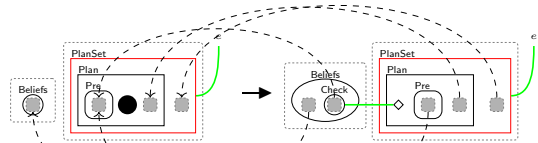
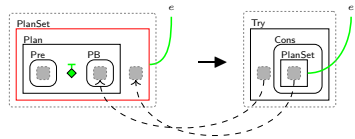
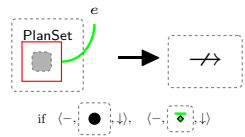
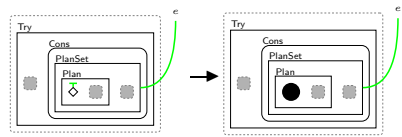
\begin{figure}
  \centering
  \begin{subfigure}[b]{\linewidth}
	\centering
	\resizebox{0.6\linewidth}{!}{
		\begin{tikzpicture}[]
  \begin{scope}[local bounding box=lhs]
    \node[big site] (s1_l) {};
    \node[ellipse, inner sep=0, draw, fit=(s1_l)] (bs) {};
    \node[anchor=south, inner sep=0.3] at (bs.north) (bs_lbl) {\tiny \sf Beliefs};
    \node[big region, fit=(bs)(bs_lbl)] (r1) {};

\node[big site, right=0.9 of s1_l] (s2_l) {};
    \node[circle, fill=black, right=0.2 of s2_l] (ct) {};
    \node[big site, right=0.1 of ct] (s3_l) {};
    \node[big site, right=0.3 of s3_l] (s4_l) {};

    \node[draw, rounded corners, fit=(s2_l)] (pre) {};
    \node[anchor=south west, inner sep=0.3] at (pre.north west) (pre_lbl) {\tiny \sf Pre};

    \node[draw, fit=(pre)(pre_lbl)(s3_l)] (pl) {};
    \node[anchor=south west, inner sep=0.3] at (pl.north west) (pl_lbl) {\tiny \sf Plan};

    \node[draw, red, fit=(pl)(pl_lbl)(s4_l)] (ps) {};
    \node[anchor=south west, inner sep=0.3] at (ps.north west) (ps_lbl) {\tiny \sf PlanSet};

    \node[big region, fit=(ps)(ps_lbl)] (r2) {};

    \node[above right=0.3 of ps] (e) {\tiny $e$};
    \draw[big edge] (ps.east) to[out=0, in =-90] (e);
  \end{scope}

  \begin{scope}[shift={(4.8,0)}, local bounding box=rhs]
    \node[big site] (s1_r) {};
    \node[big site, right=0.2 of s1_r] (s2_r) {};
    \node[ellipse, draw, inner sep=0, fit=(s2_r)] (c) {};
    \node[anchor=south, inner sep=0.3] at (c.north) (c_lbl) {\tiny \sf Check};
    \node[ellipse, inner sep=0, draw, fit=(s1_r)(c)(c_lbl)] (bs) {};
    \node[anchor=south, inner sep=0.3] at (bs.north) (bs_lbl) {\tiny \sf Beliefs};
    \node[big region, fit=(bs)(bs_lbl)] (r1) {};

\node[diamond, draw, inner sep=1.2, right=0.9 of s2_r] (cr) {};
    \node[big site, right=0.3 of cr] (s3_r) {};
    \node[big site, right=0.3 of s3_r] (s4_r) {};
    \node[big site, right=0.4 of s4_r] (s5_r) {};

    \node[draw, rounded corners, fit=(s3_r)] (pre) {};
    \node[anchor=south west, inner sep=0.3] at (pre.north west) (pre_lbl) {\tiny \sf Pre};

    \node[draw, fit=(cr)(pre)(pre_lbl)(s4_r)] (pl) {};
    \node[anchor=south west, inner sep=0.3] at (pl.north west) (pl_lbl) {\tiny \sf Plan};

    \node[draw, red, fit=(pl)(pl_lbl)(s5_r)] (ps) {};
    \node[anchor=south west, inner sep=0.3] at (ps.north west) (ps_lbl) {\tiny \sf PlanSet};

    \node[big region, fit=(ps)(ps_lbl)] (r2) {};

    \node[above right=0.3 of ps] (e) {\tiny $e$};
    \draw[big edge] (ps.east) to[out=0, in =-90] (e);

    \draw[big edge] (c.east) to[out=0, in=180] (cr.west);
  \end{scope}

  \node[] at ($(lhs.east)!0.5!(rhs.west) + (0,-0.2)$) {$\rrul$} ;

  \draw[->, dashed, looseness=1] (s1_r) to[out=-90, in=-90] (s1_l);
  \draw[->, dashed, looseness=1] (s2_r) to[out=90, in=90] (s2_l);
  \draw[->, dashed, looseness=1] (s3_r) to[out=-90, in=-90] (s2_l);
  \draw[->, dashed, looseness=1] (s4_r) to[out=90, in=90] (s3_l);
  \draw[->, dashed, looseness=1] (s5_r) to[out=90, in=90] (s4_l);
\end{tikzpicture}
 	}
	\caption{$\mathtt{select\_plan\_check}$}
	\label{fig:rr_select_check}
  \end{subfigure}

  \begin{subfigure}[b]{\linewidth}
	\centering
	\resizebox{0.4\linewidth}{!}{
		\begin{tikzpicture}[]
  \begin{scope}[local bounding box=lhs]
    \node[big site] (s1_l) {};
    \node[diamond, draw, fill=green, inner sep=1.2, right=0.3 of s1_l] (cr) {};
    \draw[big edgec] (cr.north) to[in=-90, out=90] ($(cr.north) + (0,0.08)$) {};
    \node[big site, right=0.3 of cr] (s2_l) {};

    \node[draw, rounded corners, fit=(s1_l)] (pre) {};
    \node[anchor=south west, inner sep=0.3] at (pre.north west) (pre_lbl) {\tiny \sf Pre};

    \node[draw, rounded corners, fit=(s2_l)] (pbdy) {};
    \node[anchor=south west, inner sep=0.3] at (pbdy.north west) (pbdy_lbl) {\tiny \sf PB};

    \node[draw, fit=(cr)(pre)(pre_lbl)(pbdy)(pbdy_lbl)] (pl) {};
    \node[anchor=south west, inner sep=0.3] at (pl.north west) (pl_lbl) {\tiny \sf Plan};

    \node[big site, right=0.4 of s2_l] (s3_l) {};

    \node[draw, red, fit=(pl)(pl_lbl)(s3_l)] (ps) {};
    \node[anchor=south west, inner sep=0.3] at (ps.north west) (ps_lbl) {\tiny \sf PlanSet};

    \node[above right=0.3 of ps] (e) {\tiny $e$};
    \draw[big edge] (ps.east) to[out=0, in =-90] (e);

    \node[big region, fit=(ps)(ps_lbl)] (r1) {};
  \end{scope}

  \begin{scope}[shift={(4,0)}, local bounding box=rhs]
    \node[big site] (s1_r) {};
    \node[big site, right = 0.3 of s1_r] (s2_r) {};

    \node[draw, fit=(s2_r)] (ps) {};
    \node[anchor=south west, inner sep=0.3] at (ps.north west) (ps_lbl) {\tiny \sf PlanSet};

    \node[draw, rounded corners, fit=(ps)(ps_lbl)] (cons) {};
    \node[anchor=south west, inner sep=0.3] at (cons.north west) (cons_lbl) {\tiny \sf Cons};

    \node[draw, fit=(s1_r)(cons)(cons_lbl)] (tri) {};
    \node[anchor=south west, inner sep=0.3] at (tri.north west) (tri_lbl) {\tiny \sf Try};

    \node[above right=0.3 of tri] (e) {\tiny $e$};
    \draw[big edge] (ps.east) to[out=0, in =-90] (e);

    \node[big region, fit=(tri)(tri_lbl)] (r1) {};
  \end{scope}

  \node[] at ($(lhs.east)!0.5!(rhs.west) + (0,-0.2)$) {$\rrul$} ;

  \draw[->, dashed, looseness=1] (s1_r) to[out=-90, in=-90] (s2_l);
  \draw[->, dashed, looseness=1.2] (s2_r) to[out=-90, in=-90] (s3_l);
\end{tikzpicture}
 	}
	\caption{$\mathtt{select\_plan\_T}$}
	\label{fig:rr_select_plan_T}
  \end{subfigure}

  \begin{subfigure}[b]{0.45\linewidth}
	\centering
	\resizebox{0.7\linewidth}{!}{
		\begin{tikzpicture}[]
  \begin{scope}[local bounding box=lhs]
    \node[big site] (s1_l) {};

    \node[draw, red, fit=(s1_l)] (ps) {};
    \node[anchor=south west, inner sep=0.3] at (ps.north west) (ps_lbl) {\tiny \sf PlanSet};

    \node[above right=0.3 of ps] (e) {\tiny $e$};
    \draw[big edge] (ps.east) to[out=0, in =-90] (e);

    \node[big region, fit=(ps)(ps_lbl)] (r1) {};
  \end{scope}

  \begin{scope}[shift={(2,0)}, local bounding box=rhs]
    \node[] (rf) {\large $\nrightarrow$};
    \node[above right=0.3 of rf] (e) {\tiny $e$};
    \node[big region, fit=(rf)] (r1) {};
  \end{scope}

  \begin{scope}[shift={(0,-0.7)}, local bounding box=cond, scale=0.45, transform shape]
    \node[] (if1) {if};
    \node[right=0.08 of if1] (lb1) {$\langle -, $};
    \node[circle, fill=black, right=0.2 of lb1] (ct) {};
    \node[big region, fit=(ct)] (r2) {};
    \node[right=0.15 of ct] (lb2) {$, \downarrow \rangle , $};

    \node[right=0.15 of lb2] (lb3) {$\langle -, $};
    \node[diamond, draw, fill=green, inner sep=1.2, right=0.2 of lb3] (cr) {};
    \draw[big edgec] (cr.north) to[in=-90, out=90] ($(cr.north) + (0,0.08)$) {};
    \node[big region, fit=(cr)] (cr_r) {};
    \node[right=0.15 of cr] (lb4) {$, \downarrow \rangle $};
  \end{scope}

  \node[] at ($(lhs.east)!0.5!(rhs.west) + (0,-0.2)$) {$\rrul$};

\end{tikzpicture}
 	}
	\caption{$\mathtt{select\_plan\_F}$ }
	\label{fig:rr_select_plan_F}
  \end{subfigure}
  \begin{subfigure}[b]{0.45\linewidth}
	\centering
	\resizebox{\linewidth}{!}{
		\begin{tikzpicture}[]
  \begin{scope}[local bounding box=lhs]
    \node[diamond, draw, inner sep=1.2] (cr) {};
    \draw[big edgec] (cr.north) to[in=-90, out=90] ($(cr.north) + (0,0.08)$) {};

    \node[big site, right=0.2 of cr] (s3_l) {};

    \node[draw, fit=(cr)(s3_l)] (pl) {};
    \node[anchor=south west, inner sep=0.3] at (pl.north west) (pl_lbl) {\tiny \sf Plan};

    \node[big site, right=0.2 of pl] (s4_l) {};

    \node[draw, fit=(pl)(pl_lbl)(s4_l)] (ps) {};
    \node[anchor=south west, inner sep=0.3] at (ps.north west) (ps_lbl) {\tiny \sf PlanSet};

    \node[draw, rounded corners, fit=(ps)(ps_lbl)] (cons) {};
    \node[anchor=south west, inner sep=0.3] at (cons.north west) (cons_lbl) {\tiny \sf Cons};

    \node[big site, left=0.2 of cons] (s1_l) {};

    \node[draw, fit=(s1_l)(cons)(cons_lbl)] (tri) {};
    \node[anchor=south west, inner sep=0.3] at (tri.north west) (tri_lbl) {\tiny \sf Try};

    \node[big region, fit=(tri)(tri_lbl)] (r1) {};

    \node[above right=0.3 of tri] (e) {\tiny $e$};
    \draw[big edge] (ps.east) to[out=0, in =-90] (e);
  \end{scope}

  \begin{scope}[shift={(4,0)}, local bounding box=rhs]
    \node[circle, fill=black] (ct) {};

    \node[big site, right=0.2 of ct] (s3_r) {};

    \node[draw, fit=(ct)(s3_r)] (pl) {};
    \node[anchor=south west, inner sep=0.3] at (pl.north west) (pl_lbl) {\tiny \sf Plan};

    \node[big site, right=0.2 of pl] (s4_r) {};

    \node[draw, fit=(pl)(pl_lbl)(s4_r)] (ps) {};
    \node[anchor=south west, inner sep=0.3] at (ps.north west) (ps_lbl) {\tiny \sf PlanSet};

    \node[draw, rounded corners, fit=(ps)(ps_lbl)] (cons) {};
    \node[anchor=south west, inner sep=0.3] at (cons.north west) (cons_lbl) {\tiny \sf Cons};

    \node[big site, left=0.2 of cons] (s1_r) {};

    \node[draw, fit=(s1_r)(cons)(cons_lbl)] (tri) {};
    \node[anchor=south west, inner sep=0.3] at (tri.north west) (tri_lbl) {\tiny \sf Try};

    \node[big region, fit=(tri)(tri_lbl)] (r1) {};

    \node[above right=0.3 of tri] (e) {\tiny $e$};
    \draw[big edge] (ps.east) to[out=0, in =-90] (e);
  \end{scope}
  \node[] at ($(lhs.east)!0.5!(rhs.west) + (0,-0.2)$) {$\rrul$} ;
\end{tikzpicture}
 	}
	\caption{$\rr{reset\_{planset}}$ }
	\label{fig:rr_reset_planset}
  \end{subfigure}
  \caption{Reactions for plan selection.}
  \label{fig:rr_selection}
\end{figure}

The derivation rule $select$ is modelled in a similar style to how to execute an action, beginning with the pre-condition check against the belief base before selecting an appropriate plan (if one exists). 
In detail, the reaction rule $\mathtt{select\_plan\_check}$ (in~\cref{fig:rr_select_check}) firstly finds a plan that has
not yet had the pre-conditions checked -- facilitated via an automatically-added auxiliary entity
$\ion{CheckToken}{}$ that records if a plan has already been considered -- and
initiates an operation to check the plan pre-condition. 
To ensure the automatic addition of the entity $\ion{CheckToken}{}$ to all $\ion{Plan}{}$ entities within the $\ion{Plans}{}$ perspective, an additional reaction rule is executed once at the start of a model execution to update the plan library.
As this is an implementation detail we do not add the tokens directly to the syntax encoding in the section~\cref{sec:structuralEncoding}.
After checking the pre-condition of a plan is true, the reaction rule $\mathtt{select\_plan\_T}$
(in~\cref{fig:rr_select_plan_T}) removes the selected applicable plan from the set of relevant plans, and
converts it into $\rhd$ form and keeping the rest of  plans as backups.

\begin{lemma}(Faithfulness of $select$)
  When the set of relevant plans $(|\Delta|)$ is non-empty and contains \emph{at least} one applicable plan for a given event, the \CAN derivation rule $select$ has a corresponding finite reaction sequence
$\encR{\langle\mathcal{B}, e:(|\Delta|)\rangle} \react^{+} \enc{\langle \mathcal{B}, P \rhd e:(\Delta \setminus \{\varphi : P\})\rangle}$.
\end{lemma}
\begin{proof}
  Finiteness of the plan pre-condition entailment and the use of $\ion{CheckToken}{}$ auxiliary controls ensures an applicable plan to be found in a finite number of reactions with $\react^{+}$ containing  $\xreact{\mathtt{select\_plan\_check}}^{*}\xreact{\mathtt{select\_{plan}\_{T}}}$. As the reaction $\mathtt{select\_plan\_check}$ only uses auxiliary controls, and $\mathtt{select\_plan\_T}$ performs the intention update in a single step, no new \CAN derivation rules are enabled.
\end{proof}

If no plan is applicable (a failure), the reaction rule $\mathtt{select\_plan\_F}$ (in~\cref{fig:rr_select_plan_F}) propagates a $\ion{ReduceF}{}$ up the tree. 
We use a \emph{conditional} rule to ensure the plan selection only fails if \emph{all} plans have
been checked (or there are no plans), \ie when there are no $\ion{CheckToken}{}$
entities left, and no plan that was checked is applicable.
Finally, an auxiliary reaction rule $\rr{reset\_planset}$ (\cref{fig:rr_reset_planset}) ensures that after plan selection,  the remaining unchosen (but checked) plans are re-assigned the control $\ion{CheckToken}{}$ to allow the plan to be checked again if failure recovery is required.

\subsubsection{Tree Reductions}

The remaining \CAN intention-level derivation rules specify how the AND/OR tree should
be explored. 
For example that in the sequencing structure $P_{1} ; P_{2}$ we should reduce $P_{1}$  \emph{before} $P_{2}$.

\begin{figure}
  \centering
  \begin{subfigure}[b]{0.3\linewidth}
	\centering
	\resizebox{\linewidth}{!}{
		\begin{tikzpicture}[]
  \begin{scope}[local bounding box=lhs]
    \node[big site] (s1_l) {};
    \node[big site, right=0.3 of s1_l] (s2_l) {};
    \node[draw, rounded corners, fit=(s2_l)] (cons) {};
    \node[anchor=south west, inner sep=0.3] at (cons.north west) (cons_lbl) {\tiny \sf Cons};
    \node[draw, red, fit=(s1_l)(cons)(cons_lbl)] (seq) {};
    \node[anchor=south west, inner sep=0.3] at (seq.north west) (seq_lbl) {\tiny \sf Seq};
    \node[big region, fit=(seq)(seq_lbl)] (r1) {};
  \end{scope}

  \begin{scope}[shift={(2.3,0)}, local bounding box=rhs]
    \node[big site, red, thick, fill=gray!60] (s1_l) {};
    \node[big site, right=0.3 of s1_l] (s2_l) {};
    \node[draw, rounded corners, fit=(s2_l)] (cons) {};
    \node[anchor=south west, inner sep=0.3] at (cons.north west) (cons_lbl) {\tiny \sf Cons};
    \node[draw, fit=(s1_l)(cons)(cons_lbl)] (seq) {};
    \node[anchor=south west, inner sep=0.3] at (seq.north west) (seq_lbl) {\tiny \sf Seq};
    \node[big region, fit=(seq)(seq_lbl)] (r1) {};
  \end{scope}

  \node[] at ($(lhs.east)!0.5!(rhs.west) + (0,-0.25)$) {$\rrul$} ;
\end{tikzpicture}
 	}
	\caption{$\mathtt{reduce\_seq}$}
	\label{fig:rr_reduce_seq}
  \end{subfigure}
  \begin{subfigure}[b]{0.3\linewidth}
	\centering
	\resizebox{\linewidth}{!}{
		\begin{tikzpicture}[]
  \begin{scope}[local bounding box=lhs]
    \node[big site] (s1_l) {};

    \node[draw, rounded corners, fit=(s1_l)] (cons) {};
    \node[anchor=south west, inner sep=0.3] at (cons.north west) (cons_lbl) {\tiny \sf Cons};

    \node[draw, red, fit=(s1_l)(cons)(cons_lbl)] (seq) {};
    \node[anchor=south west, inner sep=0.3] at (seq.north west) (seq_lbl) {\tiny \sf Seq};

    \node[big region, fit=(seq)(seq_lbl)] (r1) {};
  \end{scope}

  \begin{scope}[shift={(2,0)}, local bounding box=rhs]
    \node[big site, red, thick, fill=gray!60] (s1_r) {};
    \node[big region, fit=(s1_r)] (r1) {};
  \end{scope}

  \node[] at ($(lhs.east)!0.5!(rhs.west) + (0,-0.1)$) {$\rrul$} ;
\end{tikzpicture}
 	}
	\caption{$\mathtt{seq\_succ}$}
	\label{fig:rr_seq_succ}
  \end{subfigure}
  \begin{subfigure}[b]{0.3\linewidth}
	\centering
	\resizebox{\linewidth}{!}{
		\begin{tikzpicture}[]
  \begin{scope}[local bounding box=lhs]
    \node[] (rf) {\large $\nrightarrow$};
    \node[big site, right=0.2 of rf] (s1_l) {};

    \node[draw, rounded corners, fit=(s1_l)] (cons) {};
    \node[anchor=south west, inner sep=0.3] at (cons.north west) (cons_lbl) {\tiny \sf Cons};

    \node[draw, fit=(rf)(cons)(cons_lbl)] (seq) {};
    \node[anchor=south west, inner sep=0.3] at (seq.north west) (seq_lbl) {\tiny \sf Seq};

    \node[big region, fit=(seq)(seq_lbl)] (r1) {};
  \end{scope}

  \begin{scope}[shift={(2.5,0)}, local bounding box=rhs]
    \node[] (rf) {\large $\nrightarrow$};
    \node[big region, fit=(rf)] (r1) {};
  \end{scope}

  \node[] at ($(lhs.east)!0.5!(rhs.west) + (0,-0.1)$) {$\rrul$} ;
\end{tikzpicture}
 	}
	\caption{$\mathtt{seq\_fail}$}
	\label{fig:rr_seq_fail}
  \end{subfigure}
	\caption{Reactions for sequencing with priorities: $\rr{reduce\_seq} < \{ \rr{seq\_succ}, \rr{sec\_fail} \}$}
	\label{fig:rr_seq}
\end{figure}
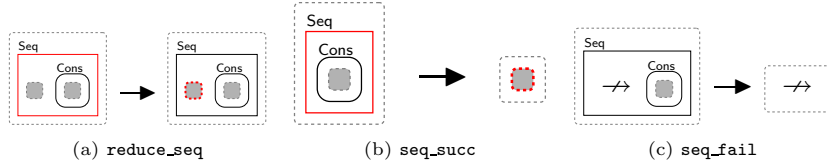

The derivation rules~$;$ and
$;_{\top}$ describe how to progress the sequencing of $P_{1} ; P_{2}$. 
The derivation rule~$;$ is encoded by the reaction rule~$\mathtt{reduce\_seq}$~(\cref{fig:rr_reduce_seq}) that pushes reduction into the first child of a sequence.
The use of a site (\ie an abstraction) in bigraphs allows this single rule to handle any type of program~$P$. 

\begin{lemma}(Faithfulness of $;$)
  $;$ has a corresponding finite reaction sequence $\encR{\langle\mathcal{B}, P_{1}; P_{2}\rangle} \react^{+} \enc{\langle \mathcal{B}', P_{1}'; P_{2}\rangle}$.
\end{lemma}
\begin{proof}
  Structural induction on
  $\encR{\mathcal{B}, P_{1}}$. The specific form of $\react^{+}$
  depends on the program $P_{1}$ (\eg $act$). 
  However, as the AND/OR tree is finite, the reductions under $P_1$ are also finite. No reductions occur under $\ion{Cons}{}$ as required.
\end{proof}

The derivation rule~$;_{\top}$ is encoded using the reaction rule~$\mathtt{seq\_succ}$ (\cref{fig:rr_seq_succ})
that matches in the case the first part of the sequence completed
successfully, \ie $\enc{nil} = 1$. 
As specified in the derivation rule in~\CAN, we not only make the children under $\ion{Cons}{}$ the new current program, but we also (try to) reduce it immediately.

\begin{lemma}(Faithfulness of $;_{\top}$)
  $;_{\top}$ has a corresponding finite reaction sequence
$\encR{\langle\mathcal{B}, nil; P_{2}\rangle} \react^{+} \enc{\langle \mathcal{B}', P_{2}'\rangle}$.
\end{lemma}
\begin{proof}
  Assume $\langle \mathcal{B}, P_2 \rangle \rightarrow$. Proof by structural induction on
  $\encR{\mathcal{B}, P_{2}} \react^{+} \enc{\langle \mathcal{B}', P_{2}'\rangle} $. 
\end{proof}
Similar to other cases, if we cannot reduce a sequence the failure is propagated
up-the-tree through the reaction rule~$\mathtt{seq\_fail}$ (\cref{fig:rr_seq_fail}).
Importantly, the reaction rule~$\rr{seq\_fail}$ -- and later failure cases -- do not require the left-hand entity to be under a $\ion{Reduce}{}$. This means the reaction can be applied as soon as a failure is discovered rather than the \emph{next} time the agent attempts to advance the intention. This matches the \CAN semantics that handle the failure of intention immediately ($\rhd_\bot$ in~\cref{fig:core_semantics} -- encoded in the next section).

When using the reactions together we need to carefully manage their ordering.
The reaction rule $\rr{reduce\_seq}$ is a \emph{generalisation} of $\rr{seq\_succ}$ and $\rr{seq\_fail}$. For example, if we get rid of $\id$ under the control $ \ion{Seq}{}$ (not $\id$ under $ \ion{Cons}{}$) on the left-hand side of reaction rule~$\rr{reduce\_seq}$, we get the  reaction rule~$\rr{seq\_succ}$.
Therefore, we enforce a priority ordering on the reaction rules as given in~\cref{fig:rr_seq} to ensure that the special case are applied when they need to. 

\subsubsection{Failure Recovery}

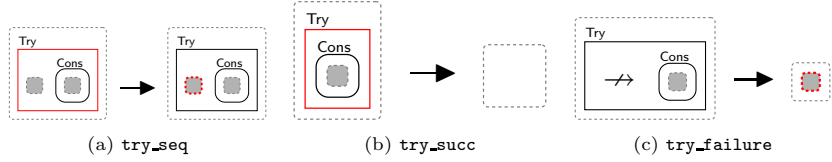
\begin{figure}
  \centering
  \begin{subfigure}[b]{0.3\linewidth}
	\centering
	\resizebox{\linewidth}{!}{
		\begin{tikzpicture}[]
  \begin{scope}[local bounding box=lhs]
    \node[big site] (s1_l) {};
    \node[big site, right = 0.3 of s1_l] (s2_l) {};

    \node[draw, rounded corners, fit=(s2_l)] (cons) {};
    \node[anchor=south west, inner sep=0.3] at (cons.north west) (cons_lbl) {\tiny \sf Cons};

    \node[draw, red, fit=(s1_l)(cons)(cons_lbl)] (tri) {};
    \node[anchor=south west, inner sep=0.3] at (tri.north west) (tri_lbl) {\tiny \sf Try};

    \node[big region, fit=(tri)(tri_lbl)] (r1) {};
  \end{scope}

  \begin{scope}[shift={(2.3,0)}, local bounding box=rhs]
    \node[big site, red, thick, fill=gray!60] (s1_r) {};
    \node[big site, right = 0.3 of s1_r] (s2_r) {};

    \node[draw, rounded corners, fit=(s2_r)] (cons) {};
    \node[anchor=south west, inner sep=0.3] at (cons.north west) (cons_lbl) {\tiny \sf Cons};

    \node[draw, fit=(s1_r)(cons)(cons_lbl)] (tri) {};
    \node[anchor=south west, inner sep=0.3] at (tri.north west) (tri_lbl) {\tiny \sf Try};
    \node[big region, fit=(tri)(tri_lbl)] (r1) {};
  \end{scope}

  \node[] at ($(lhs.east)!0.5!(rhs.west) + (0,-0.25)$) {$\rrul$} ;
\end{tikzpicture}
 	}
	\caption{$\mathtt{try\_seq}$}
	\label{fig:rr_tri_seq}
  \end{subfigure}
  \begin{subfigure}[b]{0.3\linewidth}
	\centering
	\resizebox{\linewidth}{!}{
		\begin{tikzpicture}[]
  \begin{scope}[local bounding box=lhs]
    \node[big site]  (s1_l) {};

    \node[draw, rounded corners, fit=(s1_l)] (cons) {};
    \node[anchor=south west, inner sep=0.3] at (cons.north west) (cons_lbl) {\tiny \sf Cons};

    \node[draw, red, fit=(cons)(cons_lbl)] (tri) {};
    \node[anchor=south west, inner sep=0.3] at (tri.north west) (tri_lbl) {\tiny \sf Try};

    \node[big region, fit=(tri)(tri_lbl)] (r1) {};
  \end{scope}

  \begin{scope}[shift={(2,0)}, local bounding box=rhs]
    \node[big region, minimum width=20, minimum height=20] (r1) {};
  \end{scope}

  \node[] at ($(lhs.east)!0.5!(rhs.west) + (0,-0.1)$) {$\rrul$} ;
\end{tikzpicture}
 	}
	\caption{$\mathtt{try\_succ}$}
	\label{fig:rr_tri_succ}
  \end{subfigure}
  \begin{subfigure}[b]{0.3\linewidth}
	\centering
	\resizebox{\linewidth}{!}{
		\begin{tikzpicture}[]
  \begin{scope}[local bounding box=lhs]
    \node[] (rf) {\large $\nrightarrow$};

    \node[big site, right=0.3 of rf]  (s1_l) {};

    \node[draw, rounded corners, fit=(s1_l)] (cons) {};
    \node[anchor=south west, inner sep=0.3] at (cons.north west) (cons_lbl) {\tiny \sf Cons};

    \node[draw, fit=(rf)(cons)(cons_lbl)] (tri) {};
    \node[anchor=south west, inner sep=0.3] at (tri.north west) (tri_lbl) {\tiny \sf Try};

    \node[big region, fit=(tri)(tri_lbl)] (r1) {};
  \end{scope}

  \begin{scope}[shift={(2.5,0)}, local bounding box=rhs]
    \node[reducing site]  (s1_r) {};
    \node[big region, fit=(s1_r)] (r1) {};
  \end{scope}

  \node[] at ($(lhs.east)!0.5!(rhs.west) + (0,-0.1)$) {$\rrul$} ;
\end{tikzpicture}
 	}
	\caption{$\mathtt{try\_failure}$}
	\label{fig:rr_tri_bot}
  \end{subfigure}
	\caption{Reactions for recovery with priorities: $\rr{try\_seq} < \{ \rr{try\_succ}, \rr{try\_failure} \}$.}
	\label{fig:rr_tri}
\end{figure}

 We now encode the derivation rules~$\rhd_{;}$, $\rhd_{\top}$, and $\rhd_{\bot}$ that relate to failure recovery
 
The reaction rule $\mathtt{try\_seq}$ (\cref{fig:rr_tri_seq}) encodes the derivation rule~$\rhd_{;}$ by pushing reduction into the left of the $\rhd$ operator if no failure occurs.

\begin{lemma}(Faithfulness of $\rhd_{;}$)
  $\rhd_{;}$ has a corresponding finite reaction sequence $\encR{\langle\mathcal{B}, P_{1} \rhd P_{2}\rangle} \react^{+} \enc{\langle \mathcal{B}', P_{1}' \rhd P_{2}\rangle}$.
\end{lemma}
\begin{proof}
  Structural induction on
  $\encR{\mathcal{B}, P_{1}} \react^{+} \enc{\langle \mathcal{B}', P_{1}'\rangle} $.
\end{proof}

If the selected plan was executed successfully, the reaction rule $\mathtt{try\_succ}$ (in~\cref{fig:rr_tri_succ}) encodes the derivation rule $\rhd_{\top}$ to propagate \emph{success} up-the-tree by removing the $\rhd$ structure.

\begin{lemma}(Faithfulness of $\rhd_{\top}$)
  $\langle\mathcal{B}, nil \rhd P_{2} \rangle\rangle \xrightarrow{\rhd_{\top}} \langle \mathcal{B}', nil \rhd P_{2}\rangle$, has a corresponding finite reaction sequence
$\encR{\langle\mathcal{B}, nil \rhd P_{2}\rangle} \react^{+} \enc{\langle \mathcal{B}', nil\rangle}$.
\end{lemma}
\begin{proof}
  $\react^{+}$ corresponds to $\xreact{\mathtt{try\_suc}}$. Trivial.
\end{proof}

Finally, the reaction rule $\rr{try\_failure}$ (depcited in~\cref{fig:rr_tri_bot}) encodes the derivation rule $\rhd_{\bot}$. This is the first instance where $\ion{ReduceF}{}$ is used as a \emph{premise} to denote a program which failed to progress.
To recover, the failed program is deleted and the agent tries to reduce the right-hand side of $\rhd$ (\ie by choosing from the remaining the set of relevant plans).

\begin{lemma}(Faithfulness of $\rhd_{\bot}$)
  $\rhd_{\bot}$  has a corresponding finite reaction sequence $\encR{\langle\mathcal{B}, P_{1} \rhd P_{2}\rangle} \react^{+} \enc{\langle \mathcal{B}', P_{2}'\rangle}$.
\end{lemma}
\begin{proof}
  Assume $\langle \mathcal{B}, P_{1} \rangle \nrightarrow$, and $\langle \mathcal{B}, P_{2} \rangle \rightarrow \langle \mathcal{B}, P_{2}'\rangle$. $\langle \mathcal{B}, P_{1} \rangle \nrightarrow$ implies that $\ion{ReduceF}{}$ holds as the premise required by the match.
  Structural induction on $\encR{\langle\mathcal{B}, P_{2} \rangle} \react^{+} \enc{\langle \mathcal{B}', P_{2}'\rangle}$ completes the proof. 
\end{proof}

If no backup plans apply \eg when there are no plans left to select, the failure is pushed upwards through the reaction rule $\mathtt{select\_plan\_F}$ (\cref{fig:rr_select_plan_F}).  The reaction rule $\rr{try\_failure}$ does not apply as the $\ion{Cons}{}$ entity has been removed by the point.
Finally, as with sequencing, an priority order is required as $\rr{try\_seq}$ generalises the other reactions.

\subsubsection{Agent Steps}

To complete the core semantics of \CAN, we now encode the agent-level derivation rules: $A_{event}$, $A_{step}$, and $A_{update}$.

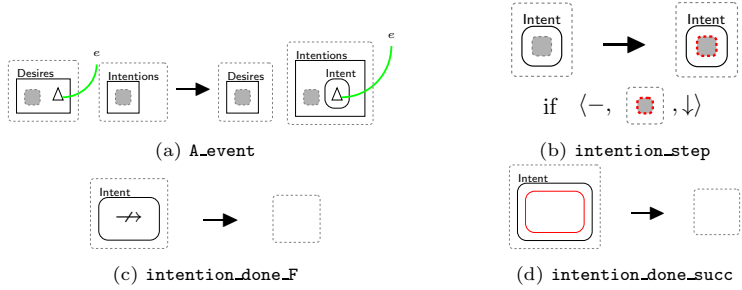
\begin{figure}
  \centering
  \begin{subfigure}[b]{0.45\linewidth}
	\centering
	\resizebox{\linewidth}{!}{
		\begin{tikzpicture}[]
  \begin{scope}[local bounding box=lhs]
    \node[big site] (s1_l) {};
    \node[event, right=0.2 of s1_l] (e1) {};

    \node[draw, fit=(s1_l)(e1)] (desires) {};
    \node[anchor=south west, inner sep=0.3] at (desires.north west) (desires_lbl) {\tiny \sf Desires};

    \node[above right=0.3 of desires] (e) {\tiny $e$};
    \draw[big edge] (e1.east) to[out=0, in =-90] (e);

    \node[big region, fit=(desires)(desires_lbl)] (r1) {};

    \node[big site, right=0.8 of e1] (s2_l) {};
    \node[draw, fit=(s2_l)] (intentions) {};
    \node[anchor=south west, inner sep=0.3] at (intentions.north west) (intentions_lbl) {\tiny \sf Intentions};

    \node[big region, fit=(intentions)(intentions_lbl)] (r1) {};

  \end{scope}

  \begin{scope}[shift={(3.2,0)}, local bounding box=rhs]
    \node[big site] (s1_r) {};

    \node[draw, fit=(s1_r)] (desires) {};
    \node[anchor=south west, inner sep=0.3] at (desires.north west) (desires_lbl) {\tiny \sf Desires};

    \node[big region, fit=(desires)(desires_lbl)] (r1) {};

    \node[big site, right=0.8 of s1_r] (s2_r) {};
    \node[event, right=0.2 of s2_r] (e1) {};

    \node[draw, rounded corners, fit=(e1)] (intent) {};
    \node[anchor=south west, inner sep=0.3] at (intent.north west) (intent_lbl) {\tiny \sf Intent};

    \node[draw, fit=(s2_r)(intent)(intent_lbl)] (intentions) {};
    \node[anchor=south west, inner sep=0.3] at (intentions.north west) (intentions_lbl) {\tiny \sf Intentions};

    \node[above right=0.3 of intentions] (e) {\tiny $e$};
    \draw[big edge] (e1.east) to[out=0, in =-90] (e);

    \node[big region, fit=(intentions)(intentions_lbl)] (r1) {};
  \end{scope}

  \node[] at ($(lhs.east)!0.5!(rhs.west) + (0,-0.2)$) {$\rrul$} ;
\end{tikzpicture}
 	}
	\caption{$\mathtt{A\_event}$}
	\label{fig:rr_agent_event}
  \end{subfigure}
  \begin{subfigure}[b]{0.45\linewidth}
	\centering
	\resizebox{0.6\linewidth}{!}{
		\begin{tikzpicture}[]
  \begin{scope}[local bounding box=lhs]
    \node[big site] (s1_l) {};
    \node[draw, rounded corners, fit=(s1_l)] (intent) {};
    \node[anchor=south west, inner sep=0.3] at (intent.north west) (intent_lbl) {\tiny \sf Intent};
    \node[big region, fit=(intent)(intent_lbl)] (r1) {};
  \end{scope}

  \begin{scope}[shift={(2,0)}, local bounding box=rhs]
    \node[reducing site] (s1_r) {};
    \node[draw, rounded corners, fit=(s1_r)] (intent) {};
    \node[anchor=south west, inner sep=0.3] at (intent.north west) (intent_lbl) {\tiny \sf Intent};
    \node[big region, fit=(intent)(intent_lbl)] (r1) {};
  \end{scope}

  \begin{scope}[shift={($(lhs.east)!0.5!(rhs.west) + (-0.9,-0.8)$)}, local bounding box=cond, scale=0.8, transform shape]
    \node[] (if1) {if};
    \node[right=0.08 of if1] (lb1) {$\langle -, $};
    \node[reducing site, right=0.25 of lb1] (ct) {};
    \node[big region, fit=(ct)] (ct_r) {};
    \node[right=0.15 of ct] (lb2) {$, \downarrow \rangle$};
  \end{scope}

  \node[] at ($(lhs.east)!0.5!(rhs.west) + (0,-0.1)$) {$\rrul$} ;
\end{tikzpicture}
 	}
	\caption{$\mathtt{intention\_step}$}
	\label{fig:rr_agent_step}
  \end{subfigure}

  \begin{subfigure}[t]{0.45\linewidth}
	\centering
	\resizebox{0.6\linewidth}{!}{
		\begin{tikzpicture}[]
  \begin{scope}[local bounding box=lhs]
    \node[] (rf) {\large $\nrightarrow$};
    \node[draw, rounded corners, fit=(rf)] (intent) {};
    \node[anchor=south west, inner sep=0.3] at (intent.north west) (intent_lbl) {\tiny \sf Intent};
    \node[big region, fit=(intent)(intent_lbl)] (r1) {};
  \end{scope}

  \begin{scope}[shift={(2.5,0)}, local bounding box=rhs]
    \node[big region, minimum height=20, minimum width=20] (r1) {};
  \end{scope}

  \node[] at ($(lhs.east)!0.5!(rhs.west) + (0,-0.1)$) {$\rrul$} ;
\end{tikzpicture}
 	}
	\caption{$\mathtt{intention\_done\_F}$}
	\label{fig:rr_agent_update}
  \end{subfigure}
  \begin{subfigure}[t]{0.45\linewidth}
	\centering
	\resizebox{0.6\linewidth}{!}{
		\begin{tikzpicture}[]
  \begin{scope}[local bounding box=lhs]
    \node[] (rf) {\phantom{\large $\nrightarrow$}};
    \node[draw, red, rounded corners, fit=(rf)] (reduce) {};
    \node[draw, rounded corners, fit=(reduce)] (intent) {};
    \node[anchor=south west, inner sep=0.3] at (intent.north west) (intent_lbl) {\tiny \sf Intent};
    \node[big region, fit=(intent)(intent_lbl)] (r1) {};
  \end{scope}

  \begin{scope}[shift={(2.5,0)}, local bounding box=rhs]
    \node[big region, minimum height=20, minimum width=20] (r1) {};
  \end{scope}

  \node[] at ($(lhs.east)!0.5!(rhs.west) + (0,-0.1)$) {$\rrul$} ;
\end{tikzpicture}
 	}
	\caption{$\mathtt{intention\_done\_succ}$}
	\label{fig:rr_agent_update_nil}
  \end{subfigure}
	\caption{Agent level reactions with priorities: $\{ \rr{A\_{event}}, \rr{intention\_step} \} < \{ \rr{intention\_done\_F}, \rr{intention\_done\_succ} \}$}
	\label{fig:rr_agent}
\end{figure}

The derivation rule $A_{event}$ allows the agent to respond to an external event by adopting it in the intention base. 
It is encoded as reaction rule $\mathtt{A\_event}$ (depicted in~\cref{fig:rr_agent_event}) that simply moves the event from being a desire to being an intention.

\begin{lemma}(Faithfulness of $A_{event}$)
  $A_{event}$ has a corresponding finite reaction sequence $\enc{\langle E^{e}, \mathcal{B}, \Gamma \rangle} \react^{+} \enc{\langle E^e \setminus e_n, \mathcal{B}', \Gamma \cup e_{n}\rangle}$.
\end{lemma}
\begin{proof}
  $\react^{+}$ corresponds to $\xreact{\mathtt{A\_event}}$. Trivial.
\end{proof}

The derivation rule $A_{step}$ allows the agent to execute a given intention one step further, i.e. one reduction step.
To encode it, we have reaction rule~$\mathtt{intention\_step}$ (in~\cref{fig:rr_agent_step}) that pushes a reduction into an intention (down-the-tree) if it is not already being reduced. This rule \emph{introduces} the reduction form~$\encR{}$ to an intention.

If the reduction below is successful we are left with a new updated $P'$ (and $\mathcal{B}'$) as required.
Unlike the \CAN derivation rule which removes the old intention and replaces it with a modified intention, ours is updated in-place.
Also multiple intentions can be reduced concurrently, \eg $\mathtt{intention\_step}$ can be applied to two different intentions. However, any single intention can only have one $\ion{Reduce}{}$ entity (as per the condition on the rule).

\begin{lemma}(Faithfulness of $A_{step}$)
  $A_{step}$ has a corresponding finite reaction sequence $\enc{\langle E^{e}, \mathcal{B}, \Gamma \rangle} \react^{+} \enc{\langle E^{e}, \mathcal{B}', \Gamma \setminus P \cup P'\rangle}$.
\end{lemma}
\begin{proof}
  Structural induction on 
  $\encR{\mathcal{B}, P} \react^{+} \enc{\langle \mathcal{B}', P'\rangle}$.
  To enable the finite reaction sequence $\encR{\mathcal{B}, P} \react^{+} \enc{\langle \mathcal{B}', P'\rangle} $, a reduction entity $\ion{Reduce}{}$ is introduced. 
  That is we move between $\Rightarrow$ and $\rightarrow$ rules.
\end{proof}

The derivation rule $A_{update}$ is encoded by reaction rules $\mathtt{intention\_done\_F}$
(\cref{fig:rr_agent_update}) and $\mathtt{intention\_done\_succ}$
(\cref{fig:rr_agent_update_nil}). 
The reaction rule $\mathtt{intention\_done\_F}$ handles the case there was a failure to progress an intention. 
That is, if after pushing a reduction into the intention (via $\mathtt{intention\_step}$), we eventually receive
$\ion{Intent}{}.\ion{ReduceF}{}$. The reaction rule~$\rr{intention\_{done}\_succ}$ is a special case of $\mathtt{intention\_done\_F}$ for the situation a intention completed successfully (returned $\ion{Intent}{}.\ion{1}{}$).
As the $A_{update}$ rule only applies on \emph{failure} to reduce an intention, $\rr{intention\_{done}\_succ}$ matches the form where we have tried to reduce an intention with the $nil$ program inside.
Importantly, this means that if an intention finishes an execution with $P = nil$, it is not until the \emph{next} attempt to reduce it that $A_{update}$ is applied.
This mirrors the \CAN semantics that cannot tell if an intention is removed because it is finished, or if it failed.

\begin{lemma}(Faithfulness of $A_{update}$)
  $A_{update}$ has a corresponding finite reaction sequence $\enc{\langle E^{e}, \mathcal{B}, \Gamma \rangle} \react^{+} \enc{\langle E^e, \mathcal{B}', \Gamma \setminus P\rangle}$.
\end{lemma}
\begin{proof} The proof is given in two cases:

\noindent \textbf{Case 1:} If $\langle \mathcal{B}, P \rangle \nrightarrow$, failure to reduce implies that $\ion{Intent}{}.\ion{ReduceF}{}$ is matched. Therefore, $\react^{+}$ corresponds to $\xreact{\mathtt{intention\_{done}\_{F}}}$. \\
    \textbf{Case 2:} If $P = nil$, then $\react^{+}$ corresponds to $\xreact{\rr{intention\_{step}}} \xreact{\mathtt{intention\_{done}\_{succ}}}$. No other rule applies as there is no way to reduce $\ion{Reduce}{}.1$ further.
\end{proof}

To ensure agent-level transition only apply if we are not currently reducing an intention both $\rr{intention\_{step}}$ and $\rr{A\_{event}}$ are in the lowest priority class.
For the same reasons as other failure recovery rules, $\rr{intention\_done\_F}$ and $\rr{intention\_done\_succ}$ are in a higher priority class so ensure they fire at the end of any failed steps.

\subsection{Proof of Correctness}

In the previous sections, we established that the \CAN derivation rules can be encoded by a finite sequence of reaction rules that do not introduce new branching (\ie \CAN derivation rules) which is not present in the starting state.  We now prove the following.

\begin{theorem}[Faithfulness]
	\label{theorem:faithful}
  For each \CAN step
  $\langle E^{e}, \mathcal{B}, \Gamma \rangle \Rightarrow \langle E'^{e}, \mathcal{B}', \Gamma' \rangle$
  there exists a \textbf{finite} sequence of reactions such that
  $\enc{\langle E^{e}, \mathcal{B}, \Gamma \rangle } \react^{+} \enc{\langle E'^{e}, \mathcal{B}', \Gamma'\rangle}$.
\end{theorem}
\begin{proof}
The proof is the collection of the proofs of Lemma 9 - 11: all three possible agent steps have corresponding finite reaction sequences and do not introduce new branching. Therefore, regardless of the step the agent takes, it is guaranteed to have a faithful encoding.
\end{proof}

\subsection{Reduction Example}\label{Example}

To show how     reduction works, in particular how failures are propagated through the AND/OR tree, we re-visit our running conference travelling example to address  external event $e_1$. Consider the following configuration: 
\begin{center}
 $	Agent = \langle \mathcal{B} = \{b_{1}, b_2, b_6, b_7\}, P = e_1 \rangle$
\end{center}
This configuration has the current belief base $\mathcal{B}$  and   current intention $P = e_1$.
The bigraph   (omitting desires and plan library) is:
\begin{center}
    $	\enc{Agent} = \ion{Beliefs}{}.(\ion{B(1)}{} \mprod \ion{B(2)}{}  \mprod \ion{B(6)}{} \mprod \ion{B(7)}{}) \pprod \ion{Intent}{}.\ion{E}{e_1} $
\end{center}

The detailed reduction step is given in~\cref{fig:reduction_example}. For succinctness, whenever appropriate, we use the mapping function to denote the part of bigraphical encoding \eg $\enc{Pl_2}$ while keeping the belief base implicitly as the background. 
The top-side of the reaction rule indicates the reaction rule that is applied and the bottom-side of the reaction rule indicates the result of application of the reaction rule, with line number~() in the beginning of each line. 
A short commentary is as follows. 
In~(1), the agent starts with an event to address. 
The reaction rule~$\mathtt{intention\_{step}}$ introduces the entity $ \ion{Reduce}{}$. 
 (2) and ~(3)  show that to reduce an event,  the event is replaced with its   relevant plans. 
  Reaction rule~$\mathtt{intention\_{step}}$ once again introduces $ \ion{Reduce}{}$     for selection of an applicable plan. 
(4) to~(6) shows the successful selection of an applicable plan,   plan $ Pl_1$. 
From~(8) to~(9), the reaction rule~$\mathtt{try\_{seq}}$ pushes reduction in the left-hand side of the $ \rhd$ symbol, and from~(9) to~(10), the reaction rule~$\mathtt{reduce\_{seq}}$ pushes the reduction into the first child of a sequence. 
 (10) to~(12)  shows the execution of an action. 
In this case, we can see that the pre-condition of the action is not met, thus producing the entity $\ion{ReduceF}{}$. 
As a consequence, this triggers   failure recovery by deleting the failed program Finally,  ~(13) to~(16) provides the successful re-selection of another applicable plan, namely plan~$Pl_2$.
\begin{figure}[htpb]
{
    \centering
	\scriptsize
	\begin{align*}
		(1) \ & \ion{Intent}{}.\ion{E}{e_1}\\
		& \xreact{\mathtt{intention\_{step}}} \\
		(2) \ & \ion{Intent}{}.\ion{Reduce}{}.\ion{E}{e_1} \\
		& \xreact{\mathtt{reduce\_{event}}} \\
		(3) \ & \ion{Intent}{}.\ion{PlanSet}{e_1}.(\ion{Plan}.(\ion{Pre}{}.\enc{\varphi_{1}} \mprod \ion{PB}{}.\ion{Seq}{}.(\enc{act_1}\mprod \ion{Cons}{}.\enc{act_2})) \mprod \enc{Pl_2})\\
		& \xreact{\mathtt{intention\_{step}}} \\
		(4) \ &\ion{Intent}{}.\ion{Reduce}{}.\ion{PlanSet}{e_1}.(\ion{Plan}.(\ion{Pre}{}.\enc{\varphi_{1}} \mprod \ion{PB}{}.\ion{Seq}{}.(\enc{act_1}\mprod \ion{Cons}{}.\enc{act_2})) \mprod \enc{Pl_2})\\
		& \xreact{\mathtt{select\_plan\_check}} \\
		(5) \ &\ion{Intent}{}.\ion{Reduce}{}.\ion{PlanSet}{e_1}.(\ion{Plan}.(\ion{CheckRes}{}.\ion{1}{} \mprod \ion{Pre}{}.\enc{\varphi_{1}} \mprod \ion{PB}{}.\ion{Seq}{}.(\enc{act_1}\mprod \ion{Cons}{}.\enc{act_2})) \mprod \enc{Pl_2})\\
		& \xreact{\mathtt{set\_ops}}^{*} \\
		(6) \ &\ion{Intent}{}.\ion{Reduce}{}.\ion{PlanSet}{e_1}.(\ion{Plan}.(\ion{CheckRes}{}.\ion{T}{}  \mprod \ion{Pre}{}.\enc{\varphi_{1}} \mprod \ion{PB}{}.\ion{Seq}{}.(\enc{act_1}\mprod \ion{Cons}{}.\enc{act_2})) \mprod \enc{Pl_2})\\
		&\xreact{\mathtt{select\_plan\_T}} \\
		(7) \ & \ion{Intent}{}.\ion{Try}{}.(\ion{Seq}{}.(\enc{act_1}\mprod \ion{Cons}{}.\enc{act_2}) \mprod \ion{Cons}{}.\ion{PlanSet}{e_1}. \enc{Pl_2})\\
		& \xreact{\mathtt{intention\_{step}}} \\
		(8) \ &\ion{Intent}{}.\ion{Reduce}{}.\ion{Try}{}.(\ion{Seq}{}.(\enc{act_1}\mprod \ion{Cons}{}.\enc{act_2}) \mprod \ion{Cons}{}.\ion{PlanSet}{e_1}. \enc{Pl_2})\\
		&\xreact{\mathtt{try\_{seq}}} \\
		(9) \ &\ion{Intent}{}.\ion{Try}{}.(\ion{Reduce}{}.\ion{Seq}{}.(\enc{act_1}\mprod \ion{Cons}{}.\enc{act_2})   \mprod \ion{Cons}{}.\ion{PlanSet}{e_1}. \enc{Pl_2})\\
		&\xreact{\mathtt{reduce\_{seq}}} \\
		(10) \ &\ion{Intent}{}.\ion{Try}{}.(\ion{Seq}{}.(\ion{Reduce}{}.\ion{Act}{}.(\ion{Pre}{}.\ion{B(3)}{} \mprod \ion{Add}{}.\ion{B(4)}{} \mprod \ion{Del}{}.\mathsf{1}) \mprod \ion{Cons}{}.\enc{act_2})  \mprod \ion{Cons}{}.\ion{PlanSet}{e_1}. \enc{Pl_2})\\
		& \xreact{\mathtt{act\_{check}}} \\
		(11) \ &\ion{Intent}{}.\ion{Try}{}.(\ion{Seq}{}.(\ion{Reduce}{}.\ion{Act}{}.(\ion{CheckRes}{}.\ion{1}{} \mprod \ion{Pre}{}.\ion{B(3)}{} \mprod \ion{Add}{}.\ion{B(4)}{} \mprod \ion{Del}{}.\mathsf{1}) \mprod \ion{Cons}{}.\enc{act_2})  \mprod \ion{Cons}{}.\ion{PlanSet}{e_1}. \enc{Pl_2})\\
		& \xreact{\mathtt{set\_ops}}^{*} \\
		(12) \ &\ion{Intent}{}.\ion{Try}{}.(\ion{Seq}{}.(\ion{Reduce}{}.\ion{Act}{}.(\ion{CheckRes}{}.\ion{F}{} \mprod \ion{Pre}{}.\ion{B(3)}{} \mprod \ion{Add}{}.\ion{B(4)}{} \mprod \ion{Del}{}.\mathsf{1}) \mprod \ion{Cons}{}.\enc{act_2})  \mprod \ion{Cons}{}.\ion{PlanSet}{e_1}. \enc{Pl_2})\\
		& \xreact{\mathtt{act\_F}} (\text{as } \mathcal{B} \nvDash b_{3}) \\
		(13) \ & \ion{Intent}{}.\ion{Try}{}.(\ion{ReduceF}{} \mprod \ion{Cons}{}.\ion{PlanSet}{e_1}.\enc{Pl_2})) \\
		& \xreact{\mathtt{try\_failure}} \\
		(14) \ & \ion{Intent}{}.\ion{Reduce}{}.\ion{PlanSet}{e_1}.\enc{Pl_2} \\
		& \xreact{\mathtt{select\_plan\_check}} \\
		(15) \ &\ion{Intent}{}.\ion{Reduce}{}.\ion{PlanSet}{e_1}.\ion{Plan}.(\ion{CheckRes}{}.\ion{1}{} \mprod \ion{Pre}{}.\enc{\varphi_{2}} \mprod \ion{PB}{}.\ion{Seq}{}.(\enc{act_{3}} \mprod \ion{Cons}{}.\ion{Seq}{}.( 	\enc{e_{2}} \mprod \ion{Cons}{}. \enc{act_{4}}))) \\
		&\xreact{\mathtt{select\_plan\_T}} \\
		(16) \ &\ion{Intent}{}.\ion{Try}{}.(\ion{Seq}{}.(\enc{act_{3}} \mprod \ion{Cons}{}.\ion{Seq}{}.( \enc{e_{2}} \mprod \ion{Cons}{}. \enc{act_{4}}))  \mprod \ion{Cons}{}.\ion{PlanSet}{e_1}. \ion{1}{})
	\end{align*}
}
\caption{Example bigraphical reduction of   event $e_{1}$ from ~\cref{fig:running_example_CAN}}
\label{fig:reduction_example}
\end{figure}

\section{Extended Features}\label{sec:extended}

The full \CAN language also supports \emph{concurrency} within
plan-bodies, and \emph{declarative goals} which allow an event to be repeatedly pursued
until specified success/failure conditions holds.  We now show how these
features are encoded as bigraph reaction rules.

\subsubsection{Concurrency}
\label{sec:concurrency}
\begin{figure}[]
	\scriptsize

	\begin{center}
		$ \dfrac{\langle \mathcal{B}, P_{1}\rangle \rightarrow \langle \mathcal{B}', P'_{1}\rangle}{\langle \mathcal{B}, (P_{1}\|P_{2})\rangle \rightarrow \langle \mathcal{B}', (P'_{1}\|P_{2})\rangle}  \|_{1}  $

	\end{center}

	\begin{center}
		$\dfrac{\langle \mathcal{B}, P_{2}\rangle \rightarrow \langle \mathcal{B}', P'_{2}\rangle}{\langle \mathcal{B}, (P_{1}\|P_{2})\rangle \rightarrow \langle \mathcal{B}', (P_{1}\|P'_{2})\rangle} \|_{2}  $
	\end{center}

	\begin{center}
		$ 	\dfrac{}{\langle \mathcal{B}, (nil\|nil) \rangle \rightarrow \langle \mathcal{B}, nil\rangle} \|_{\top}$
	\end{center}

	\caption{\CAN concurrency rules.}
	\label{fig:can_concurrency}
\end{figure}

The \CAN semantics for concurrency are given in~\cref{fig:can_concurrency}. They allow two branches within a single AND/OR tree to be reduced concurrently.
For example,   concurrency allows    an agent   to pursue two sub-tasks (i.e.~two sub-events) but the
ordering does not matter as long as they are all achieved eventually. 
However, the actual reduction is \emph{not} in parallel. Instead the agent
chooses to either reduce the left or right branch on each step (i.e.~they are interleaved). 
This matches common agent implementations that do not support multi-core processing.

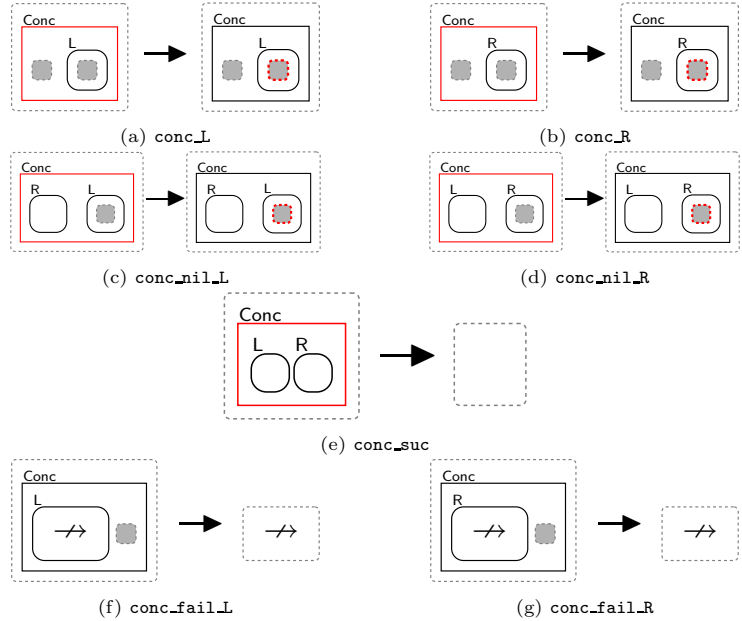
\begin{figure}
  \centering
  \begin{subfigure}[b]{0.45\linewidth}
	\centering
	\resizebox{0.8\linewidth}{!}{
		\begin{tikzpicture}[]
  \begin{scope}[local bounding box=lhs]
    \node[big site] (s1_l) {};
    \node[big site, right=0.3 of s1_l] (s2_l) {};
    \node[draw, rounded corners, fit=(s2_l)] (L) {};
    \node[anchor=south west, inner sep=0.3] at (L.north west) (L_lbl) {\tiny \sf L};
    \node[draw, red, fit=(s1_l)(L)(L_lbl)] (conc) {};
    \node[anchor=south west, inner sep=0.3] at (conc.north west) (conc_lbl) {\tiny \sf Conc};
    \node[big region, fit=(conc)(conc_lbl)] (r1) {};
  \end{scope}

  \begin{scope}[shift={(2.3,0)}, local bounding box=rhs]
    \node[big site] (s1_r) {};
    \node[big site, right=0.3 of s1_r, red, thick, fill=gray!60] (s2_r) {};
    \node[draw, rounded corners, fit=(s2_r)] (L) {};
    \node[anchor=south west, inner sep=0.3] at (L.north west) (L_lbl) {\tiny \sf L};
    \node[draw, fit=(s1_r)(L)(L_lbl)] (conc) {};
    \node[anchor=south west, inner sep=0.3] at (conc.north west) (conc_lbl) {\tiny \sf Conc};
    \node[big region, fit=(conc)(conc_lbl)] (r1) {};
  \end{scope}

  \node[] at ($(lhs.east)!0.5!(rhs.west) + (0,-0.0)$) {$\rrul$} ;
\end{tikzpicture}
 	}
	\caption{$\mathtt{conc\_{L}}$}
	\label{fig:rr_reduce_conc_L}
  \end{subfigure}
  \begin{subfigure}[b]{0.45\linewidth}
	\centering
	\resizebox{0.8\linewidth}{!}{
		\begin{tikzpicture}[]
  \begin{scope}[local bounding box=lhs]
    \node[big site] (s1_l) {};
    \node[big site, right=0.3 of s1_l] (s2_l) {};
    \node[draw, rounded corners, fit=(s2_l)] (R) {};
    \node[anchor=south west, inner sep=0.3] at (R.north west) (R_lbl) {\tiny \sf R};
    \node[draw, red, fit=(s1_l)(R)(R_lbl)] (conc) {};
    \node[anchor=south west, inner sep=0.3] at (conc.north west) (conc_lbl) {\tiny \sf Conc};
    \node[big region, fit=(conc)(conc_lbl)] (r1) {};
  \end{scope}

  \begin{scope}[shift={(2.3,0)}, local bounding box=rhs]
    \node[big site] (s1_r) {};
    \node[big site, right=0.3 of s1_r, red, thick, fill=gray!60] (s2_r) {};
    \node[draw, rounded corners, fit=(s2_r)] (R) {};
    \node[anchor=south west, inner sep=0.3] at (R.north west) (R_lbl) {\tiny \sf R};
    \node[draw, fit=(s1_r)(R)(R_lbl)] (conc) {};
    \node[anchor=south west, inner sep=0.3] at (conc.north west) (conc_lbl) {\tiny \sf Conc};
    \node[big region, fit=(conc)(conc_lbl)] (r1) {};
  \end{scope}

  \node[] at ($(lhs.east)!0.5!(rhs.west) + (0,-0.0)$) {$\rrul$} ;
\end{tikzpicture}
 	}
	\caption{$\mathtt{conc\_{R}}$}
	\label{fig:rr_reduce_conc_R}
  \end{subfigure}

  \begin{subfigure}[b]{0.45\linewidth}
	\centering
	\resizebox{0.8\linewidth}{!}{
		\begin{tikzpicture}[]
  \begin{scope}[local bounding box=lhs]
    \node[big site, right, opacity=0] (s1_l) {};
    \node[draw, rounded corners, fit=(s1_l)] (l) {};
    \node[anchor=south west, inner sep=0.3] at (l.north west) (l_lbl) {\tiny \sf R};

    \node[big site, right = 0.5 of s1_l] (s2_l) {};
    \node[draw, rounded corners, fit=(s2_l)] (R) {};
    \node[anchor=south west, inner sep=0.3] at (R.north west) (R_lbl) {\tiny \sf L};

    \node[draw, red, fit=(s1_l)(l)(l_lbl)(R)(R_lbl)] (conc) {};
    \node[anchor=south west, inner sep=0.3] at (conc.north west) (conc_lbl) {\tiny \sf Conc};
    \node[big region, fit=(conc)(conc_lbl)] (r1) {};
  \end{scope}

  \begin{scope}[shift={(2.3,0)}, local bounding box=rhs]
    \node[big site, right, opacity=0] (s1_l) {};
    \node[draw, rounded corners, fit=(s1_l)] (l) {};
    \node[anchor=south west, inner sep=0.3] at (l.north west) (l_lbl) {\tiny \sf R};

    \node[big site, red, thick, fill=gray!60, right = 0.5 of s1_l] (s2_l) {};
    \node[draw, rounded corners, fit=(s2_l)] (R) {};
    \node[anchor=south west, inner sep=0.3] at (R.north west) (R_lbl) {\tiny \sf L};

    \node[draw, fit=(s1_l)(l)(l_lbl)(R)(R_lbl)] (conc) {};
    \node[anchor=south west, inner sep=0.3] at (conc.north west) (conc_lbl) {\tiny \sf Conc};
    \node[big region, fit=(conc)(conc_lbl)] (r1) {};
  \end{scope}

  \node[] at ($(lhs.east)!0.5!(rhs.west) + (0,-0.0)$) {$\rrul$} ;
\end{tikzpicture}
 	}
	\caption{$\mathtt{conc\_nil\_{L}}$}
	\label{fig:rr_reduce_conc_nil_L}
  \end{subfigure}
  \begin{subfigure}[b]{0.45\linewidth}
	\centering
	\resizebox{0.8\linewidth}{!}{
		\begin{tikzpicture}[]
  \begin{scope}[local bounding box=lhs]
    \node[big site, right, opacity=0] (s1_l) {};
    \node[draw, rounded corners, fit=(s1_l)] (l) {};
    \node[anchor=south west, inner sep=0.3] at (l.north west) (l_lbl) {\tiny \sf L};

    \node[big site, right = 0.5 of s1_l] (s2_l) {};
    \node[draw, rounded corners, fit=(s2_l)] (R) {};
    \node[anchor=south west, inner sep=0.3] at (R.north west) (R_lbl) {\tiny \sf R};

    \node[draw, red, fit=(s1_l)(l)(l_lbl)(R)(R_lbl)] (conc) {};
    \node[anchor=south west, inner sep=0.3] at (conc.north west) (conc_lbl) {\tiny \sf Conc};
    \node[big region, fit=(conc)(conc_lbl)] (r1) {};
  \end{scope}

  \begin{scope}[shift={(2.3,0)}, local bounding box=rhs]
    \node[big site, right, opacity=0] (s1_l) {};
    \node[draw, rounded corners, fit=(s1_l)] (l) {};
    \node[anchor=south west, inner sep=0.3] at (l.north west) (l_lbl) {\tiny \sf L};

    \node[big site, red, thick, fill=gray!60, right = 0.5 of s1_l] (s2_l) {};
    \node[draw, rounded corners, fit=(s2_l)] (R) {};
    \node[anchor=south west, inner sep=0.3] at (R.north west) (R_lbl) {\tiny \sf R};

    \node[draw, fit=(s1_l)(l)(l_lbl)(R)(R_lbl)] (conc) {};
    \node[anchor=south west, inner sep=0.3] at (conc.north west) (conc_lbl) {\tiny \sf Conc};
    \node[big region, fit=(conc)(conc_lbl)] (r1) {};
  \end{scope}

  \node[] at ($(lhs.east)!0.5!(rhs.west) + (0,-0.0)$) {$\rrul$} ;
\end{tikzpicture}
 	}
	\caption{$\mathtt{conc\_nil\_{R}}$}
	\label{fig:rr_reduce_conc_nil_R}
  \end{subfigure}

  \begin{subfigure}[b]{0.45\linewidth}
	\centering
	\resizebox{0.8\linewidth}{!}{
		\begin{tikzpicture}[]

  \begin{scope}[local bounding box=lhs]
    \node[draw, rounded corners, minimum size=10] (l) {};
    \node[anchor=south west, inner sep=0.3] at (l.north west) (l_lbl) {\tiny \sf L};

    \node[draw, rounded corners, right=0.03 of l, minimum size=10] (R) {};
    \node[anchor=south west, inner sep=0.3] at (R.north west) (R_lbl) {\tiny \sf R};

    \node[draw, red, fit=(l)(R)(l_lbl)(R_lbl)] (conc) {};
    \node[anchor=south west, inner sep=0.3] at (conc.north west) (conc_lbl) {\tiny \sf Conc};
    \node[big region, fit=(conc)(conc_lbl)] (r1) {};
  \end{scope}

  \begin{scope}[shift={(2,0)}, local bounding box=rhs]
\node[draw, red, minimum width=10, minimum height=10, opacity=0] (conc) {};
    \node[anchor=south west, inner sep=0.3, opacity=0] at (conc.north west) (conc_lbl) {\tiny \sf Conc};

    \node[big region, fit=(conc)(conc_lbl)] (r1) {};
  \end{scope}

  \node[] at ($(lhs.east)!0.5!(rhs.west) + (0,-0.0)$) {$\rrul$} ;

\end{tikzpicture}
 	}
	\caption{$\mathtt{conc\_{suc}}$}
	\label{fig:rr_reduce_conc_suc}
  \end{subfigure}

  \begin{subfigure}[b]{0.45\linewidth}
	\centering
	\resizebox{0.8\linewidth}{!}{
		\begin{tikzpicture}[]
  \begin{scope}[local bounding box=lhs]
    \node[] (rf) {\large $\nrightarrow$};
    \node[big site, right=0.2 of rf] (s1_l) {};

    \node[draw, rounded corners, fit=(rf)] (l) {};
    \node[anchor=south west, inner sep=0.3] at (l.north west) (l_lbl) {\tiny \sf L};

    \node[draw, fit=(l)(l_lbl)(s1_l)] (conc) {};
    \node[anchor=south west, inner sep=0.3] at (conc.north west) (conc_lbl) {\tiny \sf Conc};

    \node[big region, fit=(conc)(conc_lbl)(s1_l)] (r1) {};
  \end{scope}

  \begin{scope}[shift={(2.5,0)}, local bounding box=rhs]
    \node[] (rf) {\large $\nrightarrow$};
    \node[big region, fit=(rf)] (r1) {};
  \end{scope}

  \node[] at ($(lhs.east)!0.5!(rhs.west) + (0,-0.0)$) {$\rrul$} ;
\end{tikzpicture}
 	}
	\caption{$\mathtt{conc\_fail\_{L}}$}
	\label{fig:rr_reduce_conc_fail_L}
  \end{subfigure}
  \begin{subfigure}[b]{0.45\linewidth}
	\centering
	\resizebox{0.8\linewidth}{!}{
		\begin{tikzpicture}[]
  \begin{scope}[local bounding box=lhs]
    \node[] (rf) {\large $\nrightarrow$};
    \node[big site, right=0.2 of rf] (s1_l) {};

    \node[draw, rounded corners, fit=(rf)] (R) {};
    \node[anchor=south west, inner sep=0.3] at (R.north west) (R_lbl) {\tiny \sf R};

    \node[draw, fit=(R)(R_lbl)(s1_l)] (conc) {};
    \node[anchor=south west, inner sep=0.3] at (conc.north west) (conc_lbl) {\tiny \sf Conc};

    \node[big region, fit=(conc)(conc_lbl)(s1_l)] (r1) {};
  \end{scope}

  \begin{scope}[shift={(2.5,0)}, local bounding box=rhs]
    \node[] (rf) {\large $\nrightarrow$};
    \node[big region, fit=(rf)] (r1) {};
  \end{scope}

  \node[] at ($(lhs.east)!0.5!(rhs.west) + (0,-0.0)$) {$\rrul$} ;
\end{tikzpicture}
 	}
	\caption{$\mathtt{conc\_fail\_{R}}$}
	\label{fig:rr_reduce_conc_fail_R}
  \end{subfigure}

	\caption{Reactions for concurrency with priorities: $\{ \rr{conc\_L}, \rr{conc\_R} \} < \{ \rr{conc\_nil\_L}, \rr{conc\_nil\_R} \} < \{ \rr{conc\_suc}, \rr{conc\_fail\_L}, \rr{con\_fail\_R} \}$}
	\label{fig:rr_concurrency}
\end{figure}

Two reaction rules $\mathtt{conc\_{L}}$ (\cref{fig:rr_reduce_conc_L}), and $\mathtt{conc\_{R}}$ (\cref{fig:rr_reduce_conc_R}) encode concurrency. As they have the same priority, these rules specify that reduction can be pushed
down \emph{either} the
left or right branch.

\begin{lemma}(Faithfulness of $\parallel_{1}$ and $\parallel_{2}$)
  $\parallel_{1}$ and $\parallel_{2}$  have a corresponding finite reaction sequence $\encR{\langle\mathcal{B}, P_{1} \parallel P_{2}\rangle} \react^{+} \enc{\langle \mathcal{B}', P_{1}' \parallel P_{2}\rangle}$ and $\encR{\langle\mathcal{B}, P_{1} \parallel P_{2}\rangle} \react^{+} \enc{\langle \mathcal{B}', P_{1} \parallel P_{2}'\rangle}$, respectively.
\end{lemma}
\begin{proof}
The proof of $\parallel_{1}$ is given in the following two cases:

\noindent \textbf{Case 1.} If $P_{2} = nil$ then $\langle \mathcal{B}, P_{2} \rangle \nrightarrow$. The reaction rule $\rr{conc\_nil\_L}$ in~\cref{fig:rr_reduce_conc_nil_L} applies. The proof completes by induction on $\encR{\mathcal{B}, P_{1}}$. \\
\textbf{Case 2.} If $P_{1} \neq nil$ then  the reaction rule~$\rr{conc\_L}$  applies. The proof completes by induction on $\encR{\mathcal{B}, P_{1}}$.
  
\noindent The proof of $\parallel_{2}$ can be given in a similar way. 
\end{proof}
Concurrent programs are considered successfully completed when both branches complete, \ie reduced to $nil$. The reaction rule $\mathtt{conc\_{suc}}$ is given in~\cref{fig:rr_reduce_conc_suc} to handles the completion of concurrent programs.

\begin{lemma}(Faithfulness of $\parallel_{\top}$)
  $\parallel_{\top}$  has a corresponding finite reaction sequence $\encR{\langle\mathcal{B}, nil \parallel nil\rangle} \react^{+} \enc{\langle \mathcal{B}', nil\rangle}$.
\end{lemma}
\begin{proof}
	$\react^{+}$ corresponds to $\xreact{\mathtt{conc\_{suc}}}$. Trivial.
\end{proof}

As always for the case of failures, additional reaction rules
$\rr{conc\_{fail}\_{L}}$~(\cref{fig:rr_reduce_conc_fail_L}) and
$\rr{conc\_{fail}\_{R}}$~(\cref{fig:rr_reduce_conc_fail_R}) propagate failure up-the-tree if either of the two concurrent branches results in a failure.
Importantly, we fail as soon as either branch fails rather than waiting for the second branch to complete (either successfully or with failure). That is, we short-circuit computation as required.

As before, the priority ordering on the reaction rules is required as some reaction rules generalise others, \eg the reaction rule $\rr{conc\_R}$ would also match reaction rule $\rr{conc\_succ}$.

\subsubsection{Declarative Goals}

\begin{figure}
	\scriptsize
	\begin{center}
		$ \dfrac{\mathcal{B} \models \varphi_{s} }{\langle \mathcal{B},  \mathit{goal}(\varphi_{s}, \mathit{P}, \varphi_{f}) \rangle  \rightarrow \langle \mathcal{B},  nil \rangle} G_{s}  $
		\qquad
		$ \dfrac{\mathcal{B} \models \varphi_{f} }{\langle \mathcal{B}, \mathit{goal}(\varphi_{s}, \mathit{P}, \varphi_{f}) \rangle  \rightarrow \langle \mathcal{B},  ?\mathit{false} \rangle}  G_{f} $
	\end{center}

	\begin{center}
		$ \dfrac{P \neq P_{1}\rhd P_{2} \ \ \  \mathcal{B} \nvDash \varphi_{s} \ \ \ \mathcal{B} \nvDash \varphi_{f}}{\langle \mathcal{B}, \mathit{goal}(\varphi_{s}, \mathit{P}, \varphi_{f}) \rangle  \rightarrow \langle \mathcal{B}, \mathit{goal}(\varphi_{s}, \mathit{P} \rhd P, \varphi_{f}) \rangle}  G_{init} $
	\end{center}

	\begin{center}
		$ \dfrac{\mathcal{B} \nvDash \varphi_{s} \ \ \ \mathcal{B} \nvDash \varphi_{f} \ \ \ \langle \mathcal{B},  P_{1}\rangle \rightarrow \langle \mathcal{B}', \mathcal{A}', P'_{1}\rangle  }{\langle \mathcal{B}, \mathit{goal}(\varphi_{s}, P_{1} \rhd P_{2}, \varphi_{f}) \rangle \rightarrow \langle \mathcal{B}', \mathcal{A}',\mathit{goal}(\varphi_{s}, P'_{1} \rhd P_{2}, \varphi_{f}) \rangle}  G_{;} $
	\end{center}

	\begin{center}
		$ \dfrac{\mathcal{B} \nvDash \varphi_{s} \ \ \ \mathcal{B} \nvDash \varphi_{f} \ \ \ \langle \mathcal{B},  P_{1}\rangle \nrightarrow }{\langle \mathcal{B}, \mathit{goal}(\varphi_{s}, P_{1} \rhd P_{2}, \varphi_{f}) \rangle \rightarrow \langle \mathcal{B},  \mathit{goal}(\varphi_{s}, P_{2} \rhd P_{2}, \varphi_{f}) \rangle}  G_{\rhd} $
	\end{center}
	\caption{Derivation rules for declarative goals.}
	\label{fig:can_declarative_goals}
\end{figure}

Declarative goals allow an agent to \emph{persistently} respond to some event
$e$ until either the success or failure conditions are met. The \CAN semantics for
declarative goals are given~in~\cref{fig:can_declarative_goals}. 
The derivation rules $G_s$ and $G_f$ deal with the cases when either the success condition~$\varphi_s$ or the failure condition~$\varphi_f$ become true. 
The derivation rule~$ G_{init}$ initialises persistence by setting the program in the declarative goal to be $ P \rhd P$, \ie if $P$ fails try $P$ again. 
The derivation rule~$ G_;$ takes care of performing a single step on an already initialised program. 
Finally, the derivation rule~$G_{\rhd}$ re-starts the original program if the current program has finished or got blocked (when neither~$\varphi_s$ nor~$\varphi_f$ becomes true). 

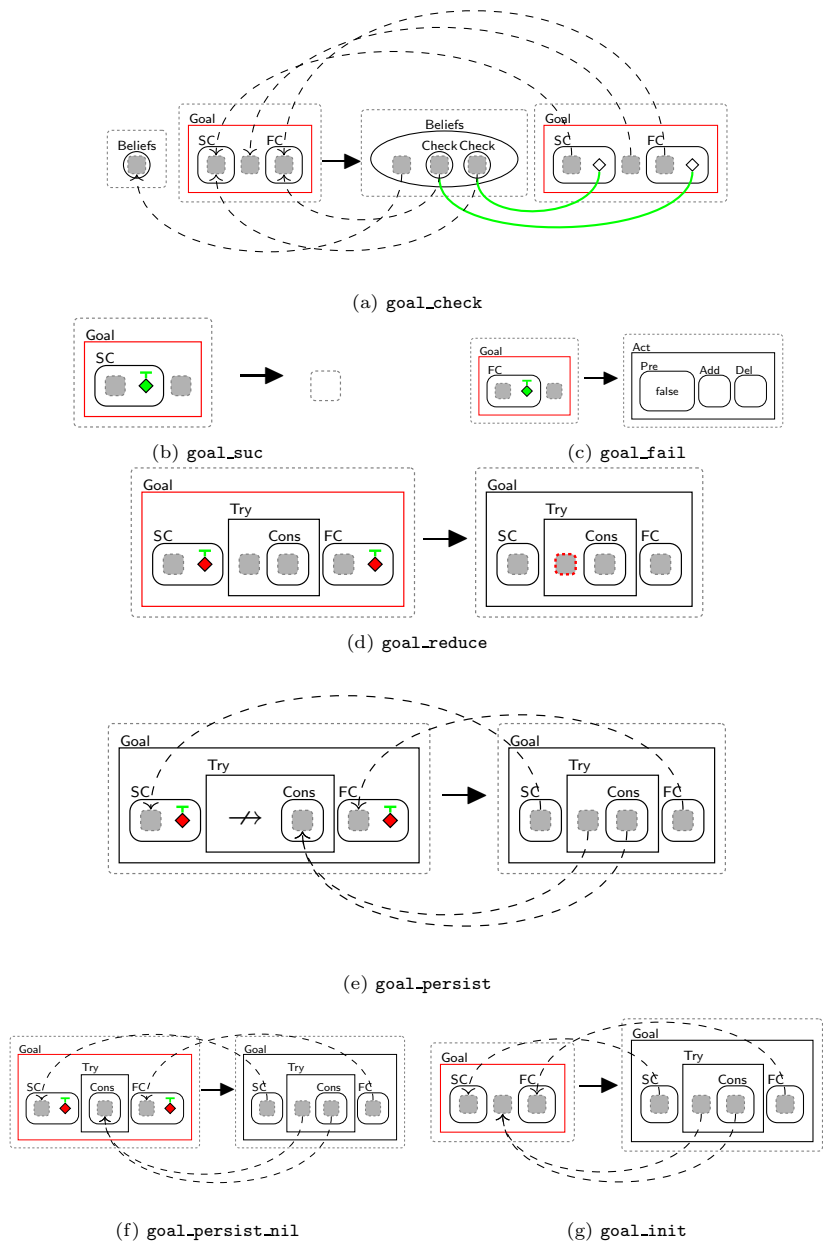
\begin{figure}
  \centering
  \begin{subfigure}[t]{\linewidth}
	\centering
	\resizebox{0.7\linewidth}{!}{
		\begin{tikzpicture}[]
  \begin{scope}[local bounding box=lhs]
    \node[big site] (s1_l) {};
    \node[ellipse, inner sep=0, draw, fit=(s1_l)] (bs) {};
    \node[anchor=south, inner sep=0.3] at (bs.north) (bs_lbl) {\tiny \sf Beliefs};
    \node[big region, fit=(bs)(bs_lbl)] (r1) {};

    \node[big site, right=0.8 of s1_l] (s2_l) {};
    \node[draw, rounded corners, fit=(s2_l)] (sc) {};
    \node[anchor=south west, inner sep=0.3] at (sc.north west) (sc_lbl) {\tiny \sf SC};

    \node[big site, right=0.2 of s2_l] (s3_l) {};

    \node[big site, right=0.2 of s3_l] (s4_l) {};
    \node[draw, rounded corners, fit=(s4_l)] (fc) {};
    \node[anchor=south west, inner sep=0.3] at (fc.north west) (fc_lbl) {\tiny \sf FC};

    \node[draw, red, fit=(sc)(sc_lbl)(s3_l)(fc)(fc_lbl)] (goal) {};
    \node[anchor=south west, inner sep=0.3] at (goal.north west) (goal_lbl) {\tiny \sf Goal};
    \node[big region, fit=(goal)(goal_lbl)] (r2) {};
  \end{scope}

  \begin{scope}[shift={(3.5,0)}, local bounding box=rhs]
    \node[big site] (s1_r) {};
    \node[big site, right=0.25 of s1_r] (s2_r) {};
    \node[ellipse, draw, inner sep=0, fit=(s2_r)] (c1) {};
    \node[anchor=south, inner sep=0.3] at (c1.north) (c1_lbl) {\tiny \sf Check};

    \node[big site, right=0.25 of s2_r] (s3_r) {};
    \node[ellipse, draw, inner sep=0, fit=(s3_r)] (c2) {};
    \node[anchor=south, inner sep=0.3] at (c2.north) (c2_lbl) {\tiny \sf Check};

    \node[ellipse, inner sep=0, draw, fit=(s1_r)(c1)(c1_lbl)(c2)(c2_lbl)] (bs) {};
    \node[anchor=south, inner sep=0.3] at (bs.north) (bs_lbl) {\tiny \sf Beliefs};
    \node[big region, fit=(bs)(bs_lbl)] (r1) {};

\node[big site, right=of s3_r] (s4_r) {};
    \node[diamond, draw, inner sep=1.2, right=0.15 of s4_r] (sc_c) {};
    \node[draw, rounded corners, fit=(s4_r)(sc_c)] (sc) {};
    \node[anchor=south west, inner sep=0.3] at (sc.north west) (sc_lbl) {\tiny \sf SC};

    \node[big site, right=0.2 of sc_c] (s5_r) {};

    \node[big site, right=0.2 of s5_r] (s6_r) {};
    \node[diamond, draw, inner sep=1.2, right=0.15 of s6_r] (fc_c) {};
    \node[draw, rounded corners, fit=(s6_r)(fc_c)] (fc) {};
    \node[anchor=south west, inner sep=0.3] at (fc.north west) (fc_lbl) {\tiny \sf FC};

    \node[draw, red, fit=(sc)(sc_lbl)(s5_r)(fc)(fc_lbl)] (goal) {};
    \node[anchor=south west, inner sep=0.3] at (goal.north west) (goal_lbl) {\tiny \sf Goal};
    \node[big region, fit=(goal)(goal_lbl)] (r2) {};

\draw[big edge] (sc_c.south) to[out=-90, in=-90] (c2.south);
    \draw[big edge, looseness=0.7] (fc_c.south) to[out=-90, in=-90] (c1.south);
  \end{scope}

\draw[->, dashed] (s1_r) to[out=-90, in=-90] (s1_l);
  \draw[->, dashed] (s2_r) to[out=-90, in=-90] (s4_l);
  \draw[->, dashed] (s3_r) to[out=-90, in=-90] (s2_l);
  \draw[->, dashed, looseness=1] (s4_r) to[out=90, in=90] (s2_l);
  \draw[->, dashed, looseness=1.15] (s5_r) to[out=90, in=90] (s3_l);
  \draw[->, dashed, looseness=1.3] (s6_r) to[out=90, in=90] (s4_l);

  \node[] at ($(lhs.east)!0.5!(rhs.west) + (0,-0.0)$) {$\rrul$} ;

\end{tikzpicture}
 	}
	\caption{$\mathtt{goal\_{check}}$}
	\label{fig:rr_goal_check}
  \end{subfigure}

  \begin{subfigure}[b]{0.45\linewidth}
	\centering
	\resizebox{0.7\linewidth}{!}{
		\begin{tikzpicture}[]
  \begin{scope}[local bounding box=lhs]
    \node[big site] (s1_l) {};
    \node[diamond, draw, inner sep=1.2, fill=green, right=0.15 of s1_l] (sc_c) {};
    \draw[big edgec] (sc_c.north) to[in=-90, out=90] ($(sc_c.north) + (0,0.08)$) {};

    \node[draw, rounded corners, fit=(s1_l)(sc_c)] (sc) {};
    \node[anchor=south west, inner sep=0.3] at (sc.north west) (sc_lbl) {\tiny \sf SC};

    \node[big site, right=0.2 of sc_c] (s2_l) {};

    \node[draw, red, fit=(sc)(sc_lbl)(s2_l)] (goal) {};
    \node[anchor=south west, inner sep=0.3] at (goal.north west) (goal_lbl) {\tiny \sf Goal};
    \node[big region, fit=(goal)(goal_lbl)] (r1) {};
  \end{scope}

  \begin{scope}[shift={(2.5,0)}, local bounding box=rhs]
    \node[big region, minimum height=10, minimum width=10] (r1) {};
  \end{scope}

  \node[] at ($(lhs.east)!0.5!(rhs.west) + (0,-0.0)$) {$\rrul$} ;

\end{tikzpicture}
 	}
	\caption{$\mathtt{goal\_{suc}}$}
	\label{fig:rr_goal_suc}
  \end{subfigure}
  \begin{subfigure}[b]{0.45\linewidth}
	\centering
	\resizebox{0.8\linewidth}{!}{
		\begin{tikzpicture}[]
  \begin{scope}[local bounding box=lhs]
    \node[big site] (s1_l) {};
    \node[diamond, draw, inner sep=1.2, fill=green, right=0.15 of s1_l] (fc_c) {};
    \draw[big edgec] (fc_c.north) to[in=-90, out=90] ($(fc_c.north) + (0,0.08)$) {};

    \node[draw, rounded corners, fit=(s1_l)(fc_c)] (fc) {};
    \node[anchor=south west, inner sep=0.3] at (fc.north west) (fc_lbl) {\tiny \sf FC};

    \node[big site, right=0.2 of fc_c] (s2_l) {};

    \node[draw, red, fit=(fc)(fc_lbl)(s2_l)] (goal) {};
    \node[anchor=south west, inner sep=0.3] at (goal.north west) (goal_lbl) {\tiny \sf Goal};
    \node[big region, fit=(goal)(goal_lbl)] (r1) {};
  \end{scope}

  \begin{scope}[shift={(2.5,0)}, local bounding box=rhs]
    \node[] (false) {\tiny \sf false};
    \node[draw, rounded corners, fit=(false)] (pre) {};
    \node[anchor=south west, inner sep=0.3] at (pre.north west) (pre_lbl) {\tiny \sf Pre};

    \node[big site, right=0.3 of false, opacity=0] (s1_r) {};
    \node[draw, rounded corners, fit=(s1_r)] (add) {};
    \node[anchor=south west, inner sep=0.3] at (add.north west) (add_lbl) {\tiny \sf Add};

    \node[big site, right=0.3 of s1_r, opacity=0] (s2_r) {};
    \node[draw, rounded corners, fit=(s2_r)] (del) {};
    \node[anchor=south west, inner sep=0.3] at (del.north west) (del_lbl) {\tiny \sf Del};

    \node[draw, fit=(add)(add_lbl)(del)(del_lbl)(pre)(pre_lbl)] (act) {};
    \node[anchor=south west, inner sep=0.3] at (act.north west) (act_lbl) {\tiny \sf Act};
    \node[big region, fit=(act)(act_lbl)] (r2) {};
  \end{scope}

  \node[] at ($(lhs.east)!0.5!(rhs.west) + (0,-0.0)$) {$\rrul$} ;

\end{tikzpicture}
 	}
	\caption{$\mathtt{goal\_{fail}}$}
	\label{fig:rr_goal_fail}
  \end{subfigure}

  \begin{subfigure}[b]{\linewidth}
	\centering
	\resizebox{0.65\linewidth}{!}{
		\begin{tikzpicture}[]
  \begin{scope}[local bounding box=lhs]
    \node[big site] (s1_l) {};
    \node[diamond, draw, inner sep=1.2, fill=red, right=0.15 of s1_l] (sc_c) {};
    \draw[big edgec] (sc_c.north) to[in=-90, out=90] ($(sc_c.north) + (0,0.08)$) {};
    \node[draw, rounded corners, fit=(s1_l)(sc_c)] (sc) {};
    \node[anchor=south west, inner sep=0.3] at (sc.north west) (sc_lbl) {\tiny \sf SC};

\node[big site, right=0.3 of sc_c] (s2_l) {};

    \node[big site, right=0.2 of s2_l] (s3_l) {};
    \node[draw, rounded corners, fit=(s3_l)] (cons) {};
    \node[anchor=south west, inner sep=0.3] at (cons.north west) (cons_lbl) {\tiny \sf Cons};

    \node[draw, fit=(s2_l)(cons)(cons_lbl)] (tri) {};
    \node[anchor=south west, inner sep=0.3] at (tri.north west) (tri_lbl) {\tiny \sf Try};

\node[big site, right=0.4 of s3_l] (s4_l) {};
    \node[diamond, draw, inner sep=1.2, fill=red, right=0.15 of s4_l] (fc_c) {};
    \draw[big edgec] (fc_c.north) to[in=-90, out=90] ($(fc_c.north) + (0,0.08)$) {};
    \node[draw, rounded corners, fit=(s4_l)(fc_c)] (fc) {};
    \node[anchor=south west, inner sep=0.3] at (fc.north west) (fc_lbl) {\tiny \sf FC};

\node[draw, red, fit=(sc)(sc_lbl)(fc)(fc_lbl)(tri)(tri_lbl)] (goal) {};
    \node[anchor=south west, inner sep=0.3] at (goal.north west) (goal_lbl) {\tiny \sf Goal};
    \node[big region, fit=(goal)(goal_lbl)] (r1) {};
  \end{scope}

  \begin{scope}[shift={(4,0)}, local bounding box=rhs]
    \node[big site] (s1_r) {};
    \node[draw, rounded corners, fit=(s1_r)] (sc) {};
    \node[anchor=south west, inner sep=0.3] at (sc.north west) (sc_lbl) {\tiny \sf SC};

\node[big site, right=0.3 of s1_r, red, thick, fill=gray!60] (s2_r) {};

    \node[big site, right=0.2 of s2_r] (s3_r) {};
    \node[draw, rounded corners, fit=(s3_r)] (cons) {};
    \node[anchor=south west, inner sep=0.3] at (cons.north west) (cons_lbl) {\tiny \sf Cons};

    \node[draw, fit=(s2_r)(cons)(cons_lbl)] (tri) {};
    \node[anchor=south west, inner sep=0.3] at (tri.north west) (tri_lbl) {\tiny \sf Try};

\node[big site, right=0.4 of s3_r] (s4_r) {};
    \node[draw, rounded corners, fit=(s4_r)] (fc) {};
    \node[anchor=south west, inner sep=0.3] at (fc.north west) (fc_lbl) {\tiny \sf FC};

\node[draw, fit=(sc)(sc_lbl)(fc)(fc_lbl)(tri)(tri_lbl)] (goal) {};
    \node[anchor=south west, inner sep=0.3] at (goal.north west) (goal_lbl) {\tiny \sf Goal};
    \node[big region, fit=(goal)(goal_lbl)] (r1) {};
  \end{scope}

  \node[] at ($(lhs.east)!0.5!(rhs.west) + (0,-0.0)$) {$\rrul$} ;

\end{tikzpicture}
 	}
	\caption{$\mathtt{goal\_{reduce}}$}
	\label{fig:rr_goal_reduce}
  \end{subfigure}

  \begin{subfigure}[b]{\linewidth}
	\centering
	\resizebox{0.7\linewidth}{!}{
		\begin{tikzpicture}[]
  \begin{scope}[local bounding box=lhs]
    \node[big site] (s1_l) {};
    \node[diamond, draw, inner sep=1.2, fill=red, right=0.15 of s1_l] (sc_c) {};
    \draw[big edgec] (sc_c.north) to[in=-90, out=90] ($(sc_c.north) + (0,0.08)$) {};
    \node[draw, rounded corners, fit=(s1_l)(sc_c)] (sc) {};
    \node[anchor=south west, inner sep=0.3] at (sc.north west) (sc_lbl) {\tiny \sf SC};

\node[right=0.3 of sc_c] (rf) {\large $\nrightarrow$};

    \node[big site, right=0.2 of rf] (s2_l) {};
    \node[draw, rounded corners, fit=(s2_l)] (cons) {};
    \node[anchor=south west, inner sep=0.3] at (cons.north west) (cons_lbl) {\tiny \sf Cons};

    \node[draw, fit=(rf)(cons)(cons_lbl)] (tri) {};
    \node[anchor=south west, inner sep=0.3] at (tri.north west) (tri_lbl) {\tiny \sf Try};

\node[big site, right=0.4 of s2_l] (s3_l) {};
    \node[diamond, draw, inner sep=1.2, fill=red, right=0.15 of s3_l] (fc_c) {};
    \draw[big edgec] (fc_c.north) to[in=-90, out=90] ($(fc_c.north) + (0,0.08)$) {};
    \node[draw, rounded corners, fit=(s3_l)(fc_c)] (fc) {};
    \node[anchor=south west, inner sep=0.3] at (fc.north west) (fc_lbl) {\tiny \sf FC};

\node[draw, fit=(sc)(sc_lbl)(fc)(fc_lbl)(tri)(tri_lbl)] (goal) {};
    \node[anchor=south west, inner sep=0.3] at (goal.north west) (goal_lbl) {\tiny \sf Goal};
    \node[big region, fit=(goal)(goal_lbl)] (r1) {};
  \end{scope}

  \begin{scope}[shift={(4.5,0)}, local bounding box=rhs]
    \node[big site] (s1_r) {};
    \node[draw, rounded corners, fit=(s1_r)] (sc) {};
    \node[anchor=south west, inner sep=0.3] at (sc.north west) (sc_lbl) {\tiny \sf SC};

\node[big site, right=0.3 of s1_r] (s2_r) {};

    \node[big site, right=0.2 of s2_r] (s3_r) {};
    \node[draw, rounded corners, fit=(s3_r)] (cons) {};
    \node[anchor=south west, inner sep=0.3] at (cons.north west) (cons_lbl) {\tiny \sf Cons};

    \node[draw, fit=(s2_r)(cons)(cons_lbl)] (tri) {};
    \node[anchor=south west, inner sep=0.3] at (tri.north west) (tri_lbl) {\tiny \sf Try};

\node[big site, right=0.4 of s3_r] (s4_r) {};
    \node[draw, rounded corners, fit=(s4_r)] (fc) {};
    \node[anchor=south west, inner sep=0.3] at (fc.north west) (fc_lbl) {\tiny \sf FC};

\node[draw, fit=(sc)(sc_lbl)(fc)(fc_lbl)(tri)(tri_lbl)] (goal) {};
    \node[anchor=south west, inner sep=0.3] at (goal.north west) (goal_lbl) {\tiny \sf Goal};
    \node[big region, fit=(goal)(goal_lbl)] (r1) {};
  \end{scope}

\draw[->, dashed] (s1_r) to[out=90, in=90] (s1_l);
  \draw[->, dashed] (s2_r) to[out=-90, in=-90] (s2_l);
  \draw[->, dashed] (s3_r) to[out=-90, in=-90] (s2_l);
  \draw[->, dashed] (s4_r) to[out=90, in=90] (s3_l);

  \node[] at ($(lhs.east)!0.5!(rhs.west) + (0,-0.0)$) {$\rrul$} ;

\end{tikzpicture}
 	}
	\caption{$\mathtt{goal\_{persist}}$}
	\label{fig:rr_goal_persist}
  \end{subfigure}

  \begin{subfigure}[b]{.45\linewidth}
	\centering
	\resizebox{1.0\linewidth}{!}{
		\begin{tikzpicture}[]
  \begin{scope}[local bounding box=lhs]
    \node[big site] (s1_l) {};
    \node[diamond, draw, inner sep=1.2, fill=red, right=0.15 of s1_l] (sc_c) {};
    \draw[big edgec] (sc_c.north) to[in=-90, out=90] ($(sc_c.north) + (0,0.08)$) {};
    \node[draw, rounded corners, fit=(s1_l)(sc_c)] (sc) {};
    \node[anchor=south west, inner sep=0.3] at (sc.north west) (sc_lbl) {\tiny \sf SC};

\node[big site, right=0.4 of sc_c] (s2_l) {};
    \node[draw, rounded corners, fit=(s2_l)] (cons) {};
    \node[anchor=south west, inner sep=0.3] at (cons.north west) (cons_lbl) {\tiny \sf Cons};

    \node[draw, fit=(cons)(cons_lbl)] (tri) {};
    \node[anchor=south west, inner sep=0.3] at (tri.north west) (tri_lbl) {\tiny \sf Try};

\node[big site, right=0.4 of s2_l] (s3_l) {};
    \node[diamond, draw, inner sep=1.2, fill=red, right=0.15 of s3_l] (fc_c) {};
    \draw[big edgec] (fc_c.north) to[in=-90, out=90] ($(fc_c.north) + (0,0.08)$) {};
    \node[draw, rounded corners, fit=(s3_l)(fc_c)] (fc) {};
    \node[anchor=south west, inner sep=0.3] at (fc.north west) (fc_lbl) {\tiny \sf FC};

\node[draw, red, fit=(sc)(sc_lbl)(fc)(fc_lbl)(tri)(tri_lbl)] (goal) {};
    \node[anchor=south west, inner sep=0.3] at (goal.north west) (goal_lbl) {\tiny \sf Goal};
    \node[big region, fit=(goal)(goal_lbl)] (r1) {};
  \end{scope}

  \begin{scope}[shift={(3.5,0)}, local bounding box=rhs]
    \node[big site] (s1_r) {};
    \node[draw, rounded corners, fit=(s1_r)] (sc) {};
    \node[anchor=south west, inner sep=0.3] at (sc.north west) (sc_lbl) {\tiny \sf SC};

\node[big site, right=0.3 of s1_r] (s2_r) {};

    \node[big site, right=0.2 of s2_r] (s3_r) {};
    \node[draw, rounded corners, fit=(s3_r)] (cons) {};
    \node[anchor=south west, inner sep=0.3] at (cons.north west) (cons_lbl) {\tiny \sf Cons};

    \node[draw, fit=(s2_r)(cons)(cons_lbl)] (tri) {};
    \node[anchor=south west, inner sep=0.3] at (tri.north west) (tri_lbl) {\tiny \sf Try};

\node[big site, right=0.4 of s3_r] (s4_r) {};
    \node[draw, rounded corners, fit=(s4_r)] (fc) {};
    \node[anchor=south west, inner sep=0.3] at (fc.north west) (fc_lbl) {\tiny \sf FC};

\node[draw, fit=(sc)(sc_lbl)(fc)(fc_lbl)(tri)(tri_lbl)] (goal) {};
    \node[anchor=south west, inner sep=0.3] at (goal.north west) (goal_lbl) {\tiny \sf Goal};
    \node[big region, fit=(goal)(goal_lbl)] (r1) {};
  \end{scope}

\draw[->, dashed] (s1_r) to[out=90, in=90] (s1_l);
  \draw[->, dashed] (s2_r) to[out=-90, in=-90] (s2_l);
  \draw[->, dashed] (s3_r) to[out=-90, in=-90] (s2_l);
  \draw[->, dashed] (s4_r) to[out=90, in=90] (s3_l);

  \node[] at ($(lhs.east)!0.5!(rhs.west) + (0,-0.0)$) {$\rrul$} ;

\end{tikzpicture}
 	}
	\caption{$\mathtt{goal\_{persist\_{nil}}}$}
	\label{fig:rr_goal_persist_nil}
  \end{subfigure}
  \begin{subfigure}[b]{.45\linewidth}
	\centering
	\resizebox{1.0\linewidth}{!}{
		\begin{tikzpicture}[]
  \begin{scope}[local bounding box=lhs]
    \node[big site] (s1_l) {};
    \node[draw, rounded corners, fit=(s1_l)] (sc) {};
    \node[anchor=south west, inner sep=0.3] at (sc.north west) (sc_lbl) {\tiny \sf SC};

    \node[big site, right=0.2 of s1_l] (s2_l) {};

    \node[big site, right=0.2 of s2_l] (s3_l) {};
    \node[draw, rounded corners, fit=(s3_l)] (fc) {};
    \node[anchor=south west, inner sep=0.3] at (fc.north west) (fc_lbl) {\tiny \sf FC};

    \node[draw, red, fit=(sc)(sc_lbl)(s2_l)(fc)(fc_lbl)] (goal) {};
    \node[anchor=south west, inner sep=0.3] at (goal.north west) (goal_lbl) {\tiny \sf Goal};
    \node[big region, fit=(goal)(goal_lbl)] (r2) {};
  \end{scope}

  \begin{scope}[shift={(2.5,0)}, local bounding box=rhs]
    \node[big site] (s1_r) {};
    \node[draw, rounded corners, fit=(s1_r)] (sc) {};
    \node[anchor=south west, inner sep=0.3] at (sc.north west) (sc_lbl) {\tiny \sf SC};

\node[big site, right=0.3 of s1_r] (s2_r) {};

    \node[big site, right=0.2 of s2_r] (s3_r) {};
    \node[draw, rounded corners, fit=(s3_r)] (cons) {};
    \node[anchor=south west, inner sep=0.3] at (cons.north west) (cons_lbl) {\tiny \sf Cons};

    \node[draw, fit=(s2_r)(cons)(cons_lbl)] (tri) {};
    \node[anchor=south west, inner sep=0.3] at (tri.north west) (tri_lbl) {\tiny \sf Try};

\node[big site, right=0.4 of s3_r] (s4_r) {};
    \node[draw, rounded corners, fit=(s4_r)] (fc) {};
    \node[anchor=south west, inner sep=0.3] at (fc.north west) (fc_lbl) {\tiny \sf FC};

\node[draw, fit=(sc)(sc_lbl)(fc)(fc_lbl)(tri)(tri_lbl)] (goal) {};
    \node[anchor=south west, inner sep=0.3] at (goal.north west) (goal_lbl) {\tiny \sf Goal};
    \node[big region, fit=(goal)(goal_lbl)] (r1) {};
  \end{scope}

  \draw[->, dashed] (s1_r) to[out=90, in=90] (s1_l);
  \draw[->, dashed] (s2_r) to[out=-90, in=-90] (s2_l);
  \draw[->, dashed] (s3_r) to[out=-90, in=-90] (s2_l);
  \draw[->, dashed] (s4_r) to[out=90, in=90] (s3_l);

  \node[] at ($(lhs.east)!0.5!(rhs.west) + (0,-0.0)$) {$\rrul$} ;

\end{tikzpicture}
 	}
	\caption{$\mathtt{goal\_init}$}
	\label{fig:rr_goal_init}
  \end{subfigure}

	\caption{Reactions for declarative goals with priorities: $\rr{goal\_init} < \{ \rr{goal\_reduce},\rr{goal\_check}, \rr{goal\_fail}, \rr{goal\_suc} \} < \{ \rr{goal\_persist}, \rr{goal\_persist\_nil} \}$. }
	\label{fig:rr_goals}
\end{figure}

To reduce the number of reaction rules for encoding declarative goals, we check both success and failure conditions
simultaneously through reaction rule $\mathtt{goal\_{check}}$
(\cref{fig:rr_goal_check}). 
As before, the entailment machinery provides atomic
checks in both cases. 
Afterwards the reaction rules $\mathtt{goal\_{suc}}$ (\cref{fig:rr_goal_suc})
and $\mathtt{goal\_{fail}}$ (\cref{fig:rr_goal_fail}) determine if the goal
should complete (either successfully or with failure). 
Strictly speaking it is possible both success/failure conditions hold simultaneously, 
however in practice it is usually assumed success/failure conditions are mutually exclusive. 

An interesting feature of the \CAN derivation rule $G_{f}$ is the use of $?\mathit{false}$ in the resulting state.
This plays a similar role to $\ion{ReduceF}{}$ by explicitly creating an
irreducible term to trigger further handling up-the-tree. Recall that the belief entailment in derivation rule $?$ can be regarded as the special case of the derivation rule $act$ (\cref{sec:structuralEncoding}). 
Therefore, we simply let $\mathtt{goal\_{fail}}$
reduce to an $\ion{Act}{}$ with a \emph{false} precondition (that always
fails) to indicate a failure. 

\begin{lemma}(Faithfulness of $G_{s}$)
  $G_{s}$ has a corresponding finite reaction sequence $\encR{\langle\mathcal{B}, goal(\varphi_{s}, P, \varphi_{f})\rangle} \react^{+} \enc{\langle \mathcal{B}, nil\rangle}$.
\end{lemma}
\begin{proof}
	$\react^{+}$ corresponds to 
	$\xreact{\mathtt{goal\_{check}}} \xreacts{\{\mathtt{set\_{ops}}\}}\xreact{\mathtt{goal\_{suc}}}$. As the checks are finite, atomic, and over auxiliary bigraphical entities the number of reactions is finite and does not introduce additional branching as required.
\end{proof}

\begin{lemma}(Faithfulness of $G_{f}$)
  $G_{f}$ has a corresponding finite reaction sequence $\encR{\langle\mathcal{B}, goal(\varphi_{s}, P, \varphi_{f})\rangle} \react^{+} \enc{\langle \mathcal{B}, ?false\rangle}$.
\end{lemma}
\begin{proof}
	$\react^{+}$ corresponds to $\xreact{\mathtt{goal\_{check}}}\xreacts{\{\mathtt{set\_{ops}}\}}\xreact{\mathtt{goal\_{fail}}}$. As the checks are finite, atomic, and over auxiliary bigraphical entities the number of reactions is finite and does not introduce additional branching as required.
\end{proof}

Similar to the derivation rule $\rhd{;}$, the derivation rule $G_{;}$ reduces the left-branch of the symbol $\rhd$. 
The reaction $\mathtt{goal\_{reduce}}$ (\cref{fig:rr_goal_reduce}) pushes the reduction down the left-branch. To ensure the ordering between rules $G_{s}$, $G_{f}$, and $G_{;}$, we explicitly match only on the case that the checks have already been performed. In other words, the goal will only be pursued if neither the success or failure condition holds. 

\begin{lemma}(Faithfulness of $G_{;}$)
  $G_{;}$, has a corresponding finite reaction sequence $\encR{\langle\mathcal{B}, goal(\varphi_{s}, P_{1} \rhd P_{2}, \varphi_{f})\rangle} \react^{+} \enc{\langle \mathcal{B}', P_{1}' \rhd P_{2}\rangle}$.
\end{lemma}
\begin{proof}
  Finite reduction on
  $\encR{\mathcal{B}, P_{1}} \react^{+} \enc{\langle \mathcal{B}', P_{1}'\rangle} $.
\end{proof}

The derivation rule $G_{\rhd}$ likewise is very similar to the derivation rule $\rhd_{\bot}$. 
Unlike in rule $\rhd_{\bot}$, however, in rule $G_{\rhd}$ we
keep the $\rhd$ structure in-place and \emph{replicate} $P_{2}$, thus giving the declarative goals their \emph{persistence}. The can reaction rule $\mathtt{goal\_{persist}}$ (\cref{fig:rr_goal_persist}) encodes this case with the match of $\ion{ReduceF}{}$ ensuring
the premise $\langle\mathcal{B}, P_{1}\rangle \nrightarrow$ holds. 
Through duplication, we decouple the failure of the \emph{plan} execution from the failure of the \emph{goal} (as specified by success/failure conditions). 
Finally, an addition reaction $\mathtt{goal\_persist\_nil}$
(\cref{fig:rr_goal_persist_nil}) enables the agent to persists even in the the case where the program
executed successfully (but the goal success/failure did not hold).
\begin{lemma}(Faithfulness of $G_{\rhd}$)
  $G_{\rhd}$ has a corresponding finite reaction sequence $\encR{\langle\mathcal{B}, goal(\varphi_{s}, P_{1} \rhd P_{2}, \varphi_{f})\rangle} \react^{+} \enc{\langle \mathcal{B}, P_{2} \rhd P_{2}\rangle}$.
\end{lemma}
\begin{proof}
	$\react^+$ corresponds to $\xreact{\mathtt{goal\_{persist}}}$ \emph{or} $\xreact{\mathtt{goal\_persist\_nil}}$. Rules are mutually exclusive.
\end{proof}

The derivation rule $G_{init}$ is encoded through reaction $\mathtt{goal\_{init}}$ that sets up the required $\rhd$ structure. To ensure this is applied at the right time, we have priority classes with $\mathtt{goal\_init} < \{ \mathtt{goal\_{persist}}, \mathtt{goal\_{reduce}} \}$ to ensure the premise $P \neq P_{1} \rhd P_{2}$ holds.

\begin{lemma}(Faithfulness of $G_{init}$)
  $G_{init}$ has a corresponding finite reaction sequence $\encR{\langle\mathcal{B}, goal(\varphi_{s}, P, \varphi_{f})\rangle} \react^{+} \enc{\langle \mathcal{B}, P \rhd P\rangle}$.
\end{lemma}
\begin{proof}
	$\react^+$ corresponds to $\xreact{\mathtt{goal\_{init}}}$. Priority classes of reactions ensure $P \neq P_{1} \rhd P_{2}$ as required.
\end{proof}

For declarative goals, due to persistence, there is no additional rule that propagates failures upwards\footnote{Meeting the failure conditions does eventually lead to failure but requires additional steps.}.

As with the previous lemmas these, extended features can be integrated easily into \cref{theorem:faithful} to prove the extended semantics is also faithful. As the theorem is over agent-level transitions the theorem itself does not change, but the set of possible finite $\react^{+}$ sequences increases.

	\section{Examples: UAVs}\label{sec:case}
	
To illustrate our modelling and verification framework,
we consider three    examples taken from UAV surveillance and retrieval mission systems. The examples cover persistent patrol, concurrent sensing, and contingency handling in object retrieval and highlight   the three distinguishing features of   \CAN  : declarative goals, concurrency, and failure recovery.

\subsection{Persistent Patrol}

\begin{figure}[htbp]
  \scriptsize
		\begin{align*}
&\textbf{BDI Agent Design for Persistent Patrol}\\			
&\texttt{1 \ // Initial beliefs}\\
&\texttt{2 \  $ \neg  $battery\_low, $ \neg  $harsh\_weather}\\
&\texttt{3 \  // External events}\\
&\texttt{4 \  e\_init1}\\
&\texttt{5 \  // Plan library}\\
&\texttt{6 \ e\_init1 : true <- goal(false, e\_patrol\_task, false)} \\
&\texttt{7 \ e\_patrol\_task : true <- goal(sc, e\_patrol, false); e\_pause}\\
&\texttt{8 \ e\_patrol : true <- patrol}\\
&\texttt{9 \ e\_pause : battery\_low <- request; wait; charge}\\ 
&\texttt{10 e\_pause : harsh\_weather<- activate\_parking} \\ 
&\text{where} \ \texttt{sc} =  \texttt{harsh\_weather} \vee \texttt{battery\_low}.  \\
& 	\textbf{Bigraphical Encoding}\\
 &  	\mathsf{big} \ \mathsf{persistent\_patrol=} \\
	&\ion{Beliefs}{}.(\ionP{B}{}{1} \mprod \ionP{B}{}{2}) \pprod \ion{Desires}{}.\ion{E}{e1} \pprod \ion{Intentions}{}.1 \\
 &\pprod \ion{Plans}{}.( \\
	&\ion{PlanSet}{e\_{init1}}.(\ion{Plan}{}.(\ion{Pre}{}.1 \mprod \ion{PB}{}.\ion{Goal}{}.(\ion{SC}{}.\ion{False}{} \mprod \ion{E}{e\_{task1}} \mprod \ion{FC}{}.\ion{False}{}))) \\
& 	\mprod \ion{PlanSet}{e\_patrol\_task}.(\ion{Plan}{}.(\ion{Pre}{}.1 \mprod  \ion{PB}{}.(\ion{Seq}{}.(\ion{Goal}{}.(\ion{SC}{}.\ionP{B}{}{3} \mprod \ion{E}{e\_{patrol}} \mid \ion{FC}{}.\ion{False}{}) \mprod \ion{Cons}{}.\ion{E}{e\_{pause}})))
		) \\
& 	\mprod \ion{PlanSet}{e\_{patrol}}.(\ion{Plan}{}.(\ion{Pre}{}.1 \mprod \ion{PB}{}.\enc{patrol})
		) \\
& 	\mprod  \ion{PlanSet}{e\_{pause}}.( \\
 & 	\mprod \ion{Plan}{}.(\ion{Pre}{}.\ionP{B}{}{4} \mprod \ion{PB}{}.(\ion{Seq}{}.(\enc{request} \mprod \ion{Cons}{}.(\ion{Seq}{}.(\enc{wait} \mprod \ion{Cons}.\enc{charge}))))) \\
 & 	\mprod \ion{Plan}{}.(\ion{Pre}{}.\ionP{B}{}{5} \mprod \ion{PB}{}.\enc{activate\_parking})))  \\
	&\text{where} \ \mathsf{B(1)} = \neg \texttt{battery\_low}, \ \mathsf{B(2)} = \neg \texttt{harsh\_weather}, \ \mathsf{B(3)} = \texttt{sc} , \\ &\mathsf{B(4)} = \texttt{battery\_low}, \text{and } \mathsf{B(5)} = \texttt{harsh\_weather}. 
	\end{align*}
		\caption{Persistent Patrol: BDI Agent Design    and  Bigraphical Encoding.}
		\label{fig:patroldesign}
	\end{figure}
 
  UAVs are used in surveillance operations,  with a UAV patrolling a pre-defined area to identify objects of interest. The UAV can request refuelling when the battery is low, and parking mode should be activated when there is harsh weather.

The  agent design and its corresponding bigraphical encoding   is in \cref{fig:patroldesign}.
The external event \texttt{e\_init1} (line 4)   initiates persistent patrol. 
 There is only one plan (line 6) relevant to  	  \texttt{e\_init1},  whose context is always true (represented by an empty region bigraph $ \ion{1}{} $), 
	thus always applicable, and whose plan-body is  declarative goal 
  \texttt{goal(false,e\_patrol\_task,false)}.    
	The event \texttt{e\_patrol\_task} is persistent because the success and failure conditions never hold, that is, we have an infinite process executing \texttt{e\_patrol\_task}.
	In practice, we require some flexibility in case of low battery or harsh weather.  The plan for \texttt{e\_patrol\_task} (line 7)     indicates  the patrol task may need to be paused     (i.e.~followed by the event \texttt{e\_pause}),    when the success condition  is true, i.e.~when \texttt{battery\_low} or \texttt{harsh\_weather} holds  (added to the belief base). 
	If the pause is required and after achieving   event \texttt{e\_pause} (lines 9-10), the event \texttt{e\_patrol\_task} will be pursued again.
	For succinct presentation, we note that the encoding of action such as \texttt{patrol} and \texttt{wait} are not shown, but can be found in our model~\cite{models}.

	\subsection{Concurrent Sensing}\label{sec:concurrentsensingtasksexample}

\begin{figure}
	\scriptsize
	\begin{align*}
		&\textbf{BDI Agent Design for Concurrent Sensing}\\	
		&\texttt{1 \  // Initial beliefs}\\
		&\texttt{2 \ ram\_free, storage\_free}\\
		&\texttt{3 \  // External events}\\
		&\texttt{4 \  e\_init2}\\
		&\texttt{5 \  // Plan library}\\
		&\texttt{6 \ e\_init2 : true <- e\_dust|| e\_photo }\\
		&\texttt{7 \ e\_dust : ram\_free $ \wedge $ storage\_free <- collect\_dust; analyse; send\_back}\\
		&\texttt{8 \ e\_photo : ram\_free $ \wedge $ storage\_free <- focus\_camera; save\_shots; zip\_shots}\\
&\textbf{Bigraphical Encoding}\\
		& \mathsf{big \ concurrent\_sensing} = \\
		&\ion{Beliefs}{}.(\ionP{B}{}{6} \mprod \ionP{B}{}{7}) \pprod \ion{Desires}{}.\ion{E}{e\_{init2}} \pprod \ion{Intentions}{}.1 \\
	&\pprod  \ion{Plans}{}.(  \\
		&\ion{PlanSet}{e\_init2}.\ion{Plan}{}.(\ion{Pre}{}.1 \mprod \ion{PB}{}.(\ion{Conc}{}.(\ion{L}{}.\ion{E}{e\_{dust}} \mprod \ion{R}{}.\ion{E}{e\_{photo}})))     \\
 &\mprod  \ion{PlanSet}{e\_{dust}}.\ion{Plan}{}.(\ion{Pre}{}.(\ionP{B}{}{6} \mprod \ionP{B}{}{7}) \mprod \ion{PB}{}.(\ion{Seq}{}.(\enc{collect\_dust} \mprod \ion{Cons}{}.(\ion{Seq}{}.(\enc{analyse} \mprod \ion{Cons}{}.\enc{send\_back}))))) \\
&\mprod \ion{PlanSet}{e\_{photo}}.\ion{Plan}{}.(\ion{Pre}{}.(\ionP{B}{}{6} \mprod \ion{B}{}{7}) \mprod \ion{PB}{}.(\ion{Seq}{}.(\enc{focus\_camera} \mprod \ion{Cons}{}.(\ion{Seq}{}.(\enc{save\_shots} \mprod \ion{Cons}{}.\enc{zip\_shots})))))) \\
 & \text{where } \mathsf{B(6)}=\texttt{ram\_free} \text{ and } \mathsf{B(7)}=\texttt{storage\_free}.
	\end{align*}
	
\caption{Concurrent Sensing:  BDI Agent Design   and   Bigraphical Encoding.}
\label{fig:concurrentsensing}
\end{figure}

UAVs may also be used for sensing tasks. In this case we consider a UAV that analyses dust particles, and performs aerial photo collection, \eg for analysis in post volcanic eruptions.

An agent design to achieve this concurrent sensing task is in \cref{fig:concurrentsensing}.
The external event \texttt{e\_init2}  (line 4)   initiates the mission and  
	 the relevant plan (line 6) has   tasks  for dust monitoring  (\texttt{e\_dust}) and   photo collection (\texttt{e\_photo})  as the concurrent programs in the plan-body. The on-board dust sensors  require high-speed RAM  to collect and analyse the data, hence condition \texttt{ram\_free}, and when the analysis is complete, results are written to storage (hence condition \texttt{storage\_free}),  and sent back to the control. Similarly, to collect aerial photos,  the UAV     reserves and focuses the camera array (\texttt{focus\_camera}), 
then   camera shots are  
	     compressed (\texttt{zip\_shot}), and sent back. Recall, for successful completion, both concurrent tasks have to   complete successfully.

	\subsection{Contingency Handling for a Retrieve Task}
  \begin{figure}
					\footnotesize
		\begin{align*}
		&\textbf{BDI Agent Design of Retrieve Task}\\	
		&\texttt{1 \ // Initial beliefs}\\
		 &\texttt{2 \  $ \neg  $sensor\_malfunc, $ \neg  $engine\_malfunc}\\
		 &\texttt{3 \  // External events}\\
		 &\texttt{4 \  e\_retrv}\\
		 &\texttt{5 \  // Plan library}\\
		 &\texttt{6 \ e\_retrv : $ \varphi $ <- take\_off; goal(at\_destination, e\_path1, fc); retrieve}\\
		 &\texttt{7 \ e\_retrv : $ \varphi $ <- take\_off; goal(at\_destination, e\_path2, fc); retrieve}\\
		&\texttt{8 \ e\_retrv : $ \varphi $ <- take\_off; goal(at\_destination, e\_path3, fc); retrieve}\\
		 &\texttt{9 \ e\_retrv : sensor\_malfunc <- return\_base}\\
		 &\texttt{10 e\_retrv : engine\_malfunc <-  activate\_parking; send\_GPS}\\
		 &\texttt{11 e\_path1 : true <-  navigate\_path\_1}\\
		&\texttt{12 e\_path2 : true <-  navigate\_path\_2}\\
		 &\texttt{13 e\_path3 : true <-  navigate\_path\_3}\\
		 &\text{where} \ \varphi  = \neg\texttt{sensor\_malfunc}  \wedge  \neg\texttt{engine\_malfunc},
		 \texttt{fc} =  \texttt{sensor\_malfunc} \vee \texttt{engine\_malfunc}  \\
&\textbf{Bigraphical Encoding}\\
	&\mathsf{big \ retrieve\_task =} \\
&\ion{Beliefs}{}.(\ionP{B}{}{8} \mprod  \ionP{B}{}{9}) \pprod \ion{Desires}{}.\ion{E}{e\_{retrv}} \pprod \ion{Intentions}{}.1  \\
	&\pprod \ion{Plans}{}.( \\
&\ion{PlanSet}{e\_{retrv}}.(  \\
 &\ion{Plan}{}.(\ion{\ion{Pre}{}}{}.(\ionP{B}{}{8} \mprod \ionP{B}{}{9}) \mprod \ion{PB}{}.(\ion{\ion{Seq}{}}{}.(\enc{\mathit{take\_off}} \mprod \ion{\ion{Cons}{}}{}.(\ion{\ion{Seq}{}}{}.(\ion{\ion{Goal}{}}{}.(\ion{\ion{SC}{}}{}.\ionP{B}{}{10} \mprod \ion{E}{e\_{path1}} \mprod  \ion{\ion{FC}{}}{}.\ion{B}{}{11}) \mprod \ion{\ion{Cons}{}}{}.\enc{retrieve})))))  \\
 &   \mprod \ion{Plan}{}.(\ion{Pre}{}.(\ionP{B}{}{8} \mprod \ionP{B}{}{9}) \mprod \ion{PB}{}.(\ion{Seq}{}.(\enc{\mathit{take\_off}} \mprod \ion{Cons}{}.(\ion{Seq}{}.(\ion{Goal}{}.(\ion{SC}{}.\ionP{B}{}{10}) \mprod \ion{E}{e\_{path2}} \mprod  \ion{FC}{}.\ionP{B}{}{11} \mprod \ion{Cons}{}.\enc{retrieve}))))) \\
&  \mprod \ion{Plan}{}.(\ion{Pre}{}.(\ionP{B}{}{8} \mprod \ionP{B}{}{9}) \mprod \ion{PB}{}.(\ion{Seq}{}.(\enc{\mathit{take\_off}} \mprod \ion{Cons}{}.(\ion{Seq}{}.(\ion{Goal}{}.(\ion{SC}{}.\ionP{B}{}{10} \mprod \ion{E}{e\_{path3}} \mprod  \ion{FC}{}.\ionP{B}{}{11}) \mprod \ion{Cons}{}.\enc{retrieve})))))  \\
&	\mprod \ion{Plan}{}.(\ion{Pre}{}.\ionP{B}{}{12}) \mprod \ion{PB}{}.\enc{return\_base})  \\
&  \mprod \ion{Plan}{}.(\ion{Pre}{}.\ionP{B}{}{13}) \mprod \ion{PB}{}.(\ion{Seq}{}.(\enc{activate\_parking} \mprod  \ion{Cons}{}.\enc{send\_GPS}))))  \\
&  \mprod  \ion{PlanSet}{e\_{path1}}.\ion{Plan}{}.(\ion{Pre}{}.1 \mprod \ion{PB}{}.(\enc{navigate\_path\_1}))  \\
&  \mprod \ion{PlanSet}{e\_{path2}}.\ion{Plan}{}.(\ion{Pre}{}.1 \mprod \ion{PB}{}.(\enc{navigate\_path\_2})) \\
&  \mprod \ion{PlanSet}{e\_{path3}}.\ion{Plan}{}.(\ion{Pre}{}.1 \mprod \ion{PB}{}.(\enc{navigate\_path\_3})))\\
 & \text{where } \mathsf{B(8)}=\neg\texttt{sensor\_malfunc}, \ \mathsf{B(9)}=\neg\texttt{engine\_malfunc}, \ \mathsf{B(10)}=\texttt{at\_destination}, \ \mathsf{B(11)}=\texttt{fc}, \\
 &\mathsf{B(12)}=\texttt{sensor\_malfunc},  \text{and }   \mathsf{B(13)}=\texttt{engine\_malfunc}.
			\end{align*}
	\caption{Retrieval Contingency: BDI Agent Design   and   Bigraphical Encoding.}
	\label{fig:retrievetask}
\end{figure}

UAVs may be used for object retrieval tasks, \eg package delivery.
An agent design for retrieval is in~\cref{fig:retrievetask}. It has one (retrieve) task, initiated by   external event \texttt{e\_retrv} (line 4), which may be affected  by engine or sensor malfunction,  
Event \texttt{e\_retrv} is handled by five relevant plans available (lines 6 to 10). 
The first 3 plans  provide different flight paths after take-off,    in which case   the failure condition   is  (subsequent)  engine or sensor malfunction.  The last 2 plans (line 9 and 10) indicate safe recovery in the event of engine or sensor malfunction.

	\section{Properties}
	\label{sec:caseproperties}

To verify the designs, we generate a transition system from the BRS representing the agents (and their semantics). The transistion system has bigraphs as states and reactions as transitions.
We can reason about  static properties using  {\em bigraph patterns}~\cite{benford2016lions}  and dynamic properties     using linear or branching time temporal logics such as Computation Tree logic (CTL)~\cite{clarke1981design}, which we use in our examples.  

\subsection{Bigraph patterns}
  Bigraph patterns are predicates on states:  if the pattern {\em matches} the current state then    the predicate is true.

We have found the  bigraph patterns most useful for reasoning about BDI agents   are often   a fragment of the right-hand side of reactions, i.e.~they check that  a desired  or anticipated  operation has taken place. 
For example, consider the state predicate: 
there is  a  declarative  goal  corresponding  to     event \texttt{e\_patrol\_task} (i.e. \texttt{goal(false, e\_patrol\_task, false)}). The   bigraph pattern is 
  $$ \mathsf{Goal.(SC.(False \mid id) \mid FC.(False \mid  id) \mid Try.id) }$$
  where $ \mathsf{SC}$  is the success condition, $ \mathsf{FC}$ the failure condition, and $ \mathsf{Try}$ the plan choice $ \rhd $.
 The presence of $ \mathsf{Try}$   indicates that   event \texttt{e\_patrol\_task} is within the given declarative goal and has been reduced to its set of relevant plans, from which an applicable plan is selected, according to the right-hand side of the reaction given in~\cref{fig:rr_select_plan_T}.
  As long as $ \mathsf{Try}$ is present (regardless of what is under it, i.e. $ \mathsf{Try.id}$),   the declarative goal is being pursued.

 \subsection{Example   properties}\label{sec:CTLproperties}

{\bf Example 1} (Persistent patrol). 
    A key property is  that  the  goal corresponding to  event \texttt{e\_patrol\_task}  is     persistent:   
 	$ \mathbf{A} [\mathbf{G} \ \mathbf{F}\varphi_{1}]$,  where $$ \varphi_{1} \defeq \mathsf{Goal.(SC.(False \mid id) \mid FC.(False \mid  id) \mid Try.id) }$$
	 As expected the property holds.

{\bf Example 2} (Concurrent sensing) 
A useful property to investigate    is whether it is possible to complete both sensing tasks regardless of their interleaving.   
Recall that in \CAN semantics, whenever an intention is completed or fails, the agent will simply remove it from the intention base ($A_{update}$~\cref{fig:rr_agent}). 
Therefore, to make sure that an intention is successfully achieved, we have to ensure that a given intention is indeed removed only \emph{after} being completed successfully. 
We denote the bigraph pattern for the successful completion of a given intention as $ \varphi_{2} \defeq \mathsf{Intent}.\mathsf{1}$, the failure of completion of an intention $ \varphi_{3} \defeq \mathsf{Intent}.\mathsf{ReduceF}$, and the removal of an intention from intention base $ \varphi_{4} \defeq \mathsf{Intentions}.\mathsf{1}$. 
 Bigraph pattern $\varphi_{4}$     specifies the removal of an intention from the intention base if and only if it is the only   intention in the   base.  If there is more than one intention,  it is impossible to reason about which intention is removed    because there are no intention identifiers in CAN. We reflect on this lack in~\cref{sec:no_meta_reasoning}.
Given the bigraph patterns above, the property is   $ \mathbf{A} [ \mathbf{F} (\varphi_{2} \wedge \mathbf{X} \varphi_{4})] $.
This property is false, and we find that   $ \mathbf{E} [ \mathbf{F} (\varphi_{3} \wedge \mathbf{X}\varphi_{4}) ] $ holds (i.e. there exists a path for which  eventually the intention is removed after being failed).
This is because concurrency can introduce    undesirable  race conditions. 
	For example,  the action \texttt{send\_back} in line 7 needs to be executed  before the action \texttt{save\_shots} is executed,   to free  required  storage.   
	This example highlights the benefits of a formal model for analysis    at design time.
	
{\bf Example 3} (Contingency handling)	
Similar to  example 2,  a desirable property is that regardless of any malfunction,  the intention for event \texttt{e\_retr} is removed after   successful completion.  The property is  $ \mathbf{A} [ \mathbf{F} (\varphi_{2} \wedge \mathbf{X} \varphi_{4})] $ and it holds. 

Before we give the results of the above property verification, we recall that our bigraph encoding introduces intermediate states that do not correspond to an agent step. Therefore, the operator  $\mathbf{X}$   (\emph{next}) has to be used carefully; some properties may require modification, because \eg the \emph{next} operator refers to the next internal state, not the next agent state.
For example, there may be belief checks  between    agent steps.  
In the  examples 2 and 3 above, no modifications were required.

\subsection{Results}
  For automatic verification we exported the transition system to the PRISM model-checker\footnote{Currently the only model-checker format supported by BigraphER.} by assuming all transitions occur with equal probability. The size of transition system generated by BigraphER and verification times are as follows: 
	
	\begin{table}[h]
    \centering
	\footnotesize
\begin{tabular}{@{}lllll@{}}
\textbf{Example}       & \textbf{States} & \textbf{Transitions} & \textbf{Build time (s)} & \textbf{Ver. time (s)} \\
  \toprule
Persistent patrol    & 239            & 287                 & 21.92              & 0.081                     \\
Concurrent sensing    & 731            & 879                & 61.16              & 0.01                      \\
Contingency handling & 644            & 922                 & 238.53             & 0.002                    
\end{tabular}
\end{table}

While  contingency handling has fewer states/transitions than the concurrent sensing example, it takes more time to generate the transition system. We attribute this to the former containing    
 more  bigraph     entities.

\section{ Reflections on \CAN}
\label{sec:reflections}
  We reflect on the insights gained into the \CAN language through the process of building the   bigraph model. 
We stress that this should not be taken as criticism of \CAN.
On the contrary, we hope to show that the explicitness of the bigraph encoding is useful, \eg to show areas of semantics with too much (resp. too little) information,  and to aid in the continuous advancement of BDI family languages, in particular, from the point of verification and validation.

\subsection{Modularity in Semantics of \CAN}
A unique characteristic of \CAN is that it has a modular operational semantics that separates how to evolve an intention (i.e. the intention-level semantics) from how to evolve the whole agent (i.e.the agent-level semantics).
This approach has its merits,
for example, we can easily extend or modify one side of the semantics (e.g. the agent-level) without altering the other one.
This was illustrated when adding the concurrency and declarative goals extensions (\cref{sec:extended}). The extensions only change intention-level steps, and as such, do not affect the overall faithfulness theorem as this is defined over agent-level steps. 

The two-levels of semantics could be useful for verification. For example, we
may consider \emph{only} the agent-level transitions which would give snapshots
of the agent state, without any information on \emph{how} choices were made. In
the bigraph model we do not make a distinction between agent-level and
intention-level transitions, and both appear in the resulting output. The
bridging of agent-level and intention-level transitions is performed by the
introduction of the $\ion{Reduce}{}$ entity. For example, the reaction
rule~$\mathtt{intention\_{step}}$ that encodes the agent-level derivation
rule~$A_{steps}$ introduces a $\ion{Reduce}{}$ entity to an intention,
requesting it to be reduced according to the intention-level semantics. The use
of instantaneous reaction rules, that do not show up in the resulting transition
system, would allow only agent-level steps to be analysed without changes to the
reaction rules themselves.

\subsection{Inconsistency of Semantics in \CAN Literature}

In the literature, there are subtle differences between definitions of the \CAN semantics. 
In particular, between~\cite{sardina:hierarchical} and~\cite{sardina:agent}.
For example, consider the $\triangleright_\bot$ rule from the two works above:
	\begin{center}
			$ \dfrac{P_{1} \neq nil \ \ \ \langle \mathcal{B},  P_{1}   \rangle \nrightarrow }{\langle \mathcal{B},  P_{1} \rhd P_{2}\rangle \rightarrow \langle \mathcal{B},  P_{2}\rangle} $ \ \ \  $ \rhd_{\bot} \text{ in~\cite{sardina:hierarchical}} $
\end{center}
\begin{center}
		$ \dfrac{P_{1} \neq nil \ \ \ \langle \mathcal{B},  P_{1}   \rangle \nrightarrow \ \ \ \langle \mathcal{B},  P_{2}\rangle \rightarrow \langle \mathcal{B}',  P'_{2}\rangle}{\langle \mathcal{B},  P_{1} \rhd P_{2}\rangle \rightarrow \langle \mathcal{B}',  P'_{2}\rangle} $ \ \ \  $ \rhd_{\bot} \text{ in~\cite{sardina:agent}}$
	\end{center}
	
In~\cite{sardina:hierarchical}, the rule $\rhd_{\bot}$ is only dependent upon the irreducibility of the program  $P_{1}$. 
However, in~\cite{sardina:agent}, not only is it dependent upon the irreducibility of  $P_{1}$, but, within the same operation, the reducibility of $P_{2}$.

This change is significant as, in the first case, we wait to do failure recovery. This can allow the current belief base to be updated before selecting a new plan (in all cases $P_2$ has the form $e:(|\Delta|)$). In the second case there is no scope to wait for belief base changes.

It is not immediately clear which approach is better in practice. One benefit of a formal model is that we can begin to unpick these questions by substituting the current $\rr{try\_failure}$ reaction for a modified version.

\subsection{Redundant Event Names}

The \CAN language includes the form $e : (|\Delta|)$ representing a set of relevant plans which can be used to address the event~$e$. This set is updated as plans are selected and executed. 
For example, when an applicable plan is selected (i.e. $ \varphi:P \in \Delta $ and $\mathcal{B} \models \varphi$), it will be removed from the set of remaining plans (i.e. $e:(\mid \Delta \setminus \{\varphi:P\} \mid)$). However, after a set of relevant plans is selected from the plan library, the event name $e$ becomes \emph{redundant} in the sense that it is never used by any \CAN semantic rules.
This is seen clearly in the bigraph model, where only $\rr{reduce\_event}$ (in~\cref{fig:rr_reduce_event}) utilises the event name link. Other rules always match the event name as open (connected to 0 or more other entities).
This suggests that the form of plans within the plan library, and those within intentions should be different, \eg $e:(|\Delta|)$ and $(|\Delta|)$.

\subsection{No Difference between Intention Success and Failure}
As a high-level planning language, \CAN remains agnostic to many important issues in practice.
One such issue is the inability to tell if an intention completed successfully, or with a failure. 
The derivation rule $A_{update} $ in~~\cref{fig:agentCANSemantics} simply removes a completed intention from the intention base, namely an intention $nil$ or one that is failed and cannot make any further transition. 
Therefore, the completion of an intention is not equivalent to the achievement of an intention. 
To verify the achievement of an intention (which in practice is the most important property to check), we also have to ensure that its completion is not due to the failure.
This is precisely how we verify the achievement of an intention in~\cref{sec:CTLproperties}. 
Therefore, in our bigraph model, we have to encode the derivation rule~$A_{update} $ into two cases, namely~$ \mathtt{intention\_done\_F}$ for failure case and~$ \mathtt{intention\_done\_succ}$ for success case. 

\subsection{Oracle for Failure}
  Failure in    \CAN semantics is denoted by $ \langle \mathcal{B},  P_{1}   \rangle \nrightarrow $  as the premiss in the related derivation rule (e.g. $\rhd_\bot$). 
Therefore, to be able to apply the rule~$\rhd_\bot$, the agent somehow can ``look-ahead" to the result of the inner-reduction, \ie there is some oracle that determines if the inner-reduction is possible.
However, in practice (\eg our bigraph encoding), no oracle exists, and the agent has to explicitly  to try progress a step to see if it reduces.
In others words, unlike the derivation rules which, to some extent, have the impression the failure occurs via one single rule~$ \langle \mathcal{B},  P_{1}   \rangle \nrightarrow $, it actually involves a strict partial execution of other rules. 
It can be clearly seen in~\cref{fig:reduction_example} where, before an action is deemed as un-executable, its pre-condition has to be actually checked to be false according to the belief base. 

\subsection{Absence of Meta-level Reasoning}\label{sec:no_meta_reasoning}

While it is possible to reason about agents when only a single intention is involved -- for example through checking a property that checks if an intention failed before it was removed (see \cref{sec:CTLproperties}) -- these approaches do not apply when there are more than one concurrent intentions.
The main issue here is that intentions lack identifiers.
If we want to stay within the semantics in \CAN, approaches to identifying specific intentions include (1) fixing the last program within any intention
to be unique to allow checking when this specific program is removed, or (2) ensuring actions add unique beliefs however this requires knowing ahead of time the actions that will be executed in success/failure cases.

Ideally, intentions would have unique identifier to aid verification.
Adding such an identifier is straightforward by replacing $P \in \Gamma$ with $ \langle \mathit{identifier}, P \rangle \in \Gamma$.
As such, to track an intention is removed, we simply have a bigraph pattern $\mathsf{Intent}.\mathsf{(identifier \mprod id)} $ where $\mathsf{identifier}$ is the identifier of the intention and $\mathsf{id}$ the site that abstracts away specific details of the intention.
We argue that by indexing the intentions and further labelling its status (e.g. active or suspended), the meta-level reasoning can be powerful for next level of agent verification, in particular, in the context of interacting with users.

Another area where keeping meta-information available is useful is to allow tracking events to the intentions that are handling those events.
In the current semantics, when an event is processed (by $A_{event}$) it is removed completely from the desires structure and replaced by the set of relevant plans within an intention.
As we execute the plans we lose track of which event $e$ generated that intention (i.e. the means-end relations).

\section{Related Work}\label{sec:related}
  Reasoning about BDI agents
through model checking    has been well
explored.
A key work in this area~\cite{bordini2006verifying} reports a translation of AgentSpeak programs
to both the Promela modelling language and Java, and shows how to apply the Spin~\cite{holzmann1997model} model checker and Java PathFinder program model checker to verify the agents.
Comparing the translation experience of these two, Java stood out as the most promising approach (as a general purpose language) compared to Promela (often used for the verification of communication protocols).
Many have built upon this Java-based verification approach.
In particular, recent work implements a BDI agent programming infrastructure as a set of Java classes -- the Agent Infrastructure Layer (AIL) \cite{dennis2008flexible}.
The Gwendolen BDI language~\cite{dennis2017gwendolen} provides the default semantics for the AIL, and is designed with verification in mind by including extra book-keeping and transition rules that purely assist verification.

The AIL has been further developed~\cite{dennis2012model} to support the verification of \emph{heterogeneous} multi-agent systems by allowing different agent programming languages to be used within the same AIL framework.
Although the AIL supports heterogeneous agents, to date the BDI programming languages implemented in the AIL~\cite{dennis2018mcapl} is tightly bounded to Gwendolen and its extensions (e.g.~\cite{dennis2016formal}) along with another language named GOAL~\cite{hindriks2000agent}.
Crucially, what these approaches verify is the \emph{implementation}  of  a given language.
The faithfulness of the implementation to the language semantics is often omitted for convenience.
Utilising Java PathFinder (and its enhanced version~\cite{dennis2012model})
has the advantage of bypassing the need of a mathematical model by deriving the model directly from the program code. However it typically suffers from  a significant performance bottleneck due to the symbolic execution of Java bytecode.

Agent properties for AIL are usually specified in linear temporal logic (LTL)~\cite{emerson1990temporal},
There is, however, an exception in \cite{dennis2018two} where the model generated by Java PathFinder is converted to the input language of PRISM~\cite{kwiatkowska2011prism} to, e.g. provide access to probabilistic property specification.
Unfortunately the conversion to PRISM does not maintain direct link between the implemented program and the model being verified, \eg it might be difficult to reflect back into the application when creating counter examples. 
 
The two main BDI languages implemented in AIL are Gwendolen and GOAL.
Unlike main-stream BDI programming languages, \eg AgentSpeak, GOAL is a pure reactive system and does not select pre-defined \emph{plans} from a library but instead selects individual \emph{actions}.
Like \CAN, Gwendolen handles declarative goals, failure recovery and concurrency with some differences.
In Gwendolen, declarative goals make statements about the beliefs the agent wishes to hold and remains a goal until the agent gains the appropriate beliefs. As such, the declarative information in Gwendolen is only carried for the initial goal of the intention, no declarative information is carried for any of its active sub-events. For example, if beliefs sought hold, the sub-event will still be executed to the end. Meanwhile, in CAN, the declarative information is carried for any stage of evolution of programs in the declarative goal. Once the success condition holds, the related program is halted immediately.  
Gwendolen does not allow goal failure conditions so is unable to decouple goal failure from plan failure.
For failure recovery, Gwendolen is explicitly programmed with the appropriate plan revision rules (as meta-level rules) which specify a prefix of the current plan to be dropped and replaced by another.
Finally, concurrency in Gwendolen is only allowed in the intention level, and, by default, is conducted in round-robin fashion to manage interleaving.

Recent work continues to extend the verification framework above, improving the efficiency of verification and supporting richer property specifications.
Program slicing has been used to increase verification efficiency by removing parts of the agent program that cannot affect the properties of interest~\cite{winikoff2018slicing}.
Meanwhile, by simplifying the structure and execution of AgentSpeak (deviating from mainstream BDI agents), it can facilitate the verification of probabilistic and time bound properties through PRISM~\cite{izzo2016stochastically}.
Finally, there is promising progress to verify the hybrid autonomous system in which the high-level is discrete logic-based framework (modelled by BDI agents) and the low-level is a continuous control system~\cite{dennis2016practical}.

Bigraphs have been shown~\cite{milner2009space} to be    suitable for  encoding  process algebras such as CCS~\cite{milner2006pure}, Mobile Process~\cite{jensen2006mobile}, $ \pi $-calculus~\cite{bundgaard2006typed}, and Actors~\cite{memo2014}. 
Recently, there is also a growing trend to specify and verify   agent-based systems via bigraphs, in particular, multi-agent systems. 
However, most of them still remain at the stage of proof of concept.
For example, the work~\cite{mansutti2014multi} proposes a methodology for modelling and simulating multi-agent systems via bigraphs.
The core idea is that the containment relation of bigraphs mirrors the administrative relations of agents while reaction rules model agent reconfigurations, e.g.~bigraph destruction translates into agent termination.
One work that is perhaps closest  to ours is~\cite{dib2015model}, which   also models BDI agents via bigraphs.
However, it  considers  multi-agent systems, and treats the internal reasoning of each BDI agent as a black box.
As a result, they provide no details regarding how the agents behave in an  environment.

\section{Future Work}\label{sec:Future Work}
The encoding of BDI agents  in bigraphs is our first step  laying out a foundation for  more advanced reasoning.
As future work, we have in mind two extensions: probabilistic reasoning, in particular, plan selection and intention trade-off, and dynamic environments.

In general, there may be several applicable plans which achieve a given event.
The agent has to select one 
and  it may be desirable to  specify what is  ``most appropriate”   at that time, which may depend upon  different, and possibly domain-specific characteristics, e.g.~cost and preference.
Additionally, the agent may be  pursuing a set of concurrent intentions,~i.e.  there is concurrency between the top-level external events. 
Similar to the plans, intention, it may be desirable to again specify  ``most appropriate” e.g.~more urgent.   

We will  develop a more nuanced approach to handling   plan selection and intention scheduling by assigning weights (to plans and events). These will be encoded by reaction rules with weights  using probabilistic bigraphs~\cite{prob2020} that export Discrete-time Markov chains (DTMCs).

Our current encoding of BDI agents is limited to a self-static environment, i.e.~the environment changes only when the agent changes it. 
We plan to develop a self-dynamic environment  and   will extend the mechanism of   failure recovery   to allow   re-selection of   previously failed plans.
This will not only increase the persistence of an agent, but also increase the likelihood of success by taking advantage of  environmental changes.

\section{Conclusion}\label{sec:conclusions}

Rational agents, such as Belief-Desire-Intention (BDI) agents, will play a key
role in future autonomous systems and it is essential we can reason about their
behaviours, and provide early, \ie design-time, indications of potential
problems, \eg deadlocks causes by shared resources.

We have presented a framework, based on Milner's bigraphs, for modelling and
verifying BDI agents specified in the \CAN language. We believe this is the
first \emph{executable semantics} of \CAN, allowing verification of abstract
agent programs, rather than verification based on a specific
 implementation of \CAN. 
The use of four perspectives in the bigraph model:  Belief, Desire, Intention and Plan, helps is to separate concerns in the encoding and offers a clear visualisation of the resulting   model. 

The two key functions are the syntax encoding  $\enc{\cdot}$ and  $\encR{}$ that enables the behavioural encoding.  The former  has the added feature of   introducing  event indices for plans, which decreases errors and aids search.  The latter is a bridge between agent-level and intention-level steps.     Bigraph parallelism  indicates  how the belief base is the environment   for reduction and conditional bigraphs allow us to prioritise reaction rules, which simplifies the encoding.   

We have shown that the encoding of \CAN
agents in bigraphs is faithful by proving any \CAN step is captured by a finite
sequence of bigraph reaction rules, and we have shown shown   the approach is practical
through three example UAV applications. In each case, generating and verifying
the model took no more than a few minutes.

This work has also highlighted many interesting features of the current semantics of \CAN, such as the inability to distinguish between the success and failure
of an intention and  lack of meta-level reasoning, and   it lays the foundation for future modelling work. We envisage
an extended model (and hence extended semantics) that features probabilistic
choice and dynamic environments -- allowing quantitative model checking of agent
programs.

	\section*{Acknowledgements}
	
	This work is supported by the Engineering and Physical Sciences Research
	Council, under PETRAS SRF grant MAGIC (EP/S035362/1) and  S4: Science of Sensor Systems Software. (EP/N007565/1)

\bibliographystyle{IEEEtran}
	\bibliography{paper}

\end{document}